\documentclass[twocolumn]{aastex63}

\usepackage{graphicx}	% Including figure files

\usepackage{multirow}

\let\tablenum=\undef % This fixes an incompatibility between aastex and siunitx
\usepackage{siunitx}
\newcommand\chem[1]{\ensuremath{\mathrm{#1}}}
\newcommand\ha{\ensuremath{\mathrm{H\alpha}}}
\newcommand\hb{\ensuremath{\mathrm{H\beta}}}
\newcounter{ionstage}
\renewcommand{\ion}[2]{\setcounter{ionstage}{#2}% 
  \ensuremath{\mathrm{#1\,\scriptstyle\Roman{ionstage}}}}
\newcommand\oiii{[\ion{O}{3}]}
\newcommand\siii{[\ion{S}{3}]}
\newcommand\nii{[\ion{N}{2}]}
\newcommand\wav[1]{\ensuremath{\lambda #1}}
\newcommand\Te{\ensuremath{T_{\mathrm{e}}}}
\newcommand\BG{\ensuremath{_{\mathrm{BG}}}}
\def\th#1#2{\ensuremath{\theta^{#1}\,\text{Ori~#2}}}

\received{XXX}
\revised{YYY}
\accepted{ZZZ}
\submitjournal{ApJ}

\shorttitle{HH~204 in the Orion Nebula}

\shortauthors{M\'endez-Delgado et al.}

\graphicspath{{./}{figures/}}

\catcode`\*=11
\makeatletter
\let\longtable*\@undefined
\let\endlongtable*\@undefined
\makeatother
\catcode`\*=12

\begin{document}

\title{Photoionized Herbig-Haro objects in the Orion Nebula through deep high-spectral resolution spectroscopy II: HH~204}

\correspondingauthor{Jos\'e E. M\'endez-Delgado}
\email{jemd@iac.es}

\author[0000-0002-6972-6411]{J. E. M\'endez-Delgado}
\affiliation{Instituto de Astrof\'isica de Canarias (IAC), E-38205 La Laguna, Spain}
\affiliation{Departamento de Astrof\'isica, Universidad de La Laguna, E-38206 La Laguna, Spain}

\author[0000-0001-6208-9109]{W. J. Henney}
\affiliation{Instituto de Radioastronom\'ia y Astrof\'isica, Universidad Nacional Aut\'onoma de M\'exico, Apartado Postal 3-72, 58090 Morelia, Michoac\'an, Mexico}

\author[0000-0002-5247-5943]{C. Esteban}
\affiliation{Instituto de Astrof\'isica de Canarias (IAC), E-38205 La Laguna, Spain}
\affiliation{Departamento de Astrof\'isica, Universidad de La Laguna, E-38206 La Laguna, Spain}

\author[0000-0002-6138-1869]{J. Garc\'ia-Rojas}
\affiliation{Instituto de Astrof\'isica de Canarias (IAC), E-38205 La Laguna, Spain}
\affiliation{Departamento de Astrof\'isica, Universidad de La Laguna, E-38206 La Laguna, Spain}

\author[0000-0003-3776-6977]{A. Mesa-Delgado}
\affiliation{Calle Camino Real 64, Icod el Alto, Los Realejos, 38414, Tenerife, Spain}

\author[0000-0002-2644-3518]{K. Z. Arellano-C\'ordova}
\affiliation{Department of Astronomy, The University of Texas at Austin, 2515 Speedway, Stop C1400, Austin, TX 78712, USA}

%\nocollaboration{1}

%\collaboration{1}{(LaTeX collaboration)}

%\nocollaboration{2}

%% Note that the \and command from previous versions of AASTeX is now
%% depreciated in this version as it is no longer necessary. AASTeX 
%% automatically takes care of all commas and "and"s between authors names.

%% AASTeX 6.3 has the new \collaboration and \nocollaboration commands to
%% provide the collaboration status of a group of authors. These commands 
%% can be used either before or after the list of corresponding authors. The
%% argument for \collaboration is the collaboration identifier. Authors are
%% encouraged to surround collaboration identifiers with ()s. The 
%% \nocollaboration command takes no argument and exists to indicate that
%% the nearby authors are not part of surrounding collaborations.

%% Mark off the abstract in the ``abstract'' environment. 
\begin{abstract}
We analyze the physical conditions, chemical composition and other  properties of the photoionized Herbig-Haro object HH~204 through Very Large Telescope (VLT) echelle spectroscopy and Hubble Space Telescope (\textit{HST}) imaging. 
We kinematically isolate the high-velocity emission of HH~204 from the emission of the background nebula and study the sub-arcsecond distribution of physical conditions and ionic abundances across the HH object.
We find that low and intermediate-ionization emission arises exclusively from gas at photoionization equilibrium temperatures,
whereas the weak high-ionization emission from HH~204 shows a significant contribution from higher temperature shock-excited gas.
%The spatial thickness of the high-temperature zone is marginally resolved with \textit{HST} but is unresolved in our spectra.
We derive separately the ionic abundances of HH~204, the emission of the Orion Nebula and the fainter Diffuse Blue Layer.
% For O$^{+}$, O$^{2+}$, N$^{+}$, Ne$^{2+}$, S$^{+}$, S$^{2+}$, Cl$^{+}$, Cl$^{2+}$, Ar$^{2+}$, Ar$^{3+}$, Fe$^{+}$, Fe$^{2+}$, Fe$^{3+}$, Ni$^{+}$, Ni$^{2+}$, Ca$^{+}$ and Cr$^{+}$ these abundances are based on collisionally excited lines (CELs), while those of He$^{+}$, O$^{+}$, O$^{2+}$ and C$^{2+}$ are based on recombination lines (RLs). 
In HH~204, the O$^{+}$ abundance determined from  Collisional Excited Lines (CELs) matches the one based on Recombination Lines (RLs), while the O$^{2+}$ abundance is very low, so that the oxygen abundance discrepancy is zero.
The ionic abundances of Ni and Fe in HH~204 have similar ionization and depletion patterns, with total abundances that are a factor of 3.5 higher than in the rest of the Orion Nebula due to dust destruction in the bowshock.
We show that a failure to resolve the kinematic components in our spectra would lead to significant error in the determination of chemical abundances (for instance, 40\% underestimate of O), mainly due to incorrect estimation of the electron density.
\end{abstract}

%% Keywords should appear after the \end{abstract} command. 
%% See the online documentation for the full list of available subject
%% keywords and the rules for their use.

%% From the front matter, we move on to the body of the paper.
%% Sections are demarcated by \section and \subsection, respectively.
%% Observe the use of the LaTeX \label
%% command after the \subsection to give a symbolic KEY to the
%% subsection for cross-referencing in a \ref command.
%% You can use LaTeX's \ref and \label commands to keep track of
%% cross-references to sections, equations, tables, and figures.
%% That way, if you change the order of any elements, LaTeX will
%% automatically renumber them.
%%
%% We recommend that authors also use the natbib \citep
%% and \citet commands to identify citations.  The citations are
%% tied to the reference list via symbolic KEYs. The KEY corresponds
%% to the KEY in the \bibitem in the reference list below. 

\section{Introduction}
\label{sec:introduction}

Collimated matter jets and Herbig-Haro objects (HHs) are phenomena associated with star formation \citep[see ][and references therein]{Mundt83,Hartigan89,Reipurth01,Nisini2005}. These objects are considered to be originated through a centrifugal-macnetic launch mechanism from Young Stellar Objects (YSOs) \citep[see ][and references therein]{Schwartz83,Strom83,Nisini2018}. These jets have a doubly important role, on the one hand, from their origin they regulate the stellar accretion by removing the angular momentum, modifying the conditions of the matter of the disk \citep[see ][and references therein]{Hartigan94, Giannini2013, Giannini2015}, and on the other hand, as it passes through the surrounding medium, they modify the physical conditions of the environment. 

Within the strong radiation field of the Orion Nebula, the HHs immersed in it are photoionized, so the emission of the gas in photoionization equilibrium of the HHs dominate the global emission over the thin cooling layer that is formed after the shock passage \citep{henney02}. This makes it possible to study the chemical composition of the gas of these HHs --which in principle must  be the same as in the Orion Nebula-- with standard methods for studying photoionized regions.

% There are several works where this object and the  neighboring HH~203 have been studied \citep[and references therein]{Doi:2004a, Henney07, Garcia-Diaz:2008a, ODell:2015a}, investigating their proper motions, structure and kinematics of the zone of the Orion Nebula where the shock takes place.

HH~204 is a HH object located in the central region of the Orion Nebula, just southeast of the Orion Bar, apparently close to the $\theta^{2} \text{ Ori A}$ star. It was observed by \citet{munch62} and classified as an HH by \citet{Canto80}. The origin of the jet is usually associated with the Orion South molecular cloud (Orion-S) \citep{odell17}, an active star formation area of the Orion Nebula. However, the source of the driving jets that feed HH~204 is not entirely clear as we discuss in this paper. HH~204 is photoionized by $\theta^{1} \text{ Ori C}$ from behind its direction of propagation, through the cavity formed by the shock \citep{odell97,odell17}. Through long-slit spectra, \citet{mesadelgado08} studied the effects of HH~204 on the gas of the Orion Nebula, finding peaks in the density and temperature distributions when crossing its surrounding area as well as increases in the emission flux of [Fe\thinspace III] lines produced by dust destruction. Using integral field spectroscopy,  \citet{nunezdiaz12} studied the influence of HH~204 in the Orion Nebula in an area $16\times 16 \text{ arcsec}^2$, finding the presence of a trapped ionization front as well as arguments in favor of the location of the object within the main body of the Orion Nebula and not in the Veil. The works by \citet{mesadelgado08}, \citet{nunezdiaz12} and \citet{odell17} show the presence of a high-$T_{\rm e}(\text{[N\thinspace II]})$ zone, attributed to shock heating. However, this effect and the coincidental fall in the total abundance of O, may be related with an underestimation of the electron density, $n_{\rm e}$, an alternative explanation that will be discussed in Sec.~\ref{sec:mixing_things}. 

This is the second article in a series dedicated to study photoionized HH objects in the Orion Nebula using high-resolution spectroscopy obtained with the  Ultraviolet and Visual Echelle Spectrograph (UVES) \citep{Dodorico00} of the Very Large Telescope (VLT) and the \textit{Hubble Space Telescope} (\textit{HST}) imaging. In this work, we analyze the physical conditions, chemical composition and dynamical properties of HH~204, separating the emission of the Orion Nebula from the HH object and other ionized gas components present in the line of sight. Previous to the present paper, there are few works dedicated to high-resolution spectroscopy of photoionized HH objects of the Orion Nebula, as HH~202~S \citep{mesadelgado09}, HH~529~II and HH~529~III \citep{Blagrave06, mendez2021}. 

This paper has the following content: in Sec.~\ref{sec:data} we describe the observational data and their  treatment. In Sec.~\ref{sec:line_int} we describe the measurement of spectral lines and the reddening correction. In Sec.~\ref{sec:gen_analysis} we derive the physical conditions and  ionic abundances of each of the observed velocity components, while in Sec.~\ref{sec:small_scale} we focus exclusively on HH~204, deriving their physical conditions, ionic abundances and some properties pixel-by-pixel along the UVES slit, as well as study the spatial distribution of the emission of HH~204 with \textit{HST} imaging. In Sec.~\ref{sec:total_abun} we estimate the total abundances of the observed gas components. In Sec.~\ref{sec:mixing_things} we study the effects of mixing three gas components of very different density along the line of sight, simulating a spectrum with lower spectral resolution. In Sec.~\ref{sec:origin-jet-that}, we investigate the origin of HH~204 and its relationship with HH~203. In Sec.~\ref{sec:dic} we discuss the main results of this work and their implications. Finally, in Sec.~\ref{sec:conc} we summarize the conclusions. In the Appendix~\ref{sec:atomic_data_fe3} we show the reliability of the [Fe\thinspace III] atomic data that we use. In Appendix~\ref{sec:sup_mat} tables of data and figures are added as support material.

\begin{figure}
\centering
\includegraphics[width=\linewidth]{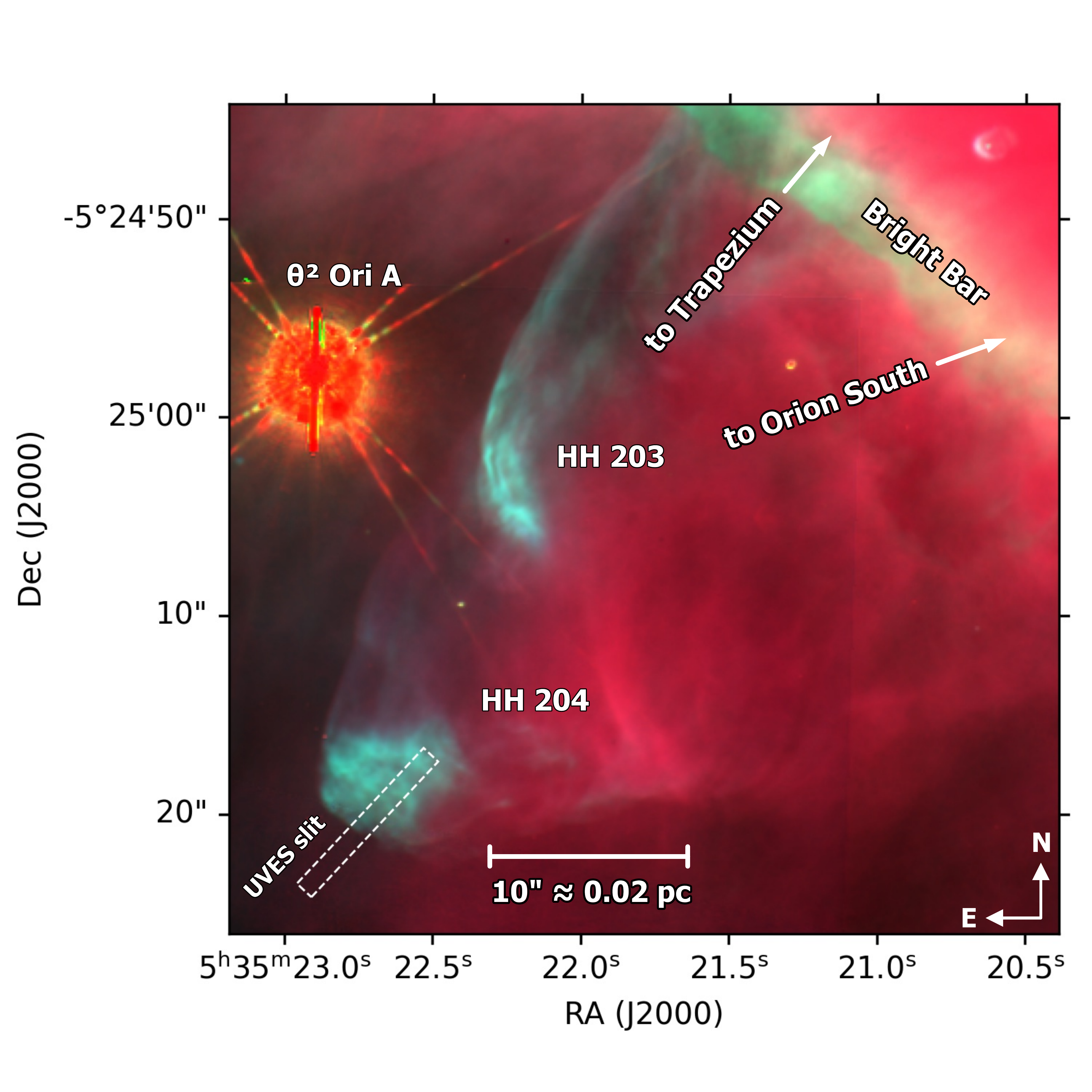}
\caption{ Location of the UVES spectrograph slit at the head of the HH~204 bow shock. The background RGB images shows the immediate environs of HH~203 and 204, derived from \textit{HST} WFPC2 observations \citep{ODell:1996a} in filters of [O\thinspace III] (red), [N\thinspace II] (green), and H$\alpha$ (blue).}
\label{fig:hh204-finding-chart-simple}
\end{figure}

\section{Observations and data reduction}
\label{sec:data}

\begin{figure*}
\centering
  \begin{minipage}{6cm}
    \centering\includegraphics[height=2cm,width=\columnwidth]{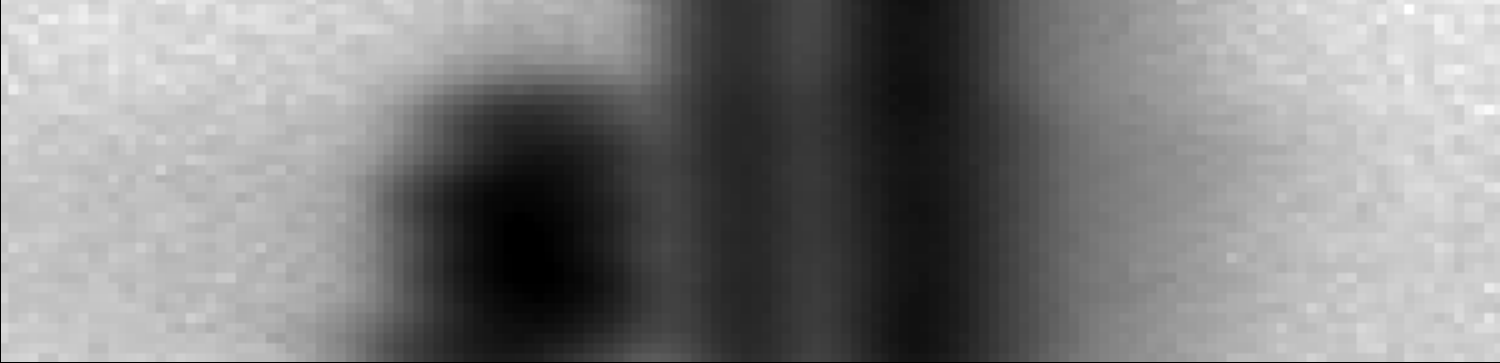}
    \centerline{(a) [O\thinspace II] $\lambda 3729$.}
    \smallskip
  \end{minipage}
  \begin{minipage}{6cm}
    \centering\includegraphics[height=2cm,width=\columnwidth]{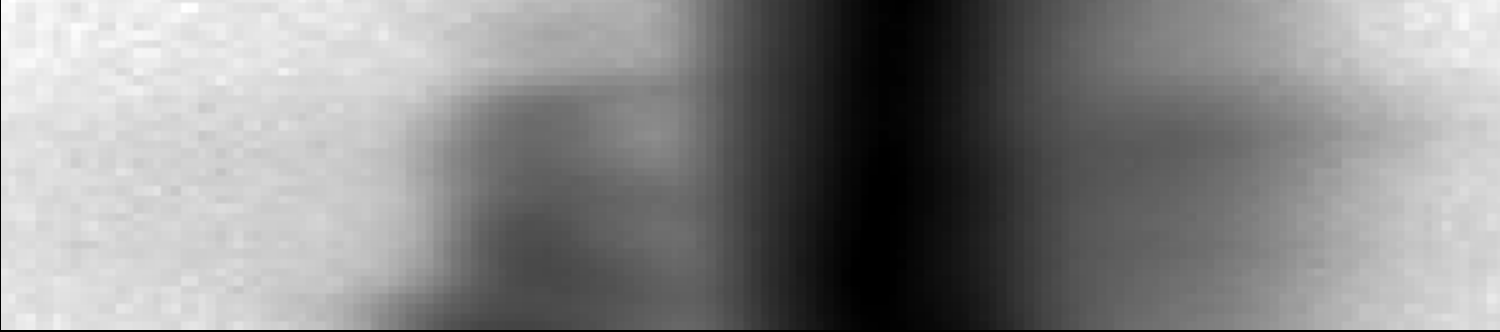}
    \centerline{(b) [O\thinspace III] $\lambda 4959$.}
    \smallskip
  \end{minipage}
 
  \begin{minipage}{6cm}
    \centering\includegraphics[height=2cm , width=\columnwidth]{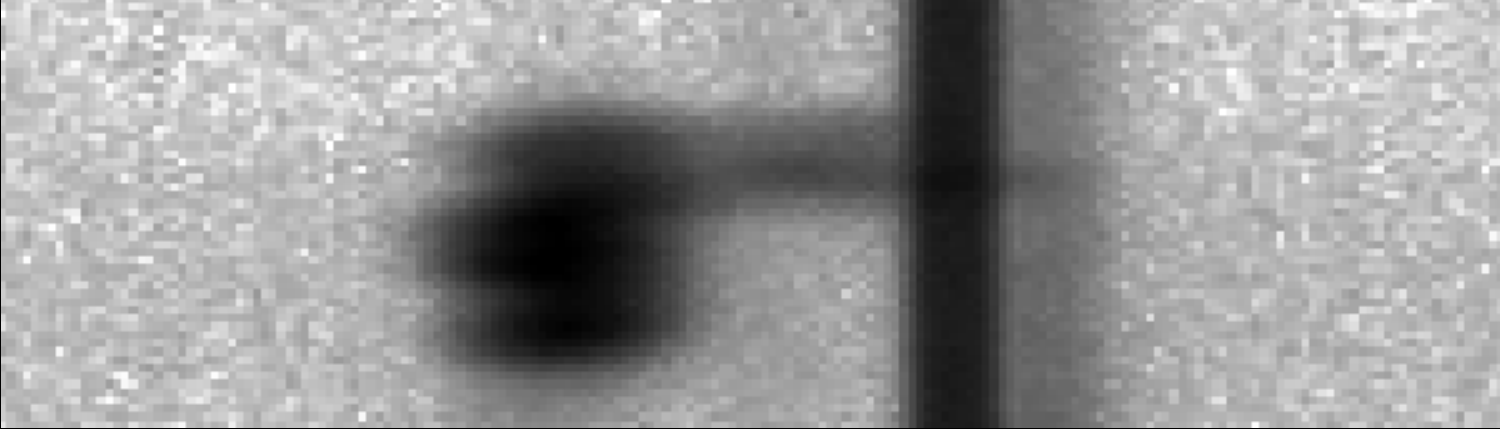}
    \centerline{(c) [O\thinspace I] $\lambda 6300$.}
    \smallskip
  \end{minipage}
  \begin{minipage}{6cm}
    \centering\includegraphics[height=2cm , width=\columnwidth]{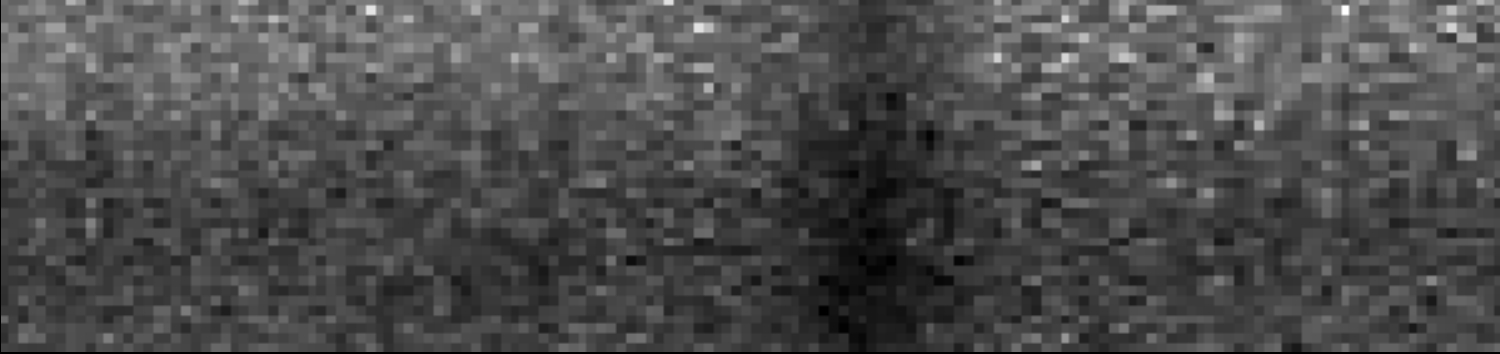}
    \centerline{(d) O\thinspace II $\lambda 4649$.}
    \smallskip
  \end{minipage}
  
  \begin{minipage}{10cm}
    \centering\includegraphics[height=4cm, width=\columnwidth]{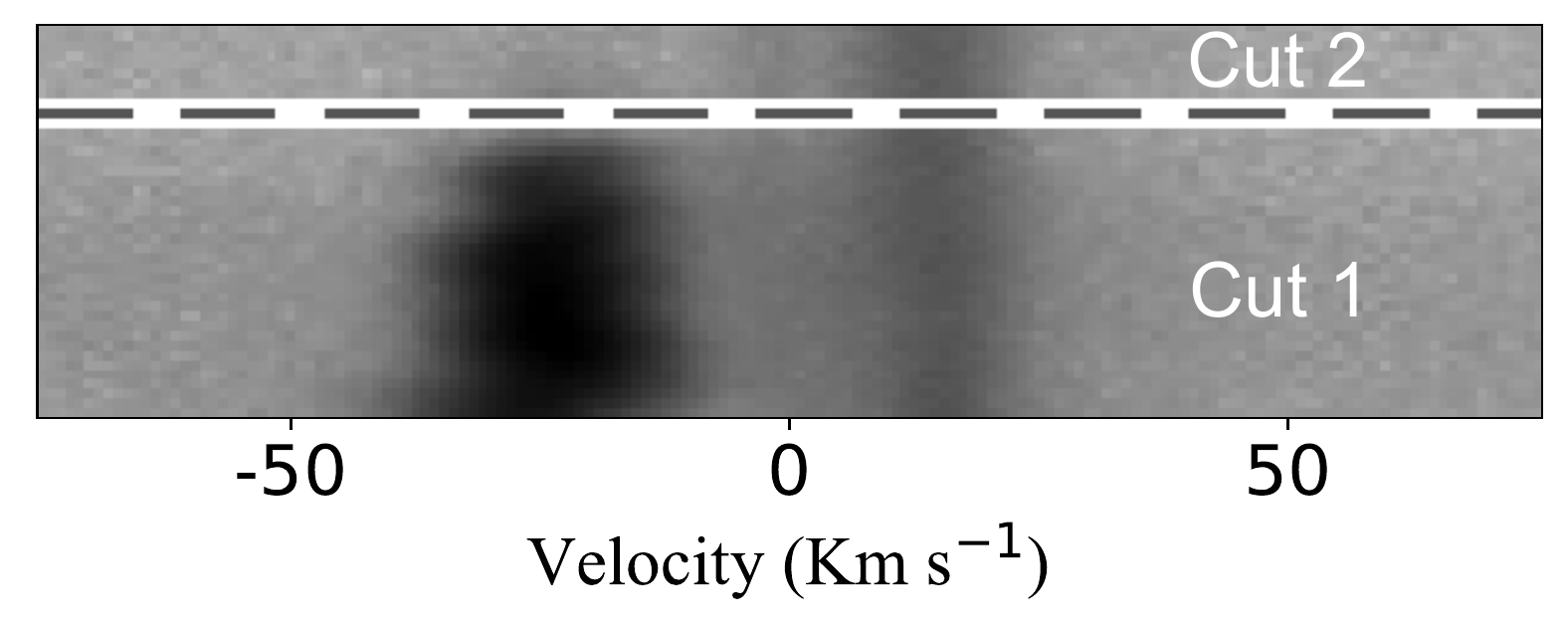}
    \centerline{(e) [Fe\thinspace III] $\lambda 4658$.} 
    \smallskip
  \end{minipage}
  \caption{\textit{Upper panels:} Sample of representative lines in the bi-dimensional spectrum. The Y axis corresponds to the spatial direction (up southeast, down northwest, see Fig.~\ref{fig:hh204-finding-chart-simple} for the spatial location of the slit) while the X axis is the spectral axis. All figures are centered at $\lambda_0$, the rest-frame reference wavelength of each line. The ``ball-shaped'' emission corresponds to HH~204. The slightly blue shifted component with respect to the nebular one is the ``Diffuse Blue Layer'' \citep{Deharveng73,garciadiaz07}, mainly noticeable in the emission of low ionization ions such as [O\thinspace II]. \textit{Bottom panel:} Emission of the [Fe\thinspace III] $\lambda 4658$ line as well as the limits and extension of the different spatial cuts selected to analyse each velocity component. Cut 1 is at the bottom, which corresponds to the westernmost one. The spatial coverage is 7.38 arcsec and 1.97 arcsec for cuts 1 and 2, respectively. The velocity scale is heliocentric.}
  \label{fig:cuts}
\end{figure*}

The observations were made during the nights of October 28 and 29, 2013 under photometric conditions using UVES in the UT2 of the Very Large Telescope (VLT) in Cerro Paranal, Chile. The slit position was centered at the coordinates RA(J2000)=05$^h$35$^m$22$^s$.72, DEC(J2000)=$-$05$^{\circ}$25$'$20.42$''$ with a position angle of 137$^{\circ}$. The slit width provides an effective spectral resolution $\lambda/\Delta \lambda \approx 6.5 \text{ km s}^{-1}$, covering the spectral range between 3100-10420\AA. Three exposures of 150s of the standard star GD71 \citep{Moehler14a, Moehler14b} were taken in the same night under similar observational conditions than the science images to achieve the flux calibration of the data. The observational settings are shown in Table~\ref{tab:observations} and the spatial coverage is presented in Fig.~\ref{fig:hh204-finding-chart-simple}. The instrumental configuration and the data reduction procedure is described in \citet[][hereinafter Paper~I]{mendez2021}. The 2D spectra (see Fig.~\ref{fig:cuts}) show three evident components: 1) the nebular one (the emission of the Orion Nebula), which is rather homogeneously distributed along the spatial axis of the slit and occupies the reddest spectral position; 2) the ``Diffuse Blue Layer'', (hereinafter DBL) a slightly blueshifted homogeneous diffuse component  \citep[previously detected by][]{Deharveng73}, that may correspond to a different H\thinspace II region along the same line of sight \citep{garciadiaz07}, and 3) HH~204, the ``ball-shaped'' blueshifted component. We define two spatial cuts -- shown in Fig.~\ref{fig:cuts} -- covering a spatial area of 7.38 arcsec for cut~1 and 1.97 arcsec for cut~2. In cut~2, we can separate the emission of the DBL and the nebular component. However, due to the strong contribution of HH~204, we can not separate those components in cut~1. In this case, we study the emission of the combined spectrum of the nebular component and the DBL. We also take advantage of the quality of the data performing a pixel-by-pixel analysis of various emission lines in order to detect small variations in physical conditions and/or the chemical composition of HH~204 along the slit.

The study of the spatial distribution of the emission of HH~204 and the gas flows that may originate it are based in the \textit{HST} WFPC2 imaging in the F502N  ($\overline{\lambda}=5012$), F547M ($\overline{\lambda}=5446$), F656N ($\overline{\lambda}=6564$), and F658N ($\overline{\lambda}=6591$) filters from program GO5469 \citep{ODell:1996a}. The spatial pixel size of these data is \SI{0.045}{arcsec}. Flux calibration and correction for contamination by continuum and non-target lines was performed using the coefficients given in \citet{ODell:2009b}.

\begin{deluxetable}{ccccc}
\tablecaption{Main parameters of UVES spectroscopic observations. \label{tab:observations}}
\tablewidth{0pt}
\tablehead{
Date & $\Delta \lambda$& Exp. time  &Seeing &Airmass\\
 & (\AA) &  (s) & (arcsec)&
}
\startdata
2013-10-29 & 3100-3885 & 5, 3$\times$180 &0.85&1.10\\
2013-10-29 & 3750-4995 & 5, 3$\times$600 & 0.70 & 1.16\\
2013-10-29 & 4785-6805 & 5, 3$\times$180 &0.85&1.10\\
2013-10-29 & 6700-10420 & 5, 3$\times$600 & 0.70 & 1.16\\
\enddata
\end{deluxetable}

\section{Line intensities and reddening}
\label{sec:line_int}

We use SPLOT task from IRAF\footnote{IRAF is distributed by National Optical Astronomy Observatory, which is operated by Association of Universities for Research in Astronomy, under cooperative agreement with the National Science Foundation} \citep{Tody93} to measure the line intensities and estimate their uncertainties as it is described in detail in Paper~I. In the case of the spectra of cut~1 and cut~2, we measure a complete set of around $\sim 500$ and $\sim 300$ emission lines, respectively, while in the case of the pixel-by-pixel measurements for HH~204, we limit the analysis to some representative lines: $\text{H9},\thinspace \text{H}\beta,\thinspace \text{H}\alpha$;  He\thinspace I $\lambda \lambda 4471,\thinspace 5876, \thinspace 6678$; [N\thinspace II] $\lambda \lambda 5755,\thinspace 6584$; O\thinspace I $\lambda 7772$; [O\thinspace I] $\lambda 6300$; [O\thinspace II] $\lambda 3726$; [O\thinspace III] $\lambda \lambda 4363,\thinspace 4959$; [Ne\thinspace III] $\lambda 3869$; [S\thinspace II] $\lambda \lambda 6716,\thinspace 6731$; [S\thinspace III] $\lambda \lambda 6312,\thinspace 9531$; [Cl\thinspace II] $\lambda 9124$; [Cl\thinspace III] $\lambda 5538$; [Ar\thinspace III] $\lambda 7136$; [Ca\thinspace II] $\lambda 7324$; [Cr\thinspace II] $\lambda 8000$; [Fe\thinspace II] $\lambda 9052$; [Fe\thinspace III] $\lambda \lambda 4658,\thinspace 4702,\thinspace 4881$; [Ni\thinspace II] $\lambda 7378$ and [Ni\thinspace III] $\lambda 7890$. The reddening correction was done using the extinction curve from \citet{Blagrave07} and the emissivity coefficients of \citet{Storey95} for H$\varepsilon$, H$\delta$, H$\gamma$, H$\beta$ and H$\alpha$ Balmer lines and the P12, P11, P10, P9 Paschen lines. The values of the extinction coefficient, $c(\text{H}\beta)$, are presented in Table~\ref{tab:c_extin}. In the case of pixel-by-pixel measurements, a value of  $c(\text{H}\beta)=0.42 \pm 0.02$ was used. An example of the spectra that can be found in the online material is shown in Table~\ref{tab:sample_of_lines}, where some lines of the spectra of cut~1 are shown.

\begin{deluxetable}{ccccc}
\tablecaption{Reddening coefficients for each component. \label{tab:c_extin}}
\tablewidth{0pt}
\tablehead{
 & \multicolumn{2}{c}{$\text{c}(\text{H}\beta)$} 
}
\startdata
 & HH~204 & Nebula + DBL\\
Cut 1 & $0.42 \pm  0.02$ & $0.31 \pm  0.03$\\
\hline
 & DBL & Nebula\\
Cut 2 & $0.42 \pm 0.09$ &$0.30 \pm 0.04$\\
\enddata
\end{deluxetable}

%\cesar{No entiendo esto de que sea heliocentrica, el punto de velocidad cero del panel (e) a que valor corresponde? respecto a la longitud de onda en laboratorio en el sistema heliocentrico (corrigiendo de la velocidad orbital terrestre)?} \eduardo{si, la velocidad 0 esta centrada en $\lambda_0$}

\section{Analysis of integrated spectra of each component}
\label{sec:gen_analysis}

\subsection{Physical Conditions}
\label{subsec:physical_cond}

\begin{figure}
\centering
\includegraphics[width=\columnwidth]{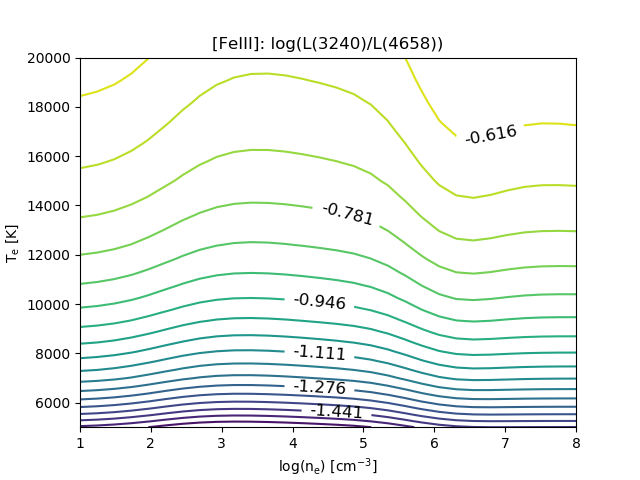}
\includegraphics[width=\columnwidth]{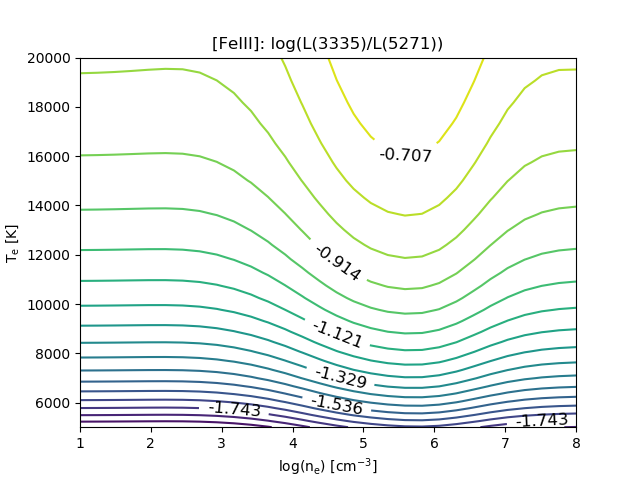}
\caption{Predicted dependence of the [Fe\thinspace III] $\lambda3240/\lambda4658$ and $\lambda3335/\lambda5271$
line intensity ratios with physical conditions.}
\label{fig:predicted_ratios_feiii}
\end{figure}

We use the version 1.1.13 of PyNeb \citep{Luridiana15} to obtain the physical conditions of the gas from the intensity ratios of collisionally excited lines (CELs) and recombination lines (RLs). PyNeb is a Python based tool to compute line emissivities and derive physical conditions and chemical abundances of ionized gas. We have used the atomic data set presented in tables~\ref{tab:atomic_data} and \ref{tab:rec_atomic_data} for the calculations made with PyNeb. We first estimate the $n_{\rm e}$ values given by each diagnostic of CELs by calculating each convergence of $T_{\rm e}-n_{\rm e}$ with the available diagnostics of electron temperature, $T_{\rm e}$, using the PyNeb task \texttt{getCrossTemDen}, as it is described in detail in Paper~I. The density and temperature diagnostics used are shown in Table~\ref{tab:pc}. Then, in the nebular and DBL components, we adopt the weighted mean\footnote{The weights were defined as the inverse of the square of the error associated to each density diagnostic.} of the available values of $n_{\rm e}$ obtained with the following diagnostics: [O\thinspace II] $\lambda$3726/$\lambda$3729, [S\thinspace II] $\lambda$6731/$\lambda$6716 and [Cl\thinspace III] $\lambda$5538/$\lambda$5518. For consistency, in the case of HH~204 we rely on the $n_{\rm e}$ derived from [Fe\thinspace III] lines since values of $10^4-10^6 \text{ cm}^{-3}$ are above the critical densities of the CELs involved in the more common diagnostics. The simultaneous estimation of $n_{\rm e}$([Fe\thinspace III]) and $T_{\rm e}$([Fe\thinspace III]) in HH~204 is achieved by a maximum-likelihood procedure, as described in Paper~I. In this procedure, different combinations of $T_{\rm e}$ and $n_{\rm e}$ are tested to obtain the abundance of Fe$^{2+}$/H$^{+}$ with several [Fe\thinspace III] lines, giving as a result the combination of $T_{\rm e}$-$n_{\rm e}$ that minimizes the dispersion between the abundances obtained with all the lines. In HH~204, we have confident detections of [Fe\thinspace III] $\lambda \lambda 3240, 3335$ lines from the $^5\text{D}-{^3}\text{D}$ transitions, whose ratios with lines from the multiplets $^5\text{D}-{^3}\text{F}$ and $^5\text{D}-{^3}\text{P}$ are highly dependent on $T_{\rm e}$ as it is shown in Fig.~\ref{fig:predicted_ratios_feiii}. We include the following lines in the maximum-likelihood calculation: [Fe\thinspace III] $\lambda \lambda$ 3240, 3335, 4658, 4702, 4734, 4881, 5011, 5271. This collection of lines allows us to obtain well-constrained values of $T_{\rm e}$([Fe\thinspace III]) and $n_{\rm e}$([Fe\thinspace III]). The intensity ratios of these selected lines are consistent with the predicted ones when using transitions coming from the same atomic level (which are independent of the physical conditions of the gas), as we  show in Table~\ref{tab:fe3_ratios_theo}. %Since these ratios are independent of the physical conditions of the gas, the data collected in the table allow us to rule out blends with other lines or telluric absorptions in the [Fe\thinspace III] lines. 
Another density indicator that can be used with our data is $n_{\rm e}$(O\thinspace II), but only for the nebular component, which is the only one where we detect RLs of multiplet 1 of O\thinspace II.

Once the representative $n_{\rm e}$ is adopted for each component, we estimate $T_{\rm e}$ through several diagnostics based on CELs as it is shown in Table~\ref{tab:pc}. In the case of $T_{\rm e}$([S\thinspace III]), telluric absortions affect the line $\lambda 9069$ in the nebular and DBL components. Thus, we adopt $I$([S\thinspace III] 9531)/$I$([S\thinspace III] 9069) = 2.47 \citep{Podobedova09} in these cases. In HH~204 we were able to separate the auroral [O\thinspace I] $\lambda 5577$ line from sky emission contamination, which permitted us to estimate $T_{\rm e}$([O\thinspace I]). In the DBL, the estimations of $T_{\rm e}$([O\thinspace II]) and $T_{\rm e}$([S\thinspace II]) are affected by some extended residual emission of HH~204 in the auroral lines that crosses the cut border, affecting the first pixels of cut~2. $T_{\rm e}$(He\thinspace I) was estimated using the average values obtained from He\thinspace I  $I(\lambda7281)/I(\lambda6678 )$, $I(\lambda7281)/I(\lambda4922 )$ and $I(\lambda7281)/I(\lambda4388)$ line intensity ratios. Finally, we define $T_{\rm e} (\text{low})$ as the weighted mean of $T_{\rm e}$([N\thinspace II]), $T_{\rm e}$([O\thinspace II]) and $T_{\rm e}$([S\thinspace II]) while $T_{\rm e} (\text{high})$ is the weighted mean of $T_{\rm e}$([O\thinspace III]) and $T_{\rm e}$([S\thinspace III]).

The resulting physical conditions for all  components are shown in Table~\ref{tab:pc}.

\begin{deluxetable*}{ccccc}
\tablecaption{Physical conditions determined from  several diagnostics. \label{tab:pc}}
\tablewidth{0pt}
\tablehead{
 & \multicolumn{2}{c}{Cut 1} & \multicolumn{2}{c}{Cut 2} \\
Diagnostic & HH~204 & Nebula + DBL & DBL  & Nebula
}
\startdata
\multicolumn{5}{c}{$n_e (\text{ cm}^{-3})$}\\
{[O\thinspace II]} $\lambda$3726/$\lambda$3729 & $15420^{+7740} _{-3850}$&$1130^{+150} _{-110}$ &$400^{+140} _{-120}$& $1480^{+190} _{-180}$\\
{[S\thinspace II]} $\lambda$6731/$\lambda$6716 & $11350^{+9920} _{-3890}$& $1350^{+290} _{-260}$ &  $300^{+140} _{-120}$&$1230^{+250} _{-230}$\\
{[Cl\thinspace III]} $\lambda$5538/$\lambda$5518 & $13370^{+1990} _{-1830}$&$1630^{+370} _{-320}$&-& $1930^{+720} _{-650}$\\
{[Fe\thinspace II]} $\lambda$9268/$\lambda$9052 &$13100 ^{+2860} _{-2990}$&-&-&\\
{[Fe\thinspace III]} $\lambda$4658/$\lambda$4702 & $13040^{+3830} _{-3130}$& $3380^{+1810} _{-1340}$&-&$3200^{+2540} _{-1540}$ \\
$n_{\rm e}$(O\thinspace II) & -&$1350 \pm 150$&-&$1050 \pm 200$\\
{[Fe\thinspace III]}$^{*}$ & $13540 \pm 1210$ &-&-&-\\
\textbf{Adopted} &  \boldmath${13540 \pm 1210 }$ & \boldmath${1230 \pm 160  }$& \boldmath${350 \pm 50   }$& \boldmath${1440 \pm 170}$ \\
 \multicolumn{5}{c}{$T_e$ (K)}\\
 $T_{\rm e}\left(\mbox{He}\thinspace \mbox{I} \right)$ & $8790 ^{+480} _{-430}$&9760:& 5650:&7980:\\ 
{[O\thinspace I]} $\lambda$5577/$\lambda \lambda$ 6300+64 & $8290^{+430} _{-320}$&-& - &-\\
{[N\thinspace II]} $\lambda$5755/$\lambda$6584  & $8760^{+170} _{-180}$&$8530^{+150} _{-190}$ &$8120^{+390} _{-360}$&$8440^{+170} _{-210}$\\
{[O\thinspace II]} $\lambda \lambda$ 3726+29/$\lambda \lambda$7319+20+30+31 &-&-&$10390^{+730} _{-640}$&$9120^{+430} _{-470}$\\
{[S\thinspace II]} $\lambda \lambda$4069+76/$\lambda \lambda$ 6716+31&$8260^{+640} _{-500}$&$11470^{+950} _{-630}$&$10440^{+1360} _{-1030}$&$9890^{+650} _{-610}$\\
{[O\thinspace III]} $\lambda$4363/$\lambda \lambda$4959+5007&$12430^{+180} _{-220}$&$8010^{+90} _{-80}$&-&$8120^{+90} _{-100}$ \\
{[S\thinspace III]} $\lambda$6312/$\lambda \lambda$9069+9531 &$9310^{+220} _{-330}$&$8180^{+190} _{-230}$&$7710^{+510} _{-400}$&$8010^{+250} _{-210}$\\
{[Fe\thinspace III]}$^{*}$ &  $8210 \pm 220$ &-&-&-\\
\textbf{\boldmath${T_e}$ (low) Adopted} &\boldmath${8760\pm 180 }$ &\boldmath${8530\pm 190 }$ &\boldmath${8120\pm 390 }$ &\boldmath${8440\pm 210 }$\\
\textbf{\boldmath${T_e}$ (high) Adopted} & \boldmath${ 12430\pm 220}$& \boldmath${ 8030\pm 60}$& \boldmath${7710\pm 510 }$&\boldmath${8110\pm 90 }$\\
\enddata
\tablecomments{$^*$ indicates that a maximum likelihood method was used.}
\end{deluxetable*}

\subsection{Ionic abundances}
\label{subsec:ionic_abundances}

We assume the appropriate values of the $n_{\rm e}$ and $T_{\rm e}$ diagnostics for each ion --assuming a three-zone approximation-- to derive the ionic abundances of the different components. We use $T_{\rm e} (\text{low})$ for N$^{+}$, O$^{+}$, S$^{+}$, Cl$^{+}$, Ca$^{2+}$, Cr$^{+}$, Fe$^{+}$, Fe$^{2+}$, Ni$^{+}$ and Ni$^{2+}$ and $T_{\rm e} (\text{[S\thinspace III]})$ for S$^{2+}$ and Cl$^{2+}$. In the case of Ne$^{2+}$, O$^{2+}$ and Ar$^{3+}$, we use $T_{\rm e} (\text{high})$. We also use $T_{\rm e} (\text{high})$ to derive the He$^{+}$, C$^{2+}$ and Ar$^{2+}$ abundances for the nebular component, but $T_{\rm e} (\text{low})$ for HH~204, as we discuss in Sec.~\ref{subsec:small_scale_pc}. We follow the same methodology described in Paper~I for abundance calculations, except in some particular cases that are discussed below together with some  abundance determinations for ions whose lines were not reported in Paper~I. %\cesar{He cambiado la ultima frase, mira a ver si es correcto lo que he puesto}

\subsection{Ionic abundances of Fe and Ni ions}
\label{subsec:ionic_abundances_fe_ni}
%We were able to observe weak lines of  multiplets and transitions from these ions which allows us to explore their abundances in detail.

In HH~204, the emission lines of [Fe\thinspace II], [Ni\thinspace II], [Fe\thinspace III] and [Ni\thinspace III] are considerably enhanced in comparison with what is observed in the nebular component. Due to the low ionization degree of HH~204, we expect that Fe$^{+}$ and Ni$^{+}$ have an important contribution to the total Fe and Ni abundances. Therefore, it seems pertinent to discuss in some detail the degree of confidence of the abundance determinations based on these two ions. 

Optical lines coming from the upper levels of the Fe$^{+}$ atom can be affected by continuum pumping \citep[][]{Lucy95,rodriguez99, verner00}. However, lower levels that produce the emission lines of multiplet $\text{a}^{4}\text{F}-\text{a}^{4}\text{P}$ are mostly populated by collisions \citep[][]{Baldwin96}. One of the strongest lines of this multiplet, [Fe\thinspace II] $\lambda 8617$ $(\text{a}^{4}\text{F}_{9/2}-\text{a}^{4}\text{P}_{5/2})$, could not be detected due to the instrumental gap of UVES in the red arm. However, weaker lines arising from the same upper level as $\lambda\lambda 9052, 9399$ ($\text{a}^{4}\text{F}_{7/2}-\text{a}^{4}\text{P}_{5/2}, \text{a}^{4}\text{F}_{5/2}-\text{a}^{4}\text{P}_{5/2} $), detected in HH~204, must be useful for the same purpose. Although the transition probabilities of the weakest detected lines coming from the $\text{a}^{4}\text{P}_{1/2}$, $\text{a}^{4}\text{P}_{3/2}$ and  $\text{a}^{4}\text{P}_{5/2}$ levels still need to be tested (since these lines may be affected by undetected telluric absorptions), there is a good agreement between the measured and predicted line ratios of [Fe\thinspace II] $\lambda9052/\lambda9399$, $\lambda8892/\lambda9227$ and $\lambda9268/\lambda9034$, as it is shown in Table~\ref{tab:fe2_ratios_theo}. In order to make a simple test of the chosen atomic data, we take advantage of the theoretical density dependence between the population of the $\text{a}^{4}\text{P}_{1/2}$ and the $\text{a}^{4}\text{P}_{5/2}$. By using the estimated $T_{\rm e}(\text{low})$ for HH~204 and the [Fe\thinspace II] $\lambda9268/\lambda9052$ intensity ratio, we obtain $n_{\rm e}(\text{[Fe\thinspace II]})=13100 ^{+2860} _{-2990} \text{ cm}^{-3}$, which is consistent with the rest of density diagnostics shown in Table~\ref{tab:pc}. In cut~2 we derive the Fe$^{+}$ abundance of the nebular component by using the uncontaminated [Fe\thinspace II] $\lambda 8892$ line. %In the Nebula+DBL spectra from cut~1, [Fe\thinspace II] lines are partially enhanced by the strong emission of HH~204. Since the [Fe\thinspace II] lines in that combined component are rather weak, any small contribution from the tail of the HH~204 emission can significantly affect the Fe$^{+}$ abundance. This is not a problem in cut~2, where the Fe$^{+}$ abundance of the nebular component can be derived from the  uncontaminated [Fe\thinspace II] $\lambda 8892$ line.

The $\text{a}^{4}\text{F}-\text{a}^{4}\text{P}_{5/2}$ transitions of [Fe\thinspace II] and the $\text{a}^{2}\text{D}-\text{a}^{2}\text{F}_{7/2}$ ones of [Ni\thinspace II] have practically the same excitation energy, giving origin to lines close in wavelength \citep{Bautista96}. However, there is an important difference between their sensitivity to fluorescence by continuum pumping due to the  multiplicity of their ground states. Photoexcitations from the Fe$^{+}$ $^6\text{D}$ ground state to the quartet levels have low probability and lines produced by  intercombination transitions from sextet to quartet levels should be very weak \citep{Bautista98}. However, \citet{rodriguez99} pointed out that the lowest quartet level,  $\text{a}^{4}\text{F}_{9/2}$, may be metastable and promote excitations to higher quartet levels. The main pumping routes starting from this level were studied by \citet{verner00} at densities  above $10^4 \text{ cm}^{-3}$, finding that this pumping populates the levels $\text{a}^{4}\text{H}$, $\text{b}^{4}\text{F}$, $\text{b}^{4}\text{P}$ and $\text{a}^{4}\text{G}$. Since transitions from any of these levels to $\text{a}^{4}\text{P}$ are rather weak, its population remains practically unaffected. Nevertheless, in the case of [Ni\thinspace II], the ground state and the participating levels are doublets which make fluorescence effects by continuum pumping more likely  \citep{Bautista96}. However, an important factor that plays against the influence of fluorescence effects in [Ni\thinspace II] in the case of HH~204 is its relatively large distance from $\theta^{1} \text{ Ori C}$ (150.4 arcsec), the main ionization source of the nebula \citep{ODell:2015a, odell17_ionizing}. In a simple procedure, following the formalism developed by \citet[][their equation 8]{Bautista96},   for a 3-level model ($\text{level 1: }\text{a}^2\text{D}_{5/2}$, $\text{level 2: } \text{a}^2\text{F}_{7/2}$ and $\text{level 3: }\text{z}^2\text{D}^{0}_{5/2}$), the critical densities $n_{\text{cf}}$ -- for which if $n_{\rm e}>n_{\text{cf}}$, collisional excitations dominate over fluorescence -- in two zones of the Orion Nebula (a and b), both excited by $\theta^{1} \text{ Ori C}$, should be related as follows:

\begin{equation}
    \label{eq:ni2}
    \frac{n_{\text{cf, a}}}{n_{\text{cf, b}}}=\left( \frac{J_{13,\text{ a}}}{J_{13,\text{ b}}}\right)\left( \frac{q_{12,\text{ b}}}{q_{12,\text{ a}}} \right),
\end{equation}

\noindent where $q_{12}$ is the Maxwellian averaged collisional strength for transitions from level 1 to 2 and $J_{13}$ is the intensity of the continuum at energies of the $1 \rightarrow 3$ transitions. If we choose the zone ``a'' as the one observed by \citet{Osterbrock92} and the zone  ``b'' as HH~204, 
we can assume ${q_{12,\text{ b}}}/{q_{12,\text{ a}}}\approx 1$, because the $T_{\rm e}$ determined by \citet{Osterbrock92} and us are very similar (9000 K and 8760 K, respectively). On the other hand, by estimating the geometrical dilution of $J_{13}$ in both areas (the zone observed by \citet{Osterbrock92} is  located at 63.98 arcsecs from $\theta^{1} \text{ Ori C}$), we get a $n_{\text{cf, a}}/n_{\text{cf, b}}\approx5.53$. By adopting the $n_{\text{cf, a}}$ estimated by \citet{Bautista96}, we obtain  $n_{\text{cf, b}}\approx 2.17 \times 10^{3} \text{ cm}^{-3}$, which is rather small compared with the density we obtain for HH~204 and therefore collisional excitation should dominate. Nevertheless, it must be considered that the apparently closer star $\theta^{2} \text{ Ori A}$  may be also a source of fluorescence for HH~204. However, by using the [Ni\thinspace II] $\lambda 7378$ ($\text{a}^{2}\text{D}_{5/2}-\text{a}^{2}\text{F}_{7/2})$ line to obtain the Ni$^{+}$ abundance and comparing with the Fe$^{+}$/H$^+$ ratio, we obtain $\text{log}(\text{Ni}^{+}/\text{Fe}^{+})=-1.27 \pm 0.06$, which is in complete agreement with the solar value of $\text{log}(\text{Ni}/\text{Fe})_{\odot}=-1.25 \pm 0.05$ \citep{lodders19}, suggesting the absence of significant fluorescence effects (as discussed before, we expect larger fluorescence effects in Ni$^+$). Therefore, we can assume that $\theta^{2} \text{ Ori A}$ is not a significant source of photon pumping of [Ni\thinspace II] lines in HH~204. We do not estimate the  Ni$^{+}$ abundances for the rest of velocity components because it requires a detailed  analysis of the fluorescence conditions in the ionized gas, which goes beyond the scope of this paper. 
%Since some  [Ni\thinspace II] lines, as $\lambda 7412$ ($^2 \text{D}_{3/2}-^2\text{F}_{5/2}$), give Ni$^+$ abundances around a factor 2.20 higher than others, continuum pumping effects should be still important in the higher levels of the Ni$^{+}$ atom in HH~204, although in a much lesser extent than in the area of the Orion Nebula observed by \citet{Osterbrock92}, where this factor reaches a value of 15.51 \citep{Lucy95}.

We derive the Fe$^{2+}$ abundance using the [Fe\thinspace III] lines indicated in Sec.~\ref{subsec:physical_cond}. It is noticeable the good agreement between $T_{\rm e}(\text{[Fe\thinspace III]})$, $T_{\rm e}(\text{[O\thinspace I]})$ and $T_{\rm e}(\text{[S\thinspace II]})$ in the case of HH~204, contrary to what was found in HH~529~II and HH~529~III, where $T_{\rm e}(\text{[Fe\thinspace III]})$ was more consistent with the temperature obtained for high ionization ions \citep{mendez2021}. This is not surprising due to the different ionization degrees of HH~204 and HH~529~II+III (see Sec.~\ref{sec:small_scale}). 
%In HH~204, the high ionization gas is a remainder flowing together with most of the gas in low ionization conditions as we will describe in detail in Sec.~\ref{sec:small_scale}, while an analogous situation happened in HH~529~II and HH~529~III with the small fraction of gas in low ionization conditions.

In Paper~I, we pointed out the inconsistency between the predicted and measured intensity ratios of [Ni\thinspace III] $^3\text{F}-^3\text{P}_2$ transitions ($\lambda \lambda 6534, 6000, 6946$) in HH~529~II, HH~529~III, HH~202~S and several zones of the Orion Nebula (see Table~D11 of Paper~I). We obtain a similar result for HH~204, $\lambda 6534/\lambda6000=1.38\pm 0.18$, which is rather far from the predicted value of 2.19 value \citep[][]{Bautista01}. This indicates that the transition probabilities of the aforementioned lines may have errors \citep[for a more detailed discussion see Appendix C in][]{mendez2021}. We have a different situation for the intensity ratios of lines arising from the $^{1}\text{D}_2$ level. After subtracting the small contribution of [Cl\thinspace III] $\lambda 8499.60$ to the measured intensity of  [Ni\thinspace III] $\lambda 8499.62$, we obtain [Ni\thinspace III] $\lambda7890/\lambda8500=2.65\pm 0.19$ in agreement with the predicted value of 2.47 \citep{Bautista01}. This indicates that, with the available atomic data, the most confident determinations of the Ni$^{2+}$ abundance can be obtained with these last lines. Thus, we will adopt the  Ni$^{2+}$ abundances determined from  [Ni\thinspace III] $\lambda 7890$ line.  Unfortunately, this line is affected by a telluric emission feature in the nebular component and, therefore, we have to rely on the [Ni\thinspace III] $\lambda6534$ line to determine the Ni$^{2+}$ abundance for this component. %\Karla{Est\'as diciendo que el cociente observado de 7889/8499.62 est\'a de acuerdo con el teorico y  adoptas 7889 para la abundancia ionica, pero despu\'es te arrepientes y usas 6533 por el efecto del cielo en la otra l\'inea. Cu\'al es el valor del cociente te\'orico por Bautista 2001?} \eduardo{no es que me arrepienta, en HH204 si uso 7890, pero esa no la puedo usar en las componentes nebulares porque tienen una linea de cielo. Ahi uso 6534, es una estimacion burda, pero no hay mucho que se pueda hacer.}

%\begin{figure}
%\centering
%\includegraphics[width=\columnwidth]{9268_9052.png}
%\caption{Predicted dependence of the [Fe\thinspace II] $\lambda9268/\lambda9052$ line intensity ratios with physical conditions.}
%\label{fig:fe2_density}
%\end{figure}

\subsection{Ionic abundances of Ca$^{+}$ and Cr$^{+}$}
\label{subsec:ionic_abundances_rare}
%The energies corresponding to  $\text{a}^{2}\text{D}-\text{a}^{2}\text{F}_{7/2}$ transitions of [Ni\thinspace II] are similar to $\text{a}^{6}\text{S}-\text{a}^{6}\text{D}$ ones of [Cr\thinspace II]. Moreover, in both cases the lower -- ground -- and upper levels have the same multiplicity, so the excitation of the upper level by starlight is likely. Therefore, it seems reasonable to assume that if fluorescence effects are important for [Ni\thinspace II] lines, they will also be important for [Cr\thinspace II] lines.

We measure some [Ca\thinspace II] and [Cr\thinspace II] lines with a good signal-to-noise ratio in HH~204. Thus, it allows to estimate Ca$^{+}$ and Cr$^{+}$ abundances. However, [Cr\thinspace II] lines may be affected by fluorescence similarly to [Ni\thinspace II] ones. As we discuss in Sec.~\ref{subsec:ionic_abundances_fe_ni}, in HH~204 collisional excitations dominate over fluorescence in the aforementioned [Ni\thinspace II] transitions and this may be also the case for [Cr\thinspace II]. With this assumption, we obtain an abundance of $12+\text{log}(\text{Cr}^{+}/\text{H}^{+})=4.28\pm 0.03$. By comparing this value with the Fe$^{+}$ and Ni$^{+}$ abundances, we obtain $\text{log}(\text{Cr}^{+}/\text{Ni}^{+})=-0.61\pm 0.05$ and $\text{log}(\text{Cr}^{+}/\text{Fe}^{+})=-1.88\pm 0.07$, in agreement with the solar values of  $\text{log}(\text{Cr} /\text{Ni})_{\odot}=-0.57\pm 0.05$ and $\text{log}(\text{Cr}/\text{Fe})_{\odot}=-1.82\pm 0.04$, respectively \citep{lodders19}. Nevertheless, the spatial distribution of the $\text{Cr}^{+}/\text{Ni}^{+}$ and $\text{Cr}^{+}/\text{Fe}^{+}$ ratios along the HH~204 jet is not completely constant, as it is described in Sec.~\ref{subsec:small_scale_ca}, which may be indicative of different ionization/depletion patterns between these elements. Unfortunately, although several [Cr\thinspace III] lines are detected, we can not derive the Cr$^{2+}$ abundance due to the lack of atomic data for this ion.

In the case of the Ca$^{+}$ abundance, we base our estimations in the [Ca\thinspace II] $\lambda 7324$ line since $\lambda 7291$ is affected by a telluric absorption in our observations. Due to its low ionization potential, much smaller than that of hydrogen, and owing to the presence of an ionization front in HH~204 \citep{nunezdiaz12}, the resulting abundance may not represent the real gaseous Ca$^{+}$ abundance in the photoionized gas of HH~204.

%The resulting abundance may be rather an upper limit of the real gaseous Ca$^{+}$ abundance since this ion is expected to be distributed in the photodissociation region due to its low ionization potential \citep{Amayo20}. 

\subsection{Ionic abundances based on RLs}
\label{subsec:ionic_abundances_rls}

For the nebular component,  the He$^{+}$ abundance is derived using $T_{\rm e} (\text{high})$ and the lines considered in Table~D14 of Paper~I, which are the least affected ones by the metastability of the $2^3\text{S}$ level. However, we have used $T_{\rm e} (\text{low})$ for HH~204. In this component, our determination of $T_{\rm e}(\text{He\thinspace I})$ is more consistent with $T_{\rm e} (\text{low})$. This is because in HH~204, [O\thinspace III] emission arises from a small localized area of higher ionized gas and $T_{\rm e} (\text{[O\thinspace III]})$ may be not representative of the He$^{+}$ volume, as we describe in Sec.~\ref{subsec:small_scale_pc}. 

C\thinspace II $\lambda 4267$ is partially blended in the two velocity components of cut~1 and therefore we base our calculations on C\thinspace II $\lambda 9903$. We use C\thinspace II $\lambda 4267$ in cut 2. C$^{2+}$ abundance estimations based on both lines are in complete agreement in cut~2. Due to the similar ionization potentials of C$^{+}$ and He$^{0}$ and the considerations outlined in the previous paragraph, $T_{\rm e}(\text{low})$ is also used for determining the C$^{2+}$ abundance in HH~204.  
%\eduardo{Due to the similar ionization potentials between C$^{2+}$ and He$^{+}$, the adequacy of $T_{\rm e}(\text{low})$ to this last ion in HH~204 suggest the similar situation for C$^{2+}$. Basically, the lower limit of the ionization ranges of both ions will be decisive due to the general low degree of ionization of the gas.}
%\jorge{Me surge una pregunta. ¿No podr\'ia ocurrir que la emisi\'on de C$^{2+}$ provenga de la misma zona de alta ionizaci\'on en la que se emite [O III]?}
%As in the case of the He$^{+}$ abundance determinations, we consider $T_{\rm e}(\text{low})$ as representative of this ion in HH~204.

Contrary to the situation presented in Paper~I, in HH~204,  O\thinspace I RLs from multiplet 1 are severely affected by telluric emission features with the exception of O\thinspace I $\lambda 7772$. We derive the O$^{+}$ abundance of the HH object using the intensity of this line and the predicted line strengths from \citet{Wiese96} following Eq.~2 of \citet{Esteban98}.

\begin{figure}
\centering
\includegraphics[width=\columnwidth]{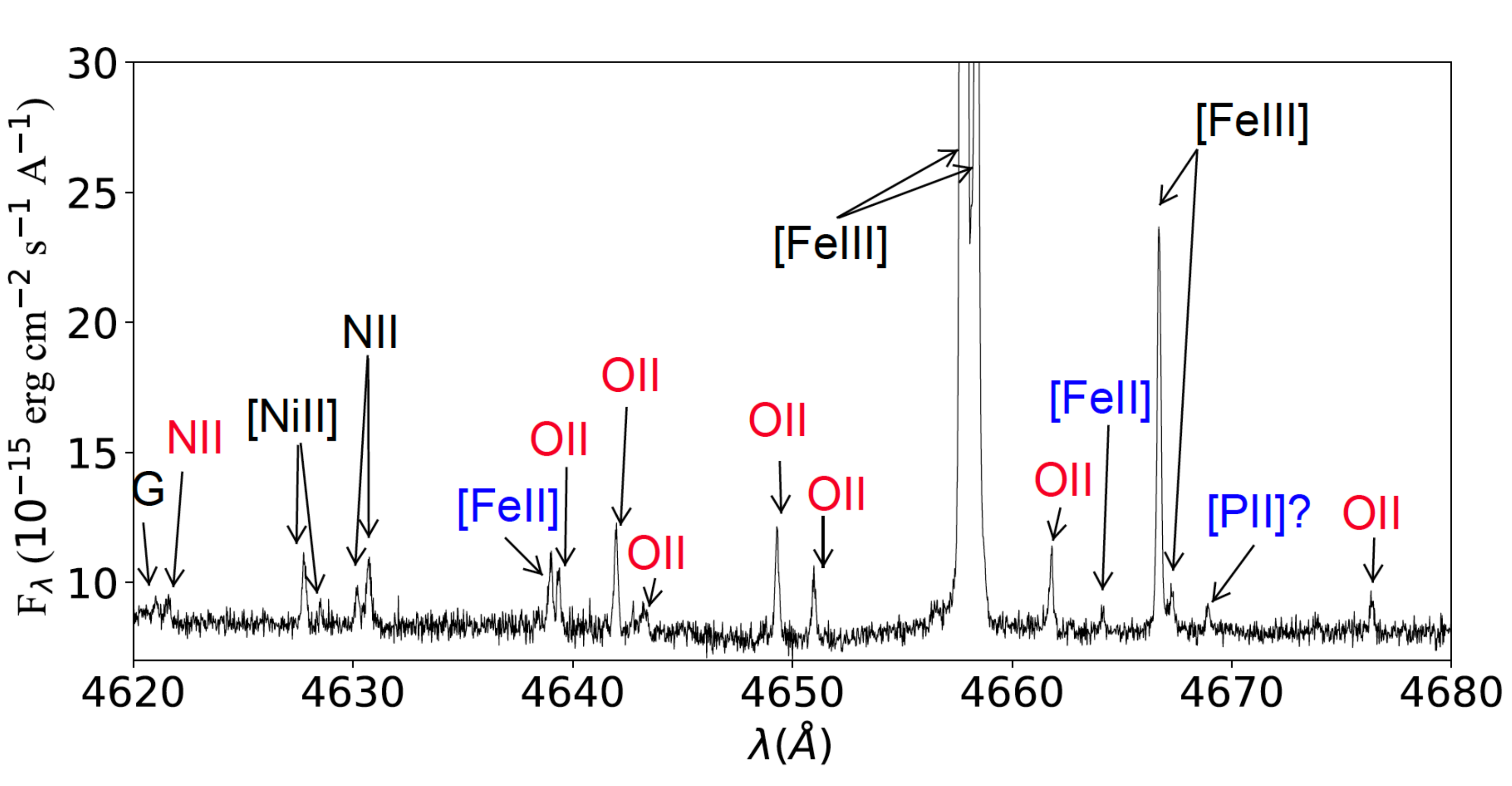}
\caption{Section of the spectrum of the spatial cut~1 covering the spectral range 4620-4680\AA. Lines emitted exclusively by HH~204 are marked in blue, while those of the Orion Nebula are marked in red. The lines observed in both components are highlighted in black. Several O\thinspace II RLs from multiplet 1 are present in the nebular component, but they are not observed in HH~204 due to its low ionization degree. The emission marked with a G is a ghost feature.}
\label{fig:range_of_rls}
\end{figure}

Estimations of the O$^{2+}$ abundance from RLs are based on the available  O\thinspace II lines of multiplet 1. These are not detected in the case of HH~204 (see Fig.~\ref{fig:range_of_rls}). We use an estimate of the upper limit to the intensity of $\lambda 4649$ line for this component.

\begin{deluxetable*}{ccccc}
\tablecaption{Chemical abundances obtained with CELs of the integrated spectra of each component. \label{tab:cels_abundances}}
\tablewidth{0pt}
\tablehead{
 & \multicolumn{2}{c}{Cut 1} & \multicolumn{2}{c}{Cut 2} \\
Ion & HH~204 & Nebula + DBL & DBL  & Nebula
}
\startdata
O$^{+}$ &  $8.62 \pm 0.05 $ & $8.14 \pm 0.05 $ & $8.26^{+0.13} _{-0.09}$&$8.18^{+0.06} _{-0.05}$\\
O$^{2+}$ & $6.34 \pm 0.02 $ & $7.96 \pm 0.02 $ & $7.33^{+0.15} _{-0.10}$&$8.04 \pm 0.02 $\\
N$^{+}$  & $7.72 \pm 0.03 $& $7.34 \pm 0.03 $&$7.40^{+0.08} _{-0.06}$&$7.29^{+0.04} _{-0.03}$\\
Ne$^{2+}$ & $5.05 \pm 0.03 $ & $7.16 \pm 0.02 $ &-&$7.23^{+0.03} _{-0.02}$\\
S$^{+}$&$6.60 \pm 0.04 $ & $5.93 \pm 0.03$ & $5.92^{+0.07} _{-0.06}$&$5.86^{+0.04} _{-0.03}$\\
S$^{2+}$& $6.80 \pm 0.03 $ & $6.84 \pm 0.03 $ &$6.85^{+0.10} _{-0.08}$&$6.89 \pm 0.04 $ \\
Cl$^{+}$ & $4.72 \pm 0.03 $&  $4.17 \pm 0.03 $&$4.08^{+0.10} _{-0.09}$&$4.05 \pm 0.04 $\\
Cl$^{2+}$ & $4.77^{+0.04} _{-0.03}$ & $4.93 \pm 0.04 $&$4.99^{+0.16} _{-0.12}$&$4.98^{+0.06} _{-0.05}$\\
Ar$^{2+}$ & $5.66 \pm 0.03 $ & $6.10 \pm 0.02 $&$5.99^{+0.10} _{-0.08}$ & $6.12 \pm 0.02 $\\
Ar$^{3+}$ & - & $3.64^{+0.13} _{-0.12}$&-&-\\
Fe$^{+}$ & $6.16 \pm 0.04 $ & - & -&$4.72 \pm 0.08 $\\
Fe$^{2+}$ & $6.49 \pm 0.02$ & $5.72 \pm 0.04$&$5.56^{+0.10} _{-0.08}$&$5.77 \pm 0.04$\\
Fe$^{3+}$ & $<5.11$ & $5.73 \pm 0.13 $&-&-\\
Ni$^{+}$ & $4.89 \pm 0.02 $ & - &-&-\\
Ni$^{2+}$ & $5.13 \pm 0.03 $ & $4.37 \pm 0.09 $&-&-\\
Ca$^{+}$ & $3.50 \pm 0.03 $ & -&-&-\\
Cr$^{+}$ & $4.28 \pm 0.03 $ & -&-&-\\
\enddata
\tablecomments{Abundances in units of 12+log(X$^{\text{n}+}$/H$^+$).}
\end{deluxetable*}

\begin{deluxetable*}{ccccccccccccc}
\tablecaption{Chemical abundances obtained with RLs of the integrated spectra of each component. \label{tab:rls_abundances}}
\tablewidth{0pt}
\tablehead{
 & \multicolumn{2}{c}{Cut 1} & \multicolumn{2}{c}{Cut 2} \\
Ion & HH~204 & Nebula + DBL & DBL  & Nebula
}
\startdata
He$^{+}$  &$10.53 \pm 0.02$&$10.85 \pm 0.03$&$10.66 \pm 0.06$&$10.92 \pm 0.04$\\
O$^{+}$ & $8.57 \pm 0.03 $& -&-&-\\
O$^{2+}$ & $<7.54$ & $8.25 \pm 0.06$&-&$8.40 \pm 0.03$ \\
C$^{2+}$  &  $7.76 \pm 0.07 $ & $8.22 \pm 0.04 $ &-&  $8.37 \pm 0.02 $\\
\enddata
\tablecomments{Abundances in units of 12+log(X$^{\text{n}+}$/H$^+$).}
\end{deluxetable*}

\section{Unveiling HH~204}
\label{sec:small_scale}

As mentioned in Sec.~\ref{sec:line_int} we measure several lines pixel by pixel along the slit. The spatial re\-so\-lu\-tion in the blue and red arms of UVES is slightly different (0.246$\arcsec$/pixel and 0.182$\arcsec$/pixel respectively). Cut~1 include 30 pixels in the blue arm and 42 in the red one. In the pixel by pixel measurements, renormalization between lines in common in each arm is not enough to dilute possible differences in the integrated flux. However, $\text{H}\beta$ is observed in the spectra of both arms and therefore we split our pixel-spectra in two parts, 27 blue-pixel-spectra and 37 red-pixel-spectra, both groups normalized with respect to $F(\text{H}\beta)$. The missing first pixels (from east to west) of cut~1 of both arms were not included since the emission of HH~204 was too faint. We proceeded as follows: based on the [Fe\thinspace III] $\lambda4658/\lambda 4702$ line ratios we derive $n_{\rm e}$ along HH~204 in the blue arm. Once the density distribution was estimated,  the calculation of $T_{\rm e}(\text{[O\thinspace III]})$ was done, also in the blue arm through the [O\thinspace III] $\lambda4363/\lambda 4959$ line ratio. The spatial distribution of $n_{\rm e}$ was linearly interpolated in the red arm to estimate $T_{\rm e}(\text{[S\thinspace III]})$ and $T_{\rm e}(\text{[N\thinspace II]})$. Once the physical conditions are determined, we estimate the ionic abundances using the same procedure followed in Sec.~\ref{subsec:ionic_abundances}. The zero point of the spatial distribution is located at coordinates: RA(J2000)=05$^h$35$^m$22$^s$.81, DEC(J2000)=$-$05$^{\circ}$25$'$21.86$''$, just at the apparent eastern -- external -- edge of the bowshock. To estimate the distance from the bowshock along the jet, we adopt an heliocentric distance of $410 \pm 10 \text{ pc}$ \citep{Binder2018} to the Orion Nebula, based on \textit{Gaia} DR2 parallaxes \citep{gaiadr2}. The integrated emission is dominated by the blue-shifted jet bullet component centered around $\sim -20 \text{ km s}^{-1}$ (in an heliocentric velocity scale) within a 1-$\sigma$ range of $\pm 10 \text{ km s}^{-1}$, being well separated from the DBL and the nebular emission.

\subsection{Small scale physical conditions}
\label{subsec:small_scale_pc}

\begin{figure}
\centering
\includegraphics[width=\columnwidth]{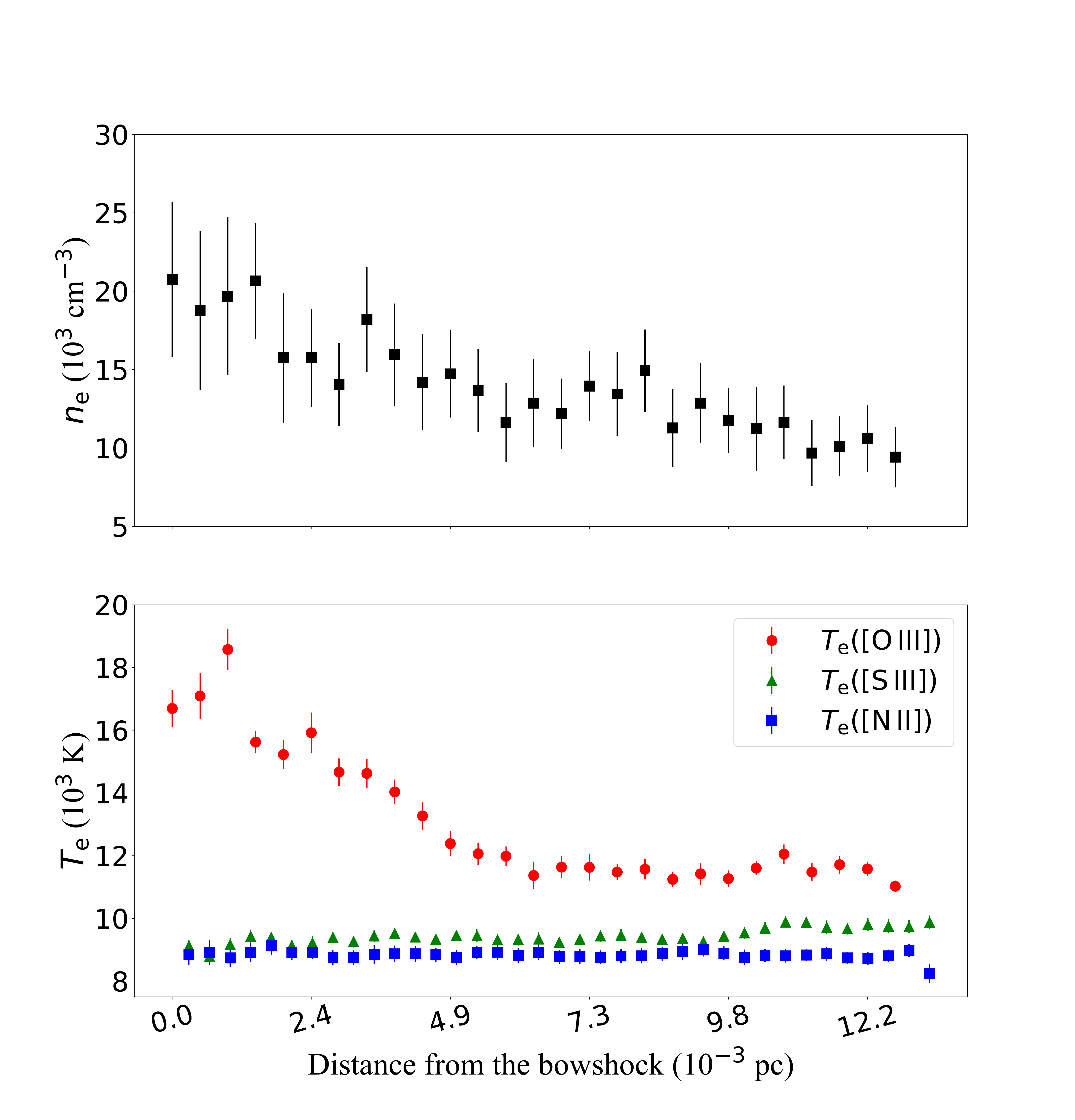}
\caption{Spatial distribution of physical conditions as a function of the distance to the eastern -- external -- edge of the bowshock of HH~204. \textit{Upper panel:} $n_{\rm e}(\text{[Fe\thinspace III]})$. The gas compresses as it approaches to the shock front, increasing its density. \textit{Bottom panel:} $T_{\rm e}$ estimates with 3 diagnostics. While $T_{\rm e}(\text{[S\thinspace III]})$ and $T_{\rm e}(\text{[N\thinspace II]})$ remain unaltered, $T_{\rm e}(\text{[O\thinspace III]})$ shows a strong increase when approaching the shock. }
\label{fig:small_scale_physical_conditions}
\end{figure}

The resulting pixel by pixel distribution of physical conditions is shown in Fig.~\ref{fig:small_scale_physical_conditions}. At the shock front, we can see that $n_{\rm e}\text{([Fe\thinspace III])}$ reaches values up to a factor of about 2 higher than at a distance of $\sim$13$ \text{ mpc}$ from the bowshock. The distribution of $T_{\rm e}(\text{[N\thinspace II]})$ is practically constant, while $T_{\rm e}(\text{[S\thinspace III]})$ decreases slightly at the edge of the bowshock. 

Conversely, $T_{\rm e}(\text{[O\thinspace III]})$ strongly increases at distances closer to the bowshock. In the presence of a shock, a photoionized gas can be heated at a temperature higher than that fixed by  photoionization equilibrium \citep{zeldovich67} (see Sec.~11 of Paper~I). After the shock passage, the gas cools down by radiative emission until reaching an equilibrium temperature, forming a cooling zone whose extension will be inversely proportional to the electron density \citep[][]{Hartigan87}. If we assume that the high-$T_{\rm e}(\text{[O\thinspace III]})$ area corresponds to the cooling zone formed after the shock, the fact that $T_{\rm e}(\text{[S\thinspace III]})$ and $T_{\rm e}(\text{[N\thinspace II]})$ are not affected in the same way, suggest that the high-ionization degree emission should come (at least partially) from a different gas volume than the one that originates the low-ionization  emission. Therefore, we suggest that we are seeing the superposition of two different emission components: one from the bow shock and one from the Mach disk (the shock internal to the jet). This model will be in discussed in Sec.~\ref{subsubsec:two_temps_model}.

\begin{figure}
\centering
\includegraphics[width=\columnwidth]{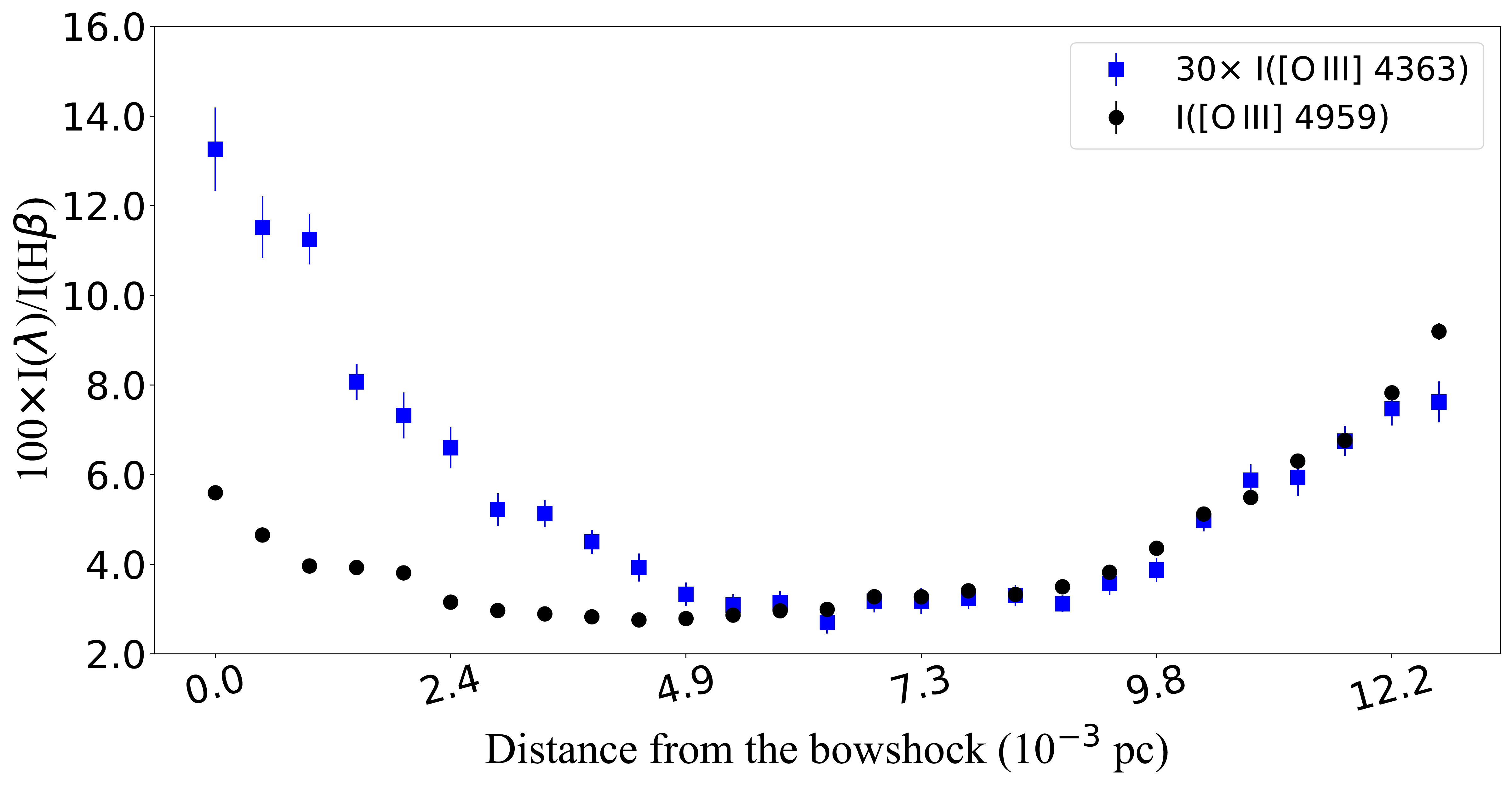}
\caption{Same as Fig.~\ref{fig:small_scale_physical_conditions} for the intensities of the [O\thinspace III] $\lambda \lambda 4363, 4959$ lines. Line intensity ratios with respect to H($\beta$) have been normalized for a clearer comparison.}
\label{fig:O3_lines}
\end{figure}

%Although part of the aforementioned high-ionization degree gas may be out of photoionization equilibrium, fortunately, its impact is negligible in the global abundance analysis of HH~204. Considering the ionization fraction $\text{O}^{2+}/(\text{O}^{+}+\text{O}^{2+})=0.005\pm 0.001$ (see Table~\ref{tab:cels_abundances}) -- which would increase to $\text{O}^{2+}/(\text{O}^{+}+\text{O}^{2+})=0.017\pm 0.003$ if we determine the O$^{2+}$ abundance using $T_{\rm e}(\text{[N\thinspace II]})$ -- we infer that the contribution of the O$^{2+}$ from the high-ionization degree gas is around $\sim 1\%$ of the oxygen abundance. A similar result is found for other metals. For example, by considering the solar Ne/O ratio recommended by \citet{lodders19} and the Ne$^{2+}$/O value of HH~204, we estimate that $\text{Ne}^{2+}/\text{Ne}\sim 0.001$. The fact that $T_{\rm e}(\text{[N\thinspace II]})$ and $T_{\rm e}(\text{[S\thinspace III]})$ are kept in balance in HH~204 along the observed pixels proves that the low and medium-ionization degree gas, which comprises more than $\sim 99\%$ of the total, is in photoionization equilibrium.

\subsection{Small-scale patterns in the ionic abundances}
\label{subsec:small_scale_ca}

\begin{figure}
\centering
\includegraphics[width=\columnwidth]{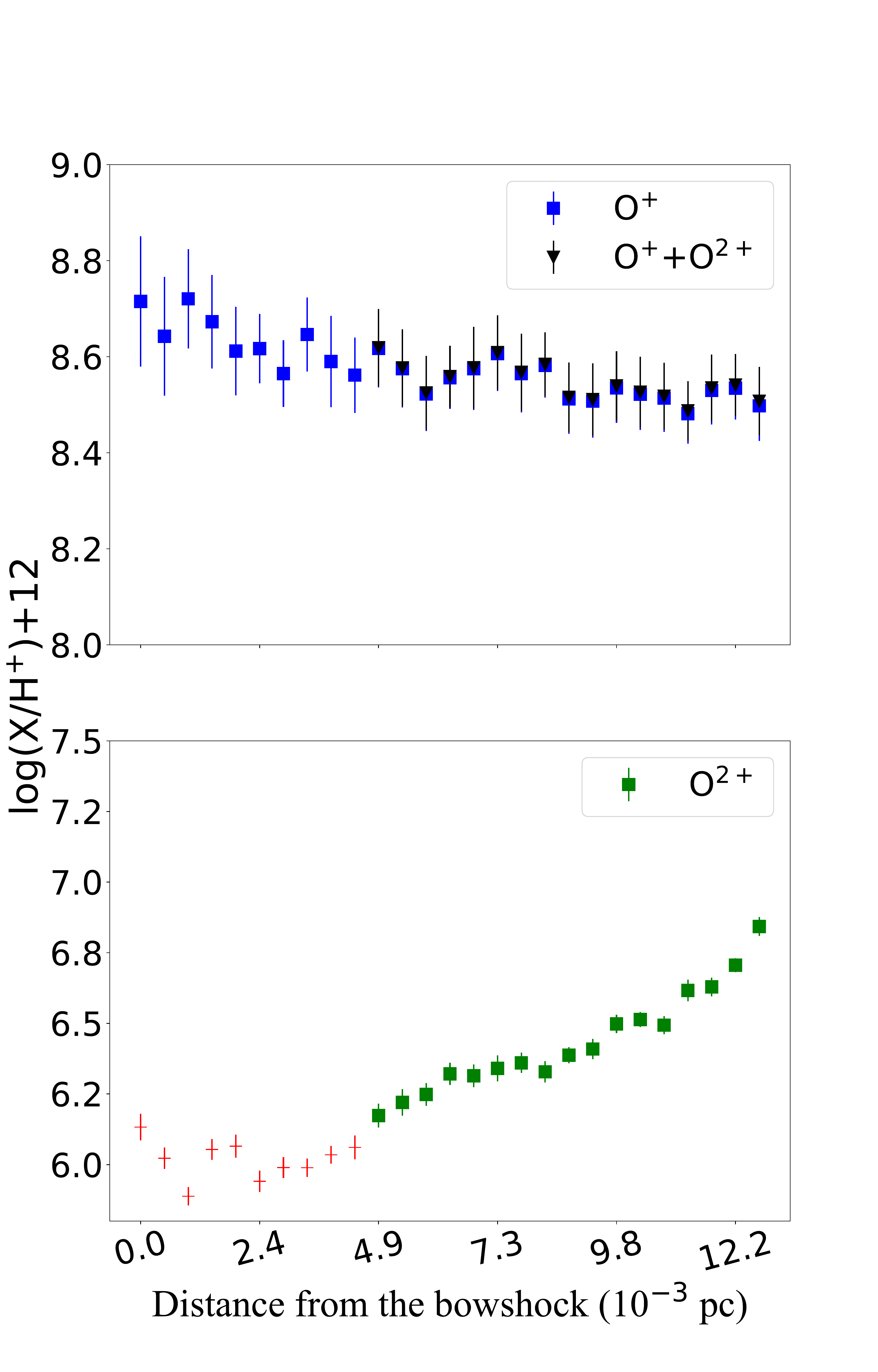}
\caption{Same as Fig.~\ref{fig:small_scale_physical_conditions} for ionic and total abundances of O. \textit{Upper panel:} O$^{+}$ and total O abundances. The total abundance of O was calculated as the sum of O$^{+}$ and O$^{2+}$ in the area where $T_{\rm e}(\text{[O\thinspace III]})$ remains constant -- distances between 4.9 and 13 mpc from the bowshock --. The contribution of O$^{2+}$ to total O abundance is negligible compared to the abundance of O$^{+}$. \textit{Bottom panel:} O$^{2+}$ abundances. The red crosses  show the zone clearly affected by the shock (see Fig.~\ref{fig:small_scale_physical_conditions}, Fig.~\ref{fig:O3_lines} and Sec.~\ref{subsec:small_scale_pc} ). The green squares indicate the area where $T_{\rm e}(\text{[O\thinspace III]})$ remains constant.} %\cesar{los simbolos rojos no se ven como cruces, pon el trazo mas fino } \jorge{Siguiendo la norma que aplican las revistas hoy en d\'ia. Hay que evitar gr\'aficas que no sean "color-blind friendly" As\'i que cambiar\'ia los colores y remarcar\'ia la diferencia en los s\'imbolos}
\label{fig:O_abundances}
\end{figure}

Fig.~\ref{fig:O_abundances} shows the spatial distribution of the ionic abundances of O. As described in Sec.~\ref{subsec:small_scale_pc}, the increase of $T_{\rm e}(\text{[O\thinspace III]})$ may be related to shock heating. Therefore, we highlight in red the O$^{2+}$ abundances in this area in the bottom panel of Fig.~\ref{fig:O_abundances}. In the upper panel, we show the O$^+$ abundances along the full distance range and the O ones in the area where $T_{\rm e}(\text{[O\thinspace III]})$ remains constant. This panel shows that practically all O is in O$^{+}$ form. It should be noted that an increase of a factor of $\sim$2 in the O$^{2+}$ abundance would represent  less than $1\%$ of the total O, well below the associated uncertainties and, therefore, this increase would be undetected in analyses lacking our spatial and spectral resolutions.

\begin{figure}
\centering
\includegraphics[width=\columnwidth]{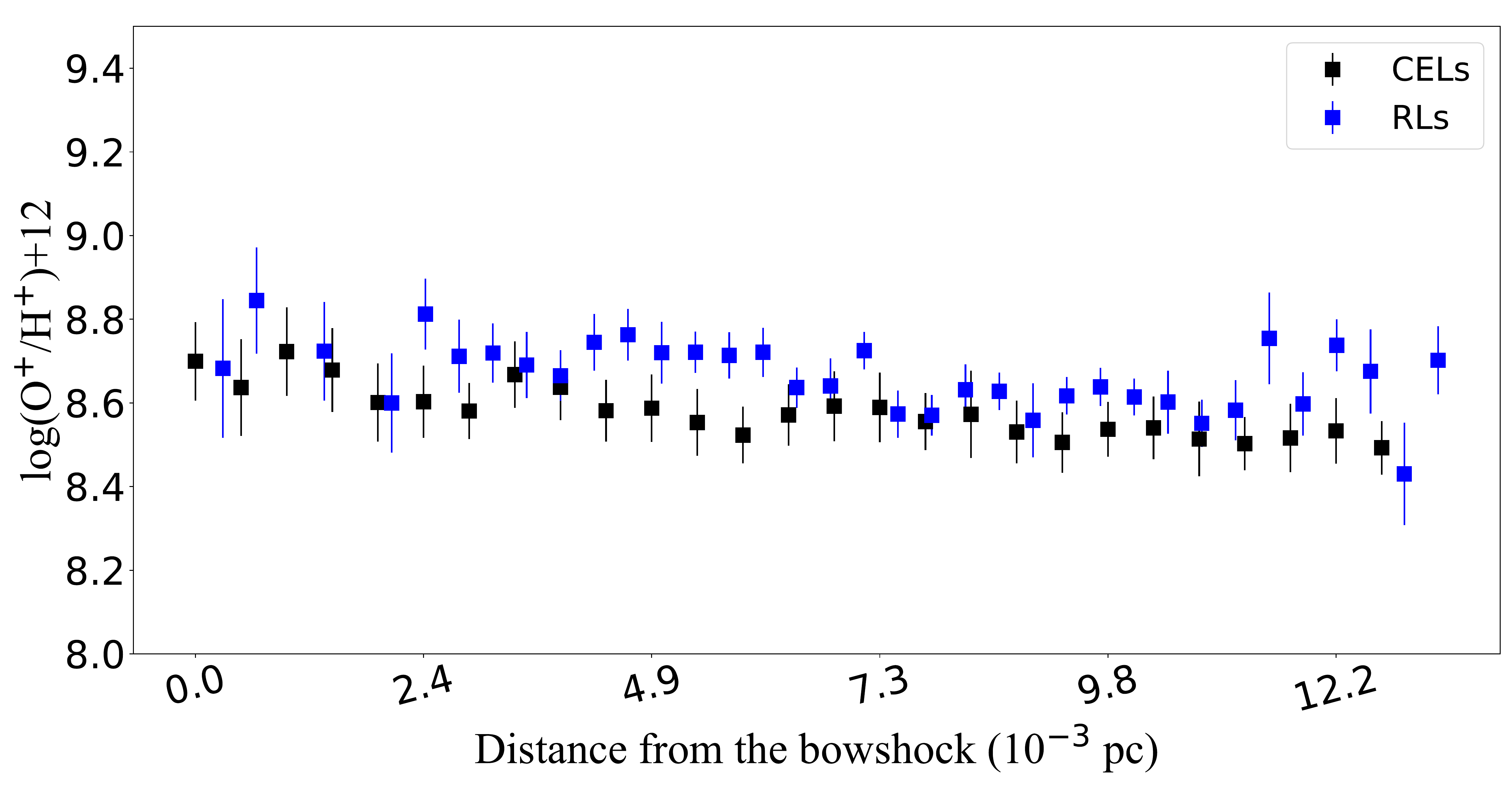}
\caption{Same as Fig.~\ref{fig:small_scale_physical_conditions} for O$^{+}$ abundances determined with CELs and RLs.}
\label{fig:O2_abundances_cels_rls}
\end{figure}

In tables~\ref{tab:cels_abundances} and \ref{tab:rls_abundances} we can see that the O$^{+}$ abundances determined from CELs and RLs for  HH~204 are the same within the errors, so we do not find an abundance discrepancy (AD) for this ion, contrary to the situation found in practically all photoionized nebulae. Fig.~\ref{fig:O2_abundances_cels_rls} indicates the absence of systematical trends of the AD in the observed areas of HH~204. Although some fluctuations seem to be present, they are very small in any case.% {\bf It should be considered that the pixel-by-pixel measurements of faint lines, such as O\thinspace I $\lambda 7771.94$, are susceptible to variations in the sensitivity of the pixels along the spatial axis, that are difficult of estimate within the error bars, while this goes unnoticed in brighter lines and in the integrated spectra along the spatial axis, as used in Sec.~\ref{sec:gen_analysis}.}
 %\jorge{Quiz\'as mencionar otra vez que esto no se ha podido hacer con el O++ debido a que el bajo grado de ionización de HH204 no ha permitido detectar las líneas de O II.}

\begin{figure}
\centering
\includegraphics[width=\columnwidth]{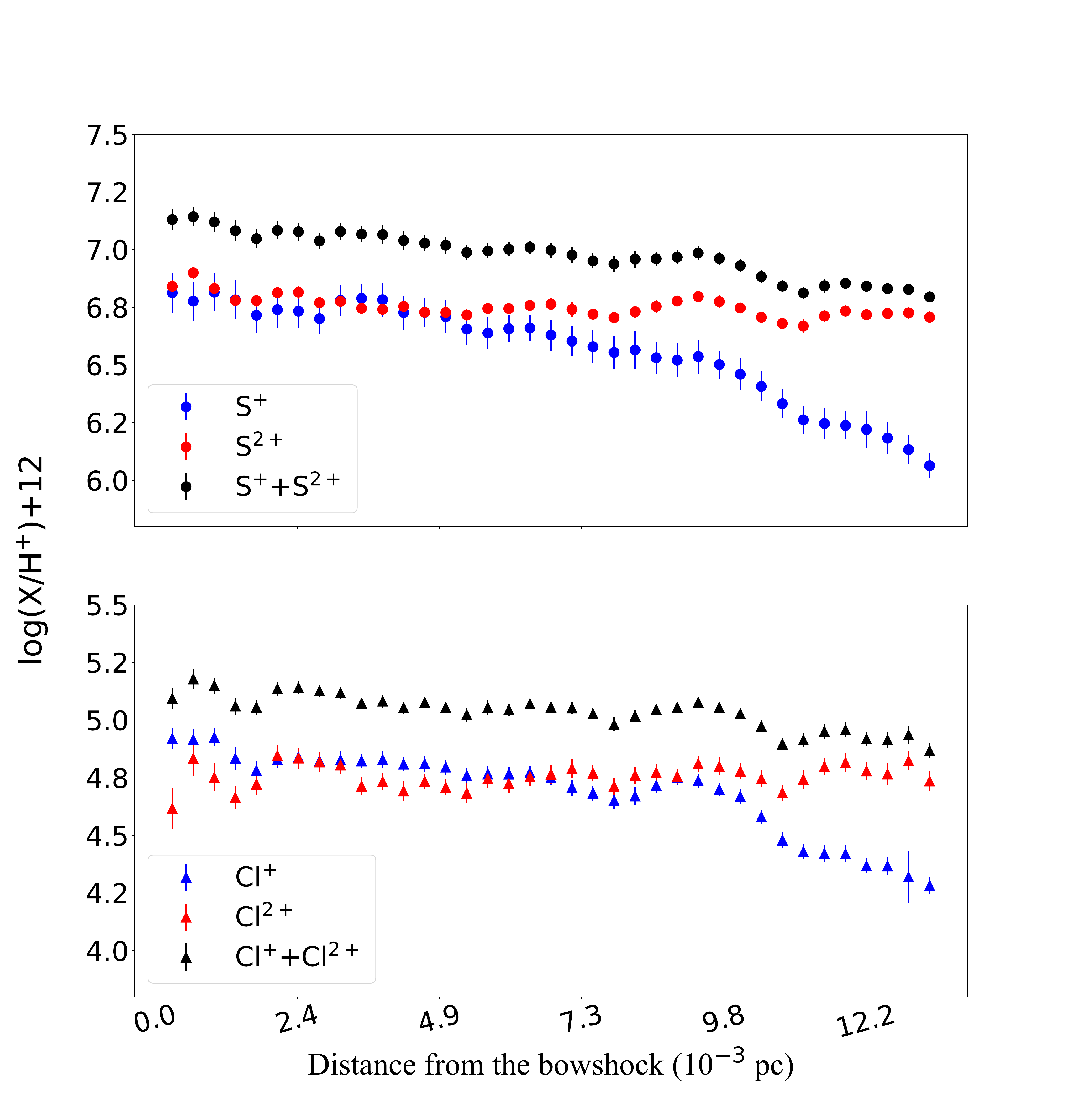}
\caption{Same as Fig.~\ref{fig:small_scale_physical_conditions} for ionic abundances of S (upper panel) and Cl (bottom panel). }
\label{fig:cl_s_abundances}
\end{figure}

In Fig.~\ref{fig:cl_s_abundances} we present the ionic abundances of Cl and S. The species of the same ionic stage of both elements show similar pixel by pixel distributions. The variations of S$^{2+}$/H$^+$ and Cl$^{2+}$/H$^+$ ratios along HH~204 are comparatively much smaller than those of S$^{+}$/H$^+$ and Cl$^{+}$/H$^+$, that show a decrease of 0.8 dex along the diagram as the distance from the bowshock increases. %This trend is consistent by a rapid decrease of the degree of ionization of the gas that decreases $n(\text{S}^{3+})$ and $n(\text{Cl}^{3+})$, favoring the abundance of the less ionized ions, which is more noticeably in $n(\text{S}^{+})$ and $n(\text{Cl}^{+})$  that initially present lower abundances. 
At distances to the bowshock smaller than $\sim$ 4.9 mpc, the abundances of S$^{+}$ and Cl$^{+}$ seem to stabilize and presumably, almost all S and Cl must be only once and twice ionized. This allows the estimation of their total abundances without an ionization correction factor (ICF).

\begin{figure}
\centering
\includegraphics[width=\columnwidth]{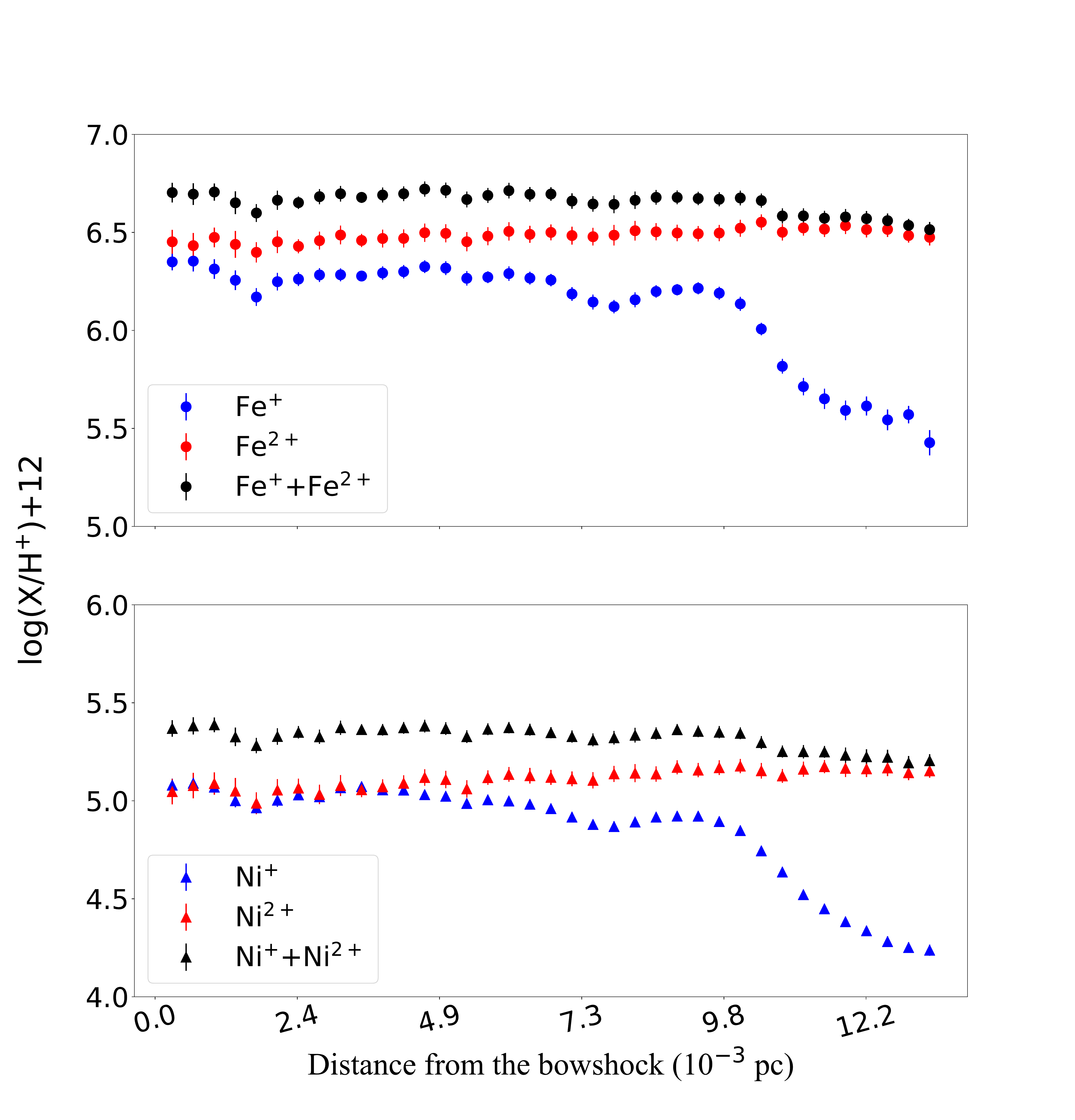}
\caption{Same as Fig.~\ref{fig:small_scale_physical_conditions} for ionic abundances of Fe (upper panel) and Ni (bottom panel). }
\label{fig:fe_ni_abundances}
\end{figure}

\begin{figure}
\centering
\includegraphics[width=\columnwidth]{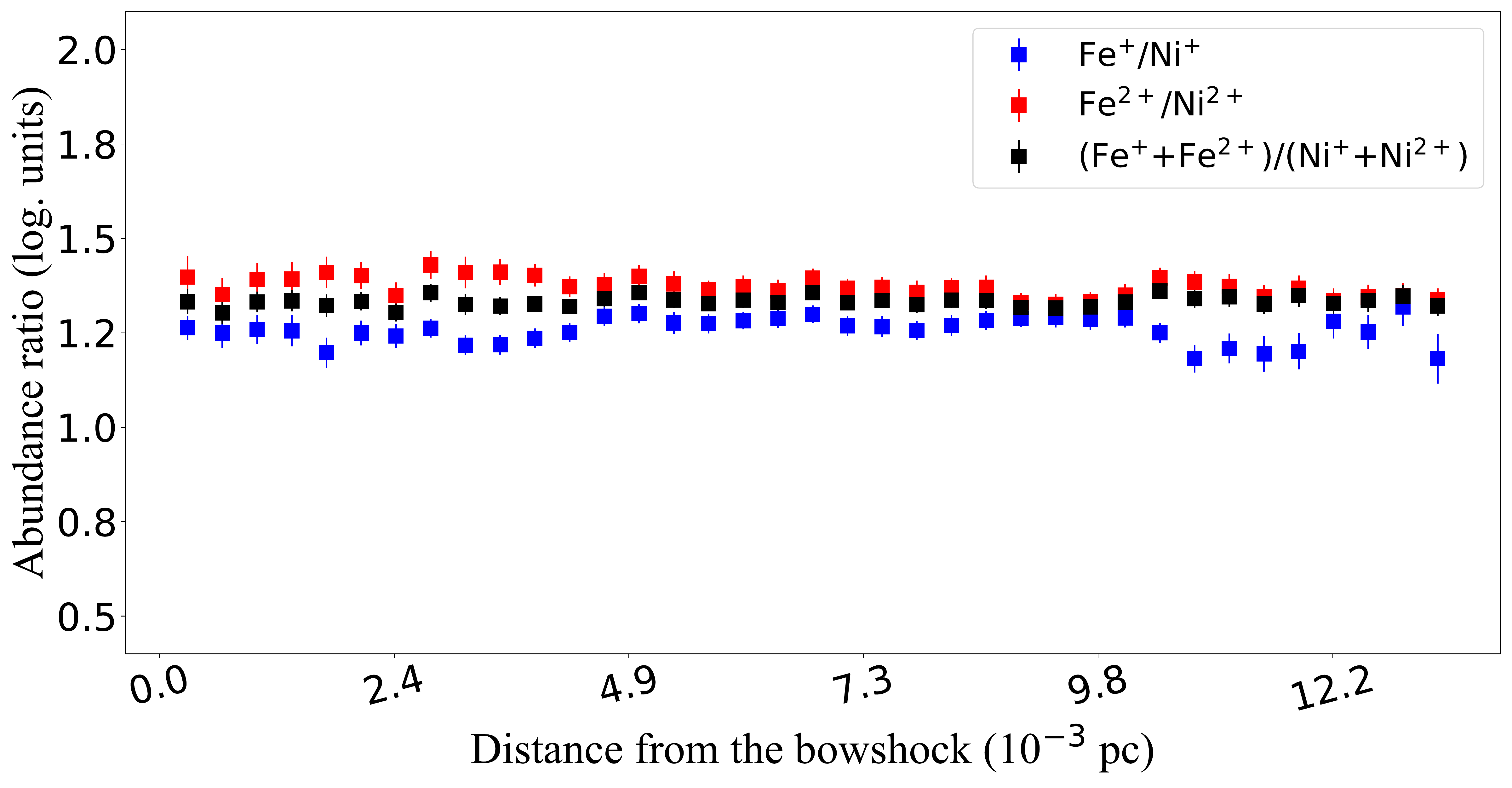}
\caption{Same as Fig.~\ref{fig:small_scale_physical_conditions} for the ratios of ionic abundances of Fe and Ni.}
\label{fig:Fe_Ni_abundance_ratios}
\end{figure}

The pixel by pixel distributions of the ionic abundances of Fe and Ni are clearly correlated, as shown in Fig.~\ref{fig:fe_ni_abundances}. Similar to that found for S and Cl (see Fig.~\ref{fig:cl_s_abundances}), close to the bowshock, the contribution of species of Fe and Ni with ionic charges higher than Fe$^{2+}$ and Ni$^{2+}$ to their total abundances should be negligible. The ratios of the ionic abundances between both elements remain constant as shown in Fig.~\ref{fig:Fe_Ni_abundance_ratios}, being $\text{log}(\text{Fe}^{+}/\text{Ni}^{+})=1.26\pm 0.03$, $\text{log}(\text{Fe}^{2+}/\text{Ni}^{2+})=1.37\pm 0.03$ and $\text{log}(\text{Fe}/\text{Ni})=1.33\pm 0.03$. Although the value of $\text{log}(\text{Fe}^{2+}/\text{Ni}^{2+})$ is slightly above the recommended solar value \citep[$\text{log}(\text{Fe}/\text{Ni})_{\odot}=1.25\pm 0.05$,][]{lodders19}, this may be the consequence of a slight systematic underestimation of Ni$^{2+}$ abundance because, as we discussed in Sec.~\ref{subsec:ionic_abundances_fe_ni}, the atomic data for this ion seems to show some inaccuracies. 

\begin{figure}
\centering
\includegraphics[width=\columnwidth]{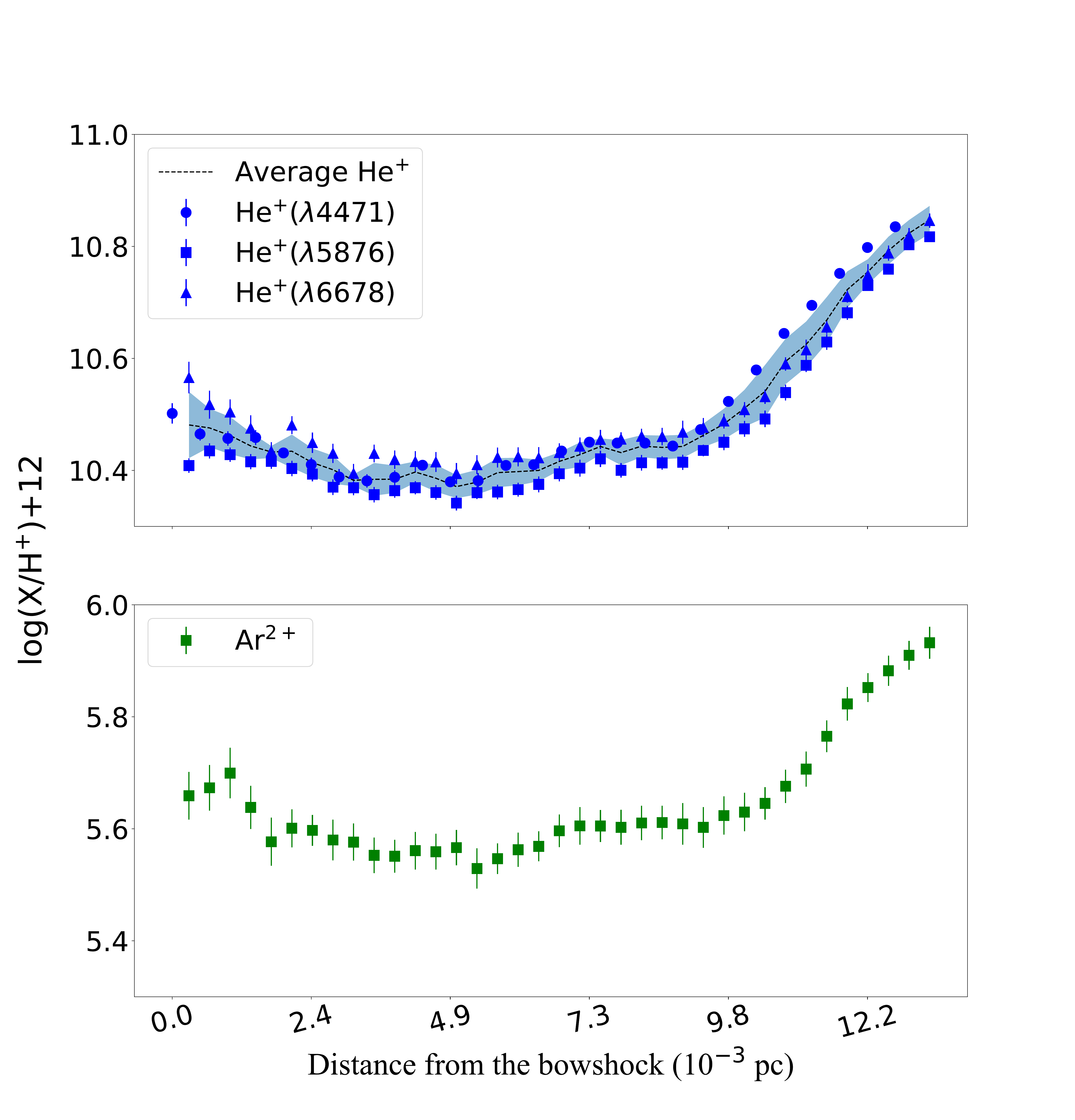}
\caption{Same as Fig.~\ref{fig:small_scale_physical_conditions} for He$^{+}$ abundances (upper panel) and Ar$^{2+}$ abundances (bottom panel). In the upper planel, the black line indicates the average He$^{+}$ abundance obtained with He\thinspace I $\lambda \lambda 4471,5876,6678$. The color band indicates the associated dispersion.}
\label{fig:he_Ar_abundances}
\end{figure}

In Fig.~\ref{fig:he_Ar_abundances} we show the similar pixel by pixel distributions of the He$^{+}$ and Ar$^{2+}$ abundances. Both quantities decrease as we approach the bowshock due to the decrease of the ionization parameter as $n_{\rm e}$ increases. A slight increase is observed at distances less than $\sim 2.4  \text{ mpc}$, probably due to the same process discussed in Sec.~\ref{subsubsec:two_temps_model} for the case of [O\thinspace III] lines. However, the impact of the shock contribution seems to be negligible for these ions. For example, the fact that $T_{\rm e}(\text{He\thinspace I})$ is consistent with $T_{\rm e}(\text{[N\thinspace II]})$ (see Sec.~\ref{subsec:physical_cond}) reflects that the population of the singlet levels, which are the ones used for determining $T_{\rm e}(\text{He\thinspace I})$, are largely unaffected.

\begin{figure}
\centering
\includegraphics[width=\columnwidth]{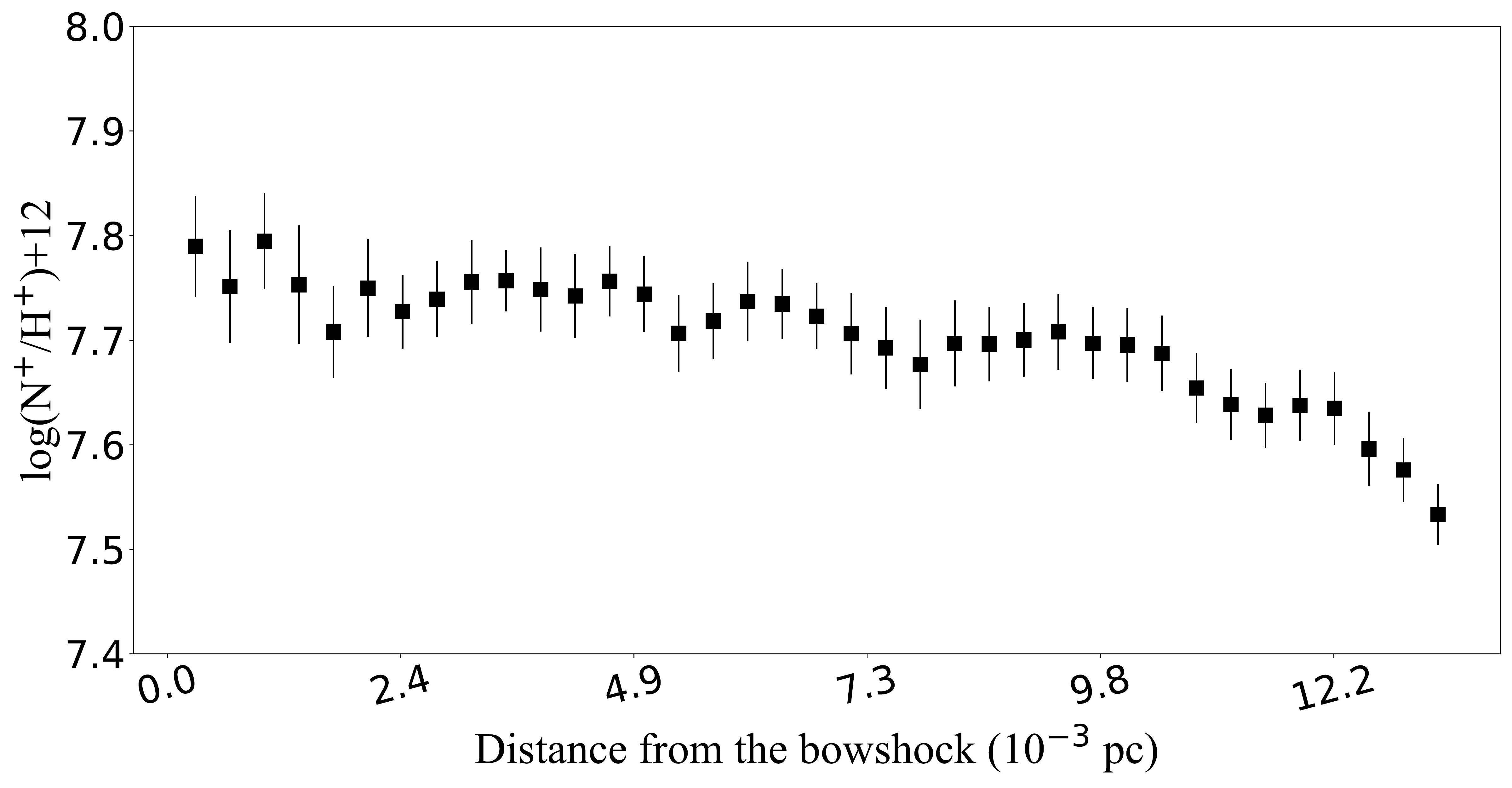}
\caption{Same as Fig.~\ref{fig:small_scale_physical_conditions} for N$^{+}$ abundances.}
\label{fig:N2_abundances}
\end{figure}

\begin{figure}
\centering
\includegraphics[width=\columnwidth]{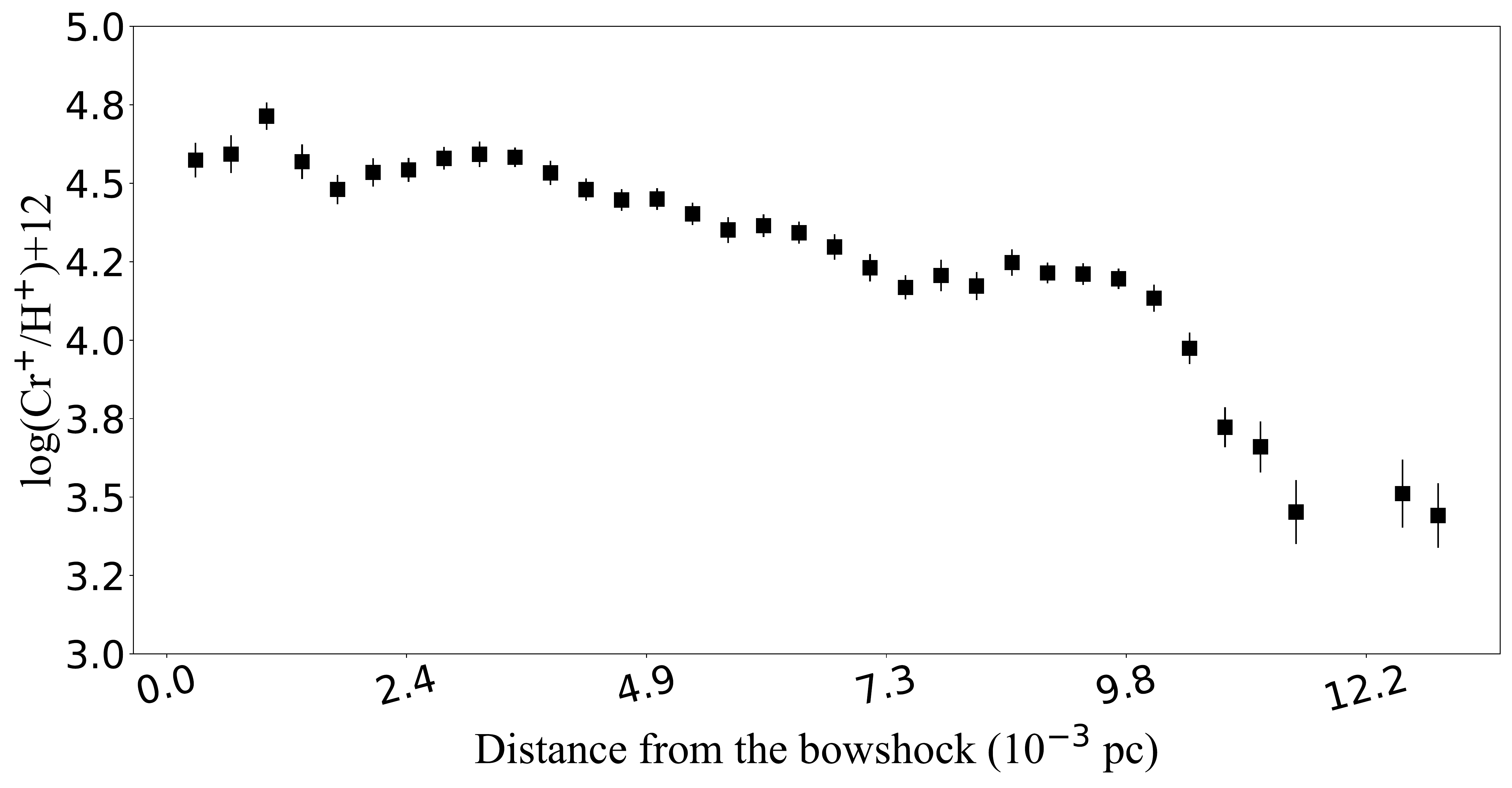}
\caption{Same as Fig.~\ref{fig:small_scale_physical_conditions} for Cr$^{+}$ abundances.}
\label{fig:Cr2_abundances}
\end{figure}

\begin{figure}
\centering
\includegraphics[width=\columnwidth]{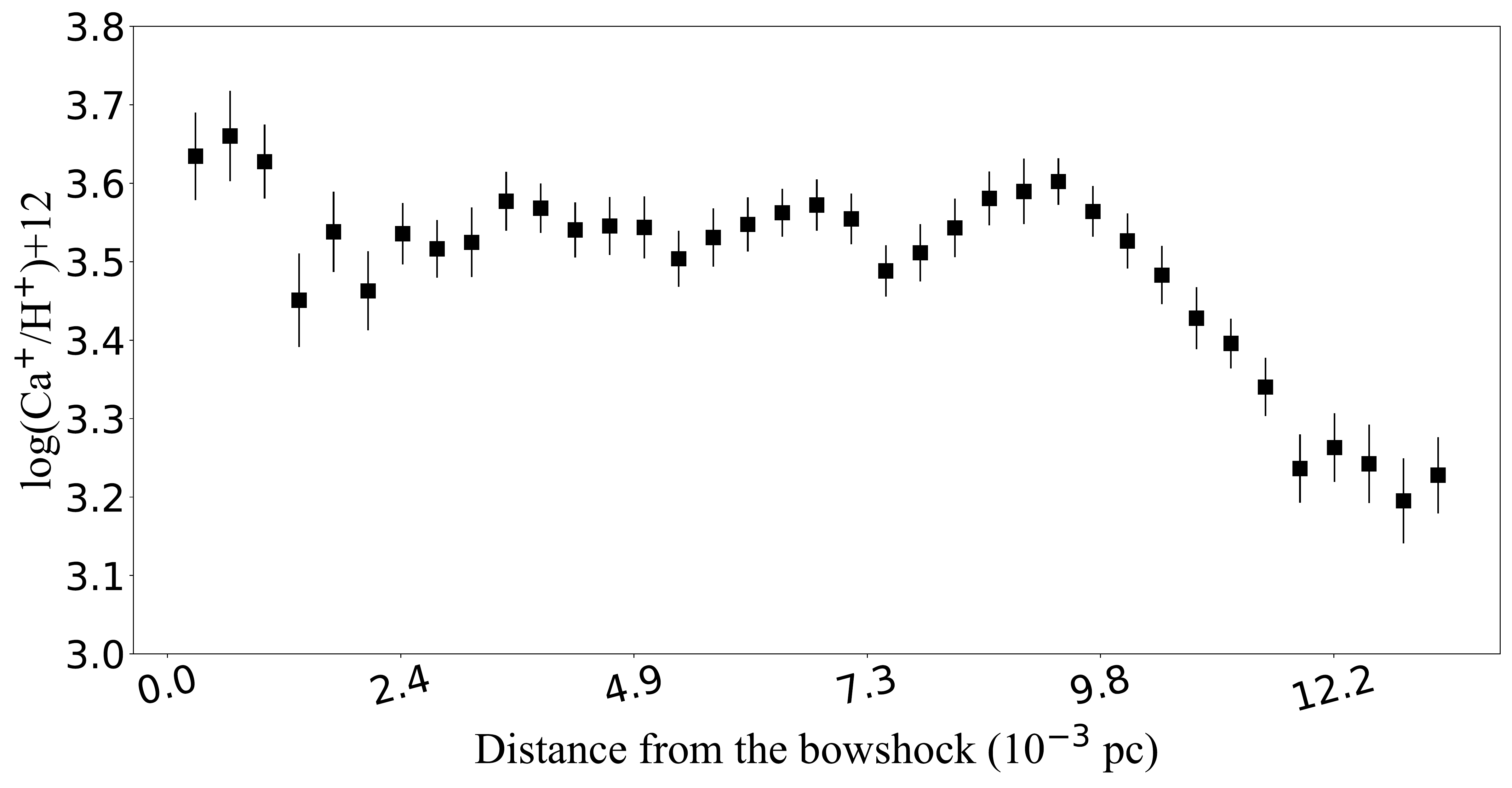}
\caption{Same as Fig.~\ref{fig:small_scale_physical_conditions} for Ca$^{+}$ abundances.}
\label{fig:Ca2_abundances}
\end{figure}

\begin{figure}
\centering
\includegraphics[width=\columnwidth]{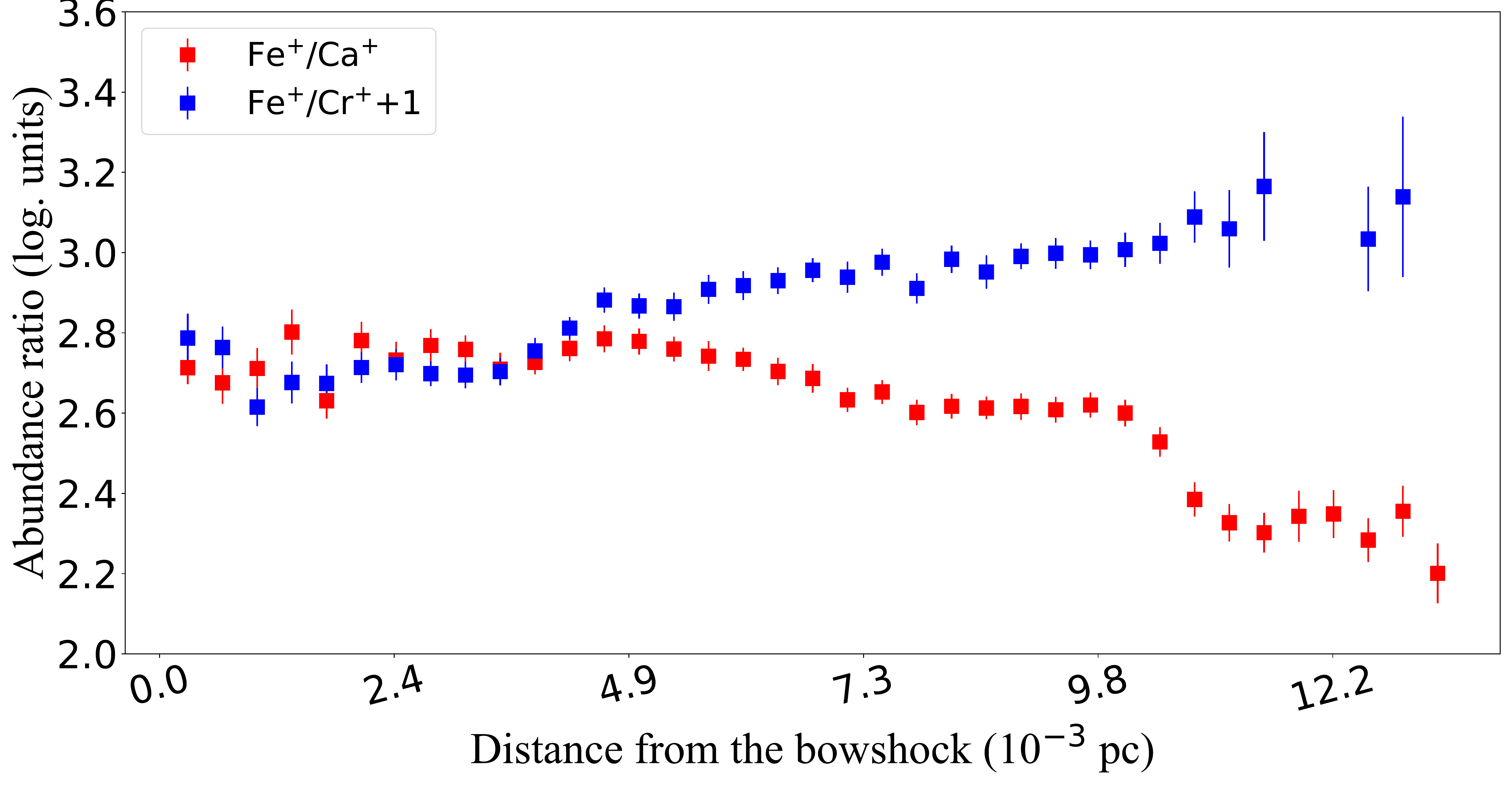}
\caption{Same as Fig.~\ref{fig:small_scale_physical_conditions} for Fe$^{+}$/Ca$^{+}$ and Fe$^{+}$/Cr$^{+}$ abundance ratios.}
\label{fig:Ca_Cr_Fe_abundance_ratios}
\end{figure}

In Fig.~\ref{fig:N2_abundances}, we show that the abundance of N$^+$ increases as we move towards the bowshock from $12+\text{log}(\text{N}^{+}/\text{H}^{+})=7.53\pm 0.03$ to an apparently constant value of $7.75\pm 0.02$. That plateau indicates that all nitrogen should be only once ionized. Figures~\ref{fig:Cr2_abundances} and \ref{fig:Ca2_abundances} show the pixel by pixel distributions of Cr$^+$ and Ca$^{+}$ abundances, respectively, which are somewhat different to the ones of Fe$^{+}$ or Ni$^{+}$ (Fig.~\ref{fig:fe_ni_abundances}). This makes that the distributions of  Fe$^{+}$/Cr$^{+}$ and Fe$^{+}$/Ca$^{+}$ ratios are not constant, contrary to what is obtained for Fe$^{+}$/Ni$^{+}$ (Fig.~\ref{fig:Ca_Cr_Fe_abundance_ratios}). In the case of the Fe$^{+}$/Cr$^{+}$ ratio, the observed trend may be related to the slight differences between their ionization energies or to different depletion patterns. The curve defined by the Fe$^{+}$/Ca$^{+}$ abundance ratio, may be due to the coexistence of this ion and H$^{0}$ in the trapped ionization front of HH~204 (see Sec.~\ref{subsec:deuterium}).

%Fig.~\ref{fig:Cr2_abundances} shows the derived Cr$^+$ abundances. There are differences between the observed trend and what is shown in Fig.~\ref{fig:fe_ni_abundances} for the Fe$^{+}$ and Ni$^{+}$ abundances, that make the trend between Cr$^{+}$/Fe$^{+}$ or Cr$^{+}$/Ni$^{+}$ abundances not constant, unlike what is obtained with Fe$^{+}$/Ni$^{+}$. \cesar{No veo la diferencia, me parece que el comportamiento es bastante similar. Igual seria mejor representar Cr+/Fe+ para ver mejor esas diferencias} This might be related to the small difference in the range of their ionization energies or to different depletion patterns. Finally, the Fig.~\ref{fig:Ca2_abundances} shows the abundances of Ca$^{+}$, which also differs from what is seen in Fe$^{+}$ and Ni$^{+}$, due to the possible coexistence of Ca$^{+}$ and H$^{0}$ in the ionization front.\cesar{Lo mismo ocurre aqui, no veo esas diferencias, di cuales son o representa Ca+/Fe+}

\subsection{Deuterium lines in HH~204}
\label{subsec:deuterium}

Deuterium emission lines were first identified in the Orion Nebula by \citet[][]{hebrard00a}. Unlike the expected isotopic shift of $-81.6 \text{ km s}^{-1}$ with respect to the hydrogen lines, they observed a shifted emission around $\sim -71 \text{ km s}^{-1}$ from H$\alpha$ and H$\beta$. The difference of $\sim 10 \text{ km s}^{-1}$ is essentially due to the fact that their emission is produced in different areas of the nebula, where the bulk of gas is moving at different radial velocities. Since the hydrogen lines are produced by recombination in the ionized area that expands towards the observer, the deuterium emission is  mainly due to fluorescence excitation by non-ionizing far-UV continuum in areas slightly beyond the ionization front, as the photon dominated region (PDR) or in the H\thinspace I-H\thinspace II interface \citep[][]{odell01_deu}. After the identification of deuterium emission lines in the Orion nebula, they were also identified in other H\thinspace II regions such as M8, M16, DEM~S~103, M20 and Sh~2-311 \citep[][]{hebrard00b,garciarojas05,garciarojas06,garciarojas07-2}. As in the Orion Nebula, the deuterium emission in these H\thinspace II regions has a narrow line width, consistent with their origin in colder areas.

\begin{figure}
\centering
\includegraphics[width=\columnwidth]{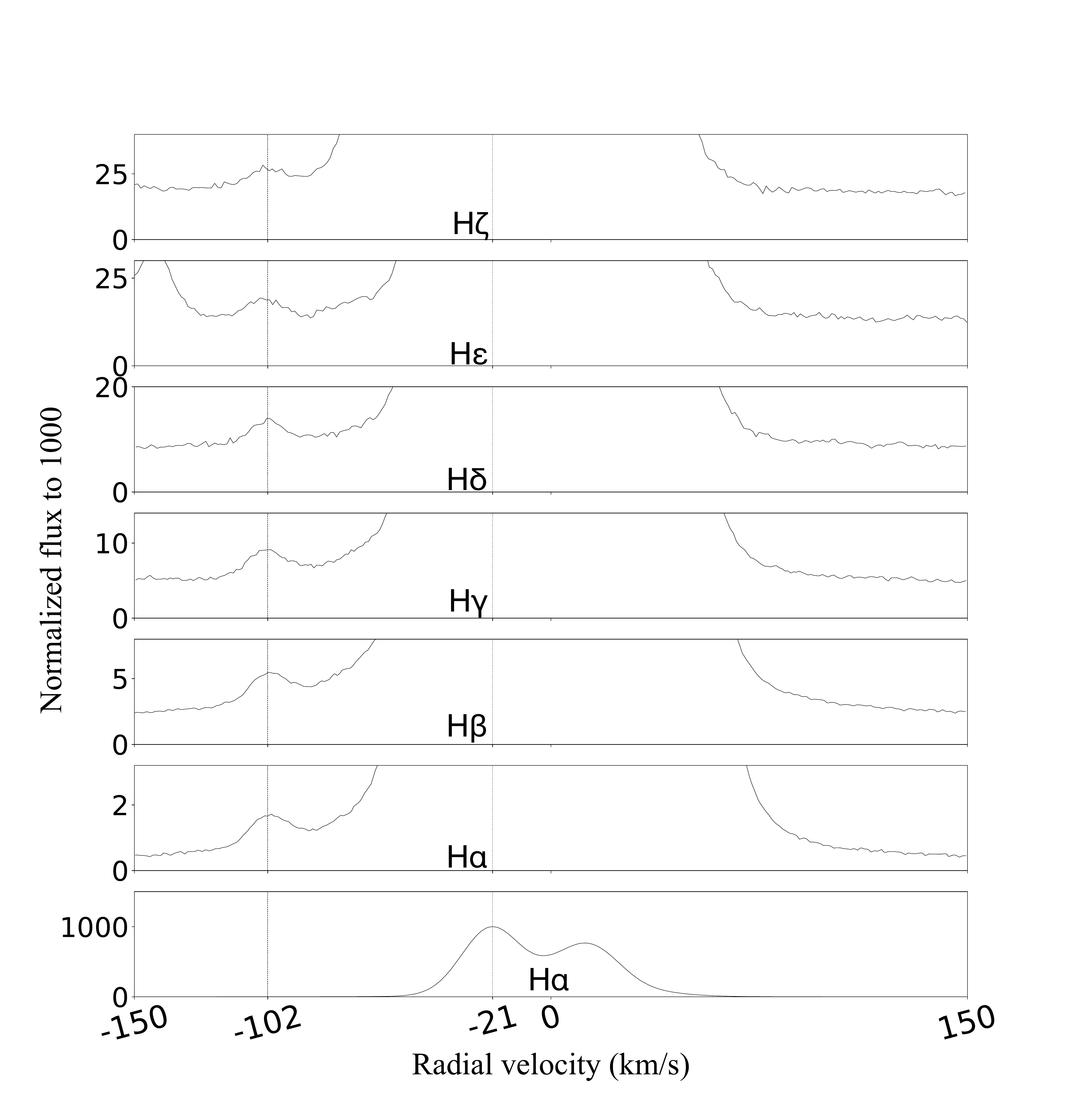}
\caption{Deuterium lines observed in the spectra of cut~1. Vertical lines indicate the position of the deuterium lines and the H\thinspace I emission from HH~204. The flux is normalized to the peak emission of HH~204 in each case. }
\label{fig:deute}
\end{figure}

In this work, we detect the emission of D$\zeta$, D$\varepsilon$, D$\delta$, D$\gamma$, D$\beta$ and D$\alpha$ as shown in Fig.~\ref{fig:deute}. In Table~\ref{tab:deute_table} we present the characteristics of these emissions, including the radial velocity of the D\thinspace I and H\thinspace I lines with respect to the laboratory wavelength of the H\thinspace I ones. The observed isotopic shift of $-81.4 \text{ km s}^{-1}$ between deuterium and hydrogen lines indicates that both kinds of lines arise from HH~204. The observed D\thinspace I/H\thinspace I intensity ratios are in good agreement with the predictions of the standard model developed by \citet[][]{odell01_deu} for the Orion Nebula, confirming the fluorescent nature of the D\thinspace I emission. Considering that the emission of deuterium occurs in areas slightly beyond the ionization front, the detection of these lines implies that the ionization front must be trapped in HH~204, moving along with it, in consistency with the results of \citet[][]{nunezdiaz12}, as well as other evidence that will be discussed in Sec.~\ref{subsec:disc_hh204}.

%\cesar{No vas a comentar que los cocientes de DI/HI son consistentes con modelo de fluorescencia de O'Dell et al. ? Es otro indicio de que tenemos un frente de ionizacion atrapado} \eduardo{si, tienes razon. Estaba dudando de meter una figura con DI/HI pero igual no es tan relevante y solo con decir que es consistente con el modelo basta. como ves?} \cesar{decir que es consistente y dar el numero seria suficiente.}

%In this context, if we compare the average FWHM of $14.44 \pm 0.86 \text{ km s}^{-1}$ of the deuterium lines with the $14.68 \pm 0.02 \text{ km s}^{-1}$ from [O\thinspace I] lines, following Eq.~2 from \citet[][]{Garcia-Diaz:2008a} 

\begin{deluxetable*}{ccccccccccccc}
\tablecaption{Characteristics of deuterium and hydrogen lines in HH~204. \label{tab:deute_table}}
\tablewidth{0pt}
\tablehead{
 & \multicolumn{2}{c}{D\thinspace I} & \multicolumn{2}{c}{H\thinspace I} \\
$\lambda_0$ & $v_r^{*}$ ($\text{km s}^{-1}$) & FWHM ($\text{km s}^{-1}$) & $v_r$ ($\text{km s}^{-1}$) & FWHM ($\text{km s}^{-1}$) & $I$(D\thinspace I)/$I$(H\thinspace I) $\times$ 1000 
}
\startdata
3889.05$^{**}$ & -103.34 & 13.80 $\pm$ 1.39 & - & - & -  \\
3970.07 & -103.10 & 14.20 $\pm$ 3.40 &  -21.54 & 24.62 $\pm$ 0.02 &  2.99 $\pm$ 0.45 \\
4101.73 & -102.83 & 13.53 $\pm$ 0.86 & -20.97 & 24.49 $\pm$ 0.01 & 2.24 $\pm$ 0.15 \\
4340.46 & -103.02 & 16.17 $\pm$ 1.40 & -20.83 & 24.66 $\pm$ 0.01 & 2.10 $\pm$ 0.16 \\
4861.32 & -102.26 & 14.31 $\pm$ 1.90 & -21.47 & 24.67 $\pm$ 0.01 &
 1.06 $\pm$ 0.11\\
6562.80 & -101.82 & 14.90 $\pm$ 0.79 & -21.88 & 24.94 $\pm$ 0.01 & 0.58 $\pm$ 0.03  \\
\enddata
\tablecomments{$^*$ With respect to the laboratory wavelength of the closest H\thinspace I line (first column).\\
$^{**}$ The H\thinspace I $\lambda$ 3889.05 emission of HH~204 is blended with the nebular one of He\thinspace I $\lambda$ 3888.65.}
\end{deluxetable*}

\subsection{Sub-arcsecond imaging of HH~204}
\label{sec:high-resol-imag}

\begin{figure*}
  \centering
  \includegraphics[width=\textwidth]{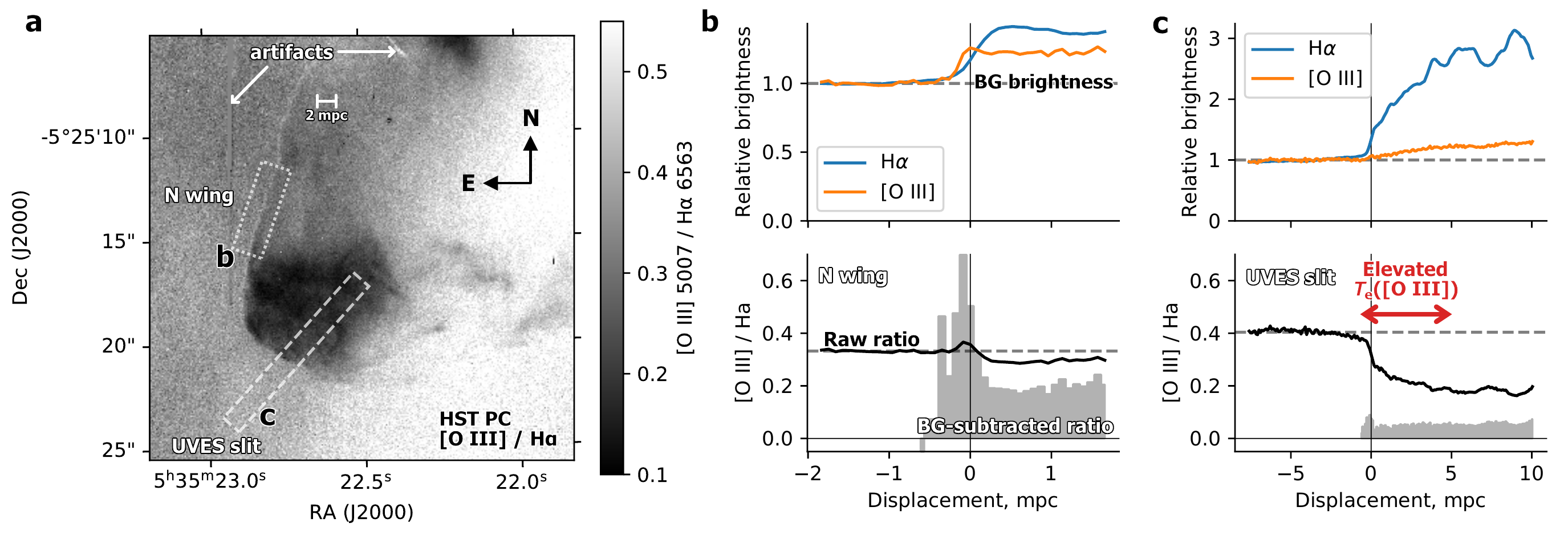}
  \caption{
    (a) Map of the line ratio \(\oiii{}\ \wav{5007} / \ha\ \wav{6563}\),
    calculated from \textit{HST} images with the PC chip of the WFPC2 camera.
    The position of the UVES spectrograph slit is outlined by a dashed box,
    while a further region of interest in the N wing of the bow shock
    is indicated by a dotted box.
    The vertically oriented ``scar'' at upper left is an artifact
    due to the bright star \th2A, located just north of the field of view.
    (b)~Average cut profiles of the \textit{HST} images for the box in the N wing
    that is outlined in panel~a.
    Upper graph shows surface brightness profiles in the two emission lines,
    normalized to the mean nebular background value outside of the shock.
    Lower graph shows the line ratio,
    with the raw ratio indicated by the black solid line
    and the background-subtracted ratio indicated by the gray histogram.
    The zero point of the displacement axis is taken to be the location
    of the maximum gradient in the \ha{} surface brightness.
    (c)~Same as panel~b, but showing average profiles of the \textit{HST} images
    along the UVES slit.
    The region of the slit that
    shows \(\Te(\oiii) > \SI{12000}{K}\) in the blueshifted component
    is indicated by the red arrow.
  }
  \label{fig:ratio-hst-oiii-ha}
\end{figure*}

Fig.~\ref{fig:ratio-hst-oiii-ha}a shows the ratio of surface brightnesses, \(R(\oiii) = S(\oiii\ \wav{5007}) / S(\ha\ \wav{6563})\), calculated from \textit{HST} WFPC2 observations in the F502N, F547M, F656N, and F658N filters from program GO5469 \citep{ODell:1996a}. It can be seen that the line ratio in the background nebula shows a pronounced gradient from \(R(\oiii) \approx 0.3\) in the north-east to \(R(\oiii) \approx 0.5\) in the south-west.%
\footnote{  For comparison with results from our UVES spectra, and using the average reddening for the HH~204 region \citep{Weilbacher:2015a}, the conversion is \(\wav{4959} / \hb \approx 1.1 R(\oiii)\).}
Inside the bow shock, the ratio is significantly smaller, for instance falling from \(\simeq 0.4\) to \(\simeq 0.2\) along the length of the UVES slit.

However, the most interesting feature of the \(R(\oiii)\) image is the slight \emph{increase} in the ratio that is seen in a thin layer along the leading edge of the bow shock. This is most clearly visible in the northern wing of HH~204, such as the area highlighted by a dotted outline box in the figure. Average profiles across the shock for this region are shown in Fig.~\ref{fig:ratio-hst-oiii-ha}b. The lower panel shows that the raw ratio (solid black line) increases only slightly above its value in the background nebula, which is because the brightness increase across the bow shock is only a small fraction of the background brightness, as can be appreciated in the upper panel. In order to isolate the emission of the shocked gas from that of the nebula, we calculate the background-subtracted line ratio:
\begin{equation}
  \label{eq:ratio-bg-sub}
  R'(\oiii) = \frac{S(\oiii) - S\BG(\oiii)}{S(\ha) - S\BG(\ha)}
\end{equation}
under the assumption that \(S\BG\) for each line is constant along the profile. The result is shown as a gray histogram in the lower panel of the figure, which reveals a sharp peak of width \(\approx \SI{0.3}{mpc}\) that reaches a maximum value \(R'(\oiii) \approx 2 R\BG(\oiii)\) and is centered on a displacement of \(\approx \SI{-0.1}{mpc}\). The origin of the displacement axis is set to the peak in the spatial gradient of the \ha{} surface brightness, corresponding to the outer edge of the dense shocked shell. The negative displacement of the \(R'(\oiii)\) peak means that this occurs
\emph{outside} the dense shell, closer to the shock front itself. 

Fig.~\ref{fig:ratio-hst-oiii-ha}c shows the same quantities calculated along a cut that coincides with our UVES slit at the head of HH~204. In this case, \(R'(\oiii)\) is always significantly less than \(R\BG(\oiii)\), but it does still show a small local peak with a position and width that is similar to the more impressive one in the northern wing. These peaks in \(R'(\oiii)\) occur over a much smaller scale than any of spatial gradients that we find in our UVES slit spectra and are only detectable because of the high spatial resolution of the \textit{HST}.%
\footnote{
  Pixel size of \SI{0.045}{arcsec},
  which well samples the PSF width at \ha{} of \SI{0.083}{arcsec}.
}
For example, the increase in \(\Te(\oiii)\) that we detect in the blue-shifted emission near the shock front (Fig.~\ref{fig:small_scale_physical_conditions}) occurs over a scale of \SI{5}{mpc}, indicated by the red arrow in the figure, which is more than 10 times larger than the width of the \(R'(\oiii)\) peak.

What is the origin of the narrow peak in the \(\oiii/\ha\) ratio that is seen just outside the shocked shell? When a shock propagates into low-ionization gas (predominantly \chem{O^+}), there are three zones where enhanced \oiii{} emission might be expected \citep{Cox:1985a, Sutherland:2017a}: first, the radiative precursor in the pre-shock gas; second, the non-equilibrium collisional ionization zone immediately after the shock; third, the radiative relaxation zone where the post-shock gas cools back down to the photoionization equilibrium temperature of \(\sim \SI{e4}{K}\). The first of these can be ruled out in the case of HH~204 because the pre-shock photoionization of \chem{O^+} would require shock velocities greater than \SI{150}{km.s^{-1}},
observed proper motion and radial velocities imply a shock velocity around \SI{100}{km.s^{-1}}. The second zone has a high temperature(\( > \SI{50 000}{K}\) for shock velocities \(>\SI{55}{km.s^{-1}}\)) but is severely under-ionized, resulting in line emissivities that are far in excess of the equilibrium values in a very thin layer. The third zone, in which oxygen is recombining through the \chem{O^{++}} stage while cooling through the range \SI{30000}{K} to \SI{10000}{K} is predicted to be somewhat thicker and with a higher electron density, yielding a greater contribution to the total \oiii{} emission. Given the electron density that we derive of \SI{13540}{cm^{-3}} (Table~3), and assuming a shock velocity \(< \SI{70}{km.s^{-1}}\),
the cooling length should be approximately \SI{0.05}{mpc}, or \SI{0.025}{arcsec}, which is a few times smaller than the \textit{HST} resolution. However, this analysis applies only to the head of the bow shock. In the wings, the shock is not perpendicular to the upstream gas velocity,
but is oblique at an angle \(\alpha\). This yields a post-shock equilibrium density that is smaller by a factor of \(\cos^2\alpha\), and a cooling length that is larger by the same factor.
Hence, the cooling length is expected to be resolved for \(\alpha\) smaller than about \ang{45}, which is consistent with our observations of the narrow peak in the \(\oiii/\ha\) ratio in the north wing. The reason that the same behaviour is not seen in the opposite wing is probably that the ambient nebular emission is much more highly ionized there, which masks the effect.

%Note that the length scales over which we see changes in $T_{\rm e}(\text{[O\thinspace III]})$ from the UVES data are much larger than this (\(\approx \SI{5}{mpc}\), see Fig.~\ref{fig:O3_lines}), and so cannot be ascribed to the cooling zone behavior described above. \eduardo{ Instead, we suggest that we are seeing the superposition of two different emission components: one from the bow shock and one from the Mach disk (the shock internal to the jet).
%The jet shock will have a lower Mach number, \(\Mach\), than the bow shock so long as the unshocked jet is denser than the ambient medium, as appears to be the case in HH~204.
%The thickness of the equilibrium (cooled) shocked shell is proportional to \(\Mach^{-1}\) and so should be larger for the shell behind the Mach disk, which thus dominates the emission in much of the slit (displacements \num{5} to \SI{12}{mpc} in Fig.~\ref{fig:O3_lines}). At smaller displacements, the contribution of the bow shock gets progressively larger, thus explaining the increase in $T_{\rm e}(\text{[O\thinspace III]})$. The expected length scale for this variation is roughly the radius of curvature of the bow shock, which agrees with the observed \SI{5}{mpc}. The only reason that this effect is visible at all is that \oiii{} emission from the equilibrium shell is so weak. For lower ionization lines, the contribution of the bow shock to the total brightness is always negligible, even for positions close to the shock.}

\section{Total Abundances}
\label{sec:total_abun}

In the case of the nebular and the DBL components, total abundances of O, Cl, and S were estimated by simply adding the abundances of all their observed ions. Although there may be some contribution of S$^{3+}$ and Cl$^{3+}$, the ICFs of \citet{stasinska78} and \citet{Esteban15}, respectively, predict negligible amounts of those species. In the case of N, Ne, Ar and C, we adopt the same ICFs used by \citet{arellanocorodova20}. For Fe, we use the two ICFs proposed by  \citet{rodriguez05}. Since the real value of Fe should be between the predictions of both ICFs \citep{rodriguez05}, in Table~\ref{tab:total_abundances} we present those determinations as lower and upper limits of the Fe abundance. 

In the case of HH~204, based on the results of Sec.~\ref{sec:small_scale}, we decided not to derive total abundances of elements for which we only observe highly ionized ions, such as He, Ne, Ar and C, due to the low ionization degree of the gas and the large contribution of the ICFs. In the cases of O, N, Cl, S, Fe and Ni, we can determine their total abundances without ICFs. %Although adding the ionic abundances of Table~\ref{tab:cels_abundances}, obtained from the entire emission of HH~204 in cut~1, can be a good estimation of the total abundances, 
As seen in Sec.~\ref{sec:small_scale}, the spatial distribution of the abundances of the once and twice ionized ions of Cl, S, Fe and Ni reach constant values at positions close to the bowshock, where the degree of ionization becomes very low. In this zone, the contribution of three -- or more -- times ionized ions of these elements should be negligible. A similar situation occurs with N, where the contribution of N$^{2+}$ is expected to be very small close to the bowshock. Therefore, in Table~\ref{tab:total_abundances} we present the total abundances  obtained by adding the mean abundances of the once and twice ionized ions of Cl, S, Fe and Ni for distances less than $4.9  \text{ mpc}$ from the bowshock. In the case of O and N, we only consider the abundance of once ionized ions in the same range of distances. At these distances, the pixel by pixel values of the O$^{+}$ abundance determined from RLs have large errors (see  Fig.~\ref{fig:O2_abundances_cels_rls}), because of the faintness of O\thinspace I $\lambda 7772$ line. In this case, we use the O$^{+}$ abundance obtained from the integrated spectrum presented in Table~\ref{tab:cels_abundances} to determine the total O abundance based on RLs.

\begin{deluxetable*}{ccccccccccccc}
\tablecaption{Total abundances. \label{tab:total_abundances}}
\tablewidth{0pt}
\tablehead{
 & \multicolumn{2}{c}{Cut 1} & \multicolumn{2}{c}{Cut 2} \\
Element & HH~204 & Nebula + DBL & DBL  & Nebula
}
\startdata
O & $8.62 \pm 0.05$  & $8.36 \pm 0.03$ & $8.31 \pm 0.12$& $8.42 \pm 0.04$\\
O$^{*}$ & $8.57 \pm 0.03$  & - & -& -\\
N & $7.75\pm 0.02$ & $7.56 ^{+0.04} _{-0.03}$ & $7.45 ^{+0.09} _{-0.08}$& $7.53 \pm 0.05$\\
Ne & - & $7.56 \pm 0.04$&-& $7.61 \pm 0.05$\\
S &  $7.07\pm 0.03$  & $6.90 \pm 0.03$ & $6.90 \pm 0.09$& $6.94 \pm 0.04$\\
Cl & $5.10 \pm 0.04$ &$5.00 \pm 0.03$ & $5.04 \pm 0.14$& $5.03 \pm 0.05$\\
Ar & - &$6.14 \pm 0.02$ & $6.09 \pm 0.10$& $6.17 \pm 0.02$\\
Fe & $6.67\pm 0.03$  & 5.91-6.09 & 5.64-6.19 & 5.97-6.13\\
Ni & $5.35 \pm 0.03$  & - & -& - \\
C$^{*}$ &- & $8.49 \pm 0.05$ & - & $8.64 \pm 0.04$\\
\enddata
\tablecomments{Abundances in units of 12+log(X/H).\\
$^*$ Based on RLs.}
\end{deluxetable*}

\section{The effects of lowering the spatial and spectral resolution}
\label{sec:mixing_things}

%In Sec.~\ref{sec:small_scale}, we have carried out a pixel by pixel analysis of the physical conditions and ionic abundances of the ionized gas in HH~204, separating the different kinematical components. In this section, we perform a completely different kind of analysis, simulating a spectrum with lower spatial and spectral resolution. Contrary to the case studied in Paper~I, where the components of HH~529~II and HH~529~III are fainter than the emission of the Orion Nebula, HH~204 is a quite brighter object, with an intensity similar to that of the nebular component. Therefore, the contribution of HH~204 to the integrated low-spectral resolution spectrum is significant. We define a new spectrum by adding the flux of all the velocity components, which includes the emission of HH~204, the DBL and the emission of the Orion Nebula along the whole UVES slit.

In this section we simulate a spectrum with lower spatial and spectral resolution by adding the flux of all the velocity components, which includes the emission of HH~204, the DBL and the emission of the Orion Nebula along the whole UVES slit. Following the reddening correction procedure described in Sec.~\ref{sec:line_int}, we obtain $c(\text{H}\beta)=0.36\pm 0.02$ for this integrated spectrum.

\begin{figure}
\centering
\includegraphics[width=\columnwidth]{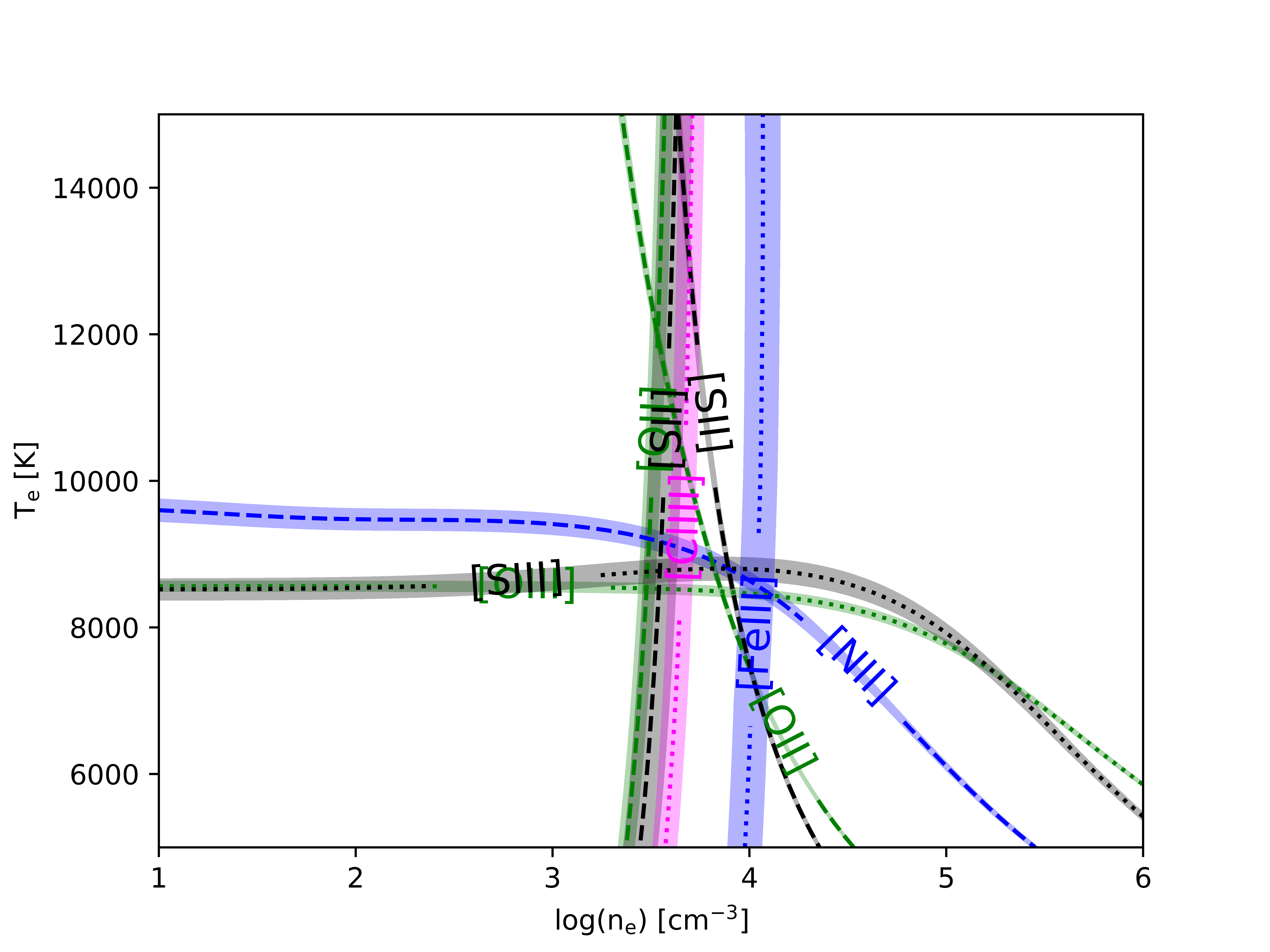}
\caption{Plasma diagnostics of the spectrum defined by adding all the observed velocity components along the whole UVES slit.}
\label{fig:plasma_mixed}
\end{figure}

In Fig.~\ref{fig:plasma_mixed}, we present the resulting plasma diagnostics of the low-resolution spectrum. This diagram can be compared with those of the individual components, shown in Fig.~\ref{fig:plasma}. If one only has the information provided by this degraded spectrum, and applies the classic procedure of averaging $n_{\rm e}(\text{[O\thinspace II]})$,  $n_{\rm e}(\text{[S\thinspace II]})$ and  $n_{\rm e}(\text{[Cl\thinspace III]})$ -- excluding $n_{\rm e}(\text{[Fe\thinspace III]})$, since the sometimes discrepant values given by this diagnostic are generally interpreted as the effect of incorrect atomic data --  we would obtain $n_{\rm e}= 3430 \pm 580 $. Using this value of density, we would obtain  $T_{\rm e}(\text{[O\thinspace II]})=12140^{+950} _{-930}$ K, $T_{\rm e}(\text{[S\thinspace II]})=19220^{+9020} _{-2530}$ K, $T_{\rm e}(\text{[N\thinspace II]})=9200\pm 200$ K, $T_{\rm e}(\text{[S\thinspace III]})=8740^{+230} _{-200}$ K and $T_{\rm e}(\text{[O\thinspace III]})=8530^{+100} _{-120}$ K. It must be noted that the resulting $T_{\rm e}(\text{[N\thinspace II]})$ is higher than the ones obtained for each individual component analyzed in Sec.~\ref{subsec:physical_cond}. Moreover, $T_{\rm e}(\text{[O\thinspace II]})$ and $T_{\rm e}(\text{[S\thinspace II]})$, the most density-dependent diagnostics, show much higher values. However, their effect on abundance determinations could be somehow mitigated, as their associated uncertainties are very high and the use of a weighted mean of the different temperature indicators would reduce their contribution. $T_{\rm e}(\text{[N\thinspace II]})$ has always much lower uncertainties and is generally the preferred temperature diagnosis for low-ionization degree ions.

\begin{figure}
\centering
\includegraphics[width=\columnwidth]{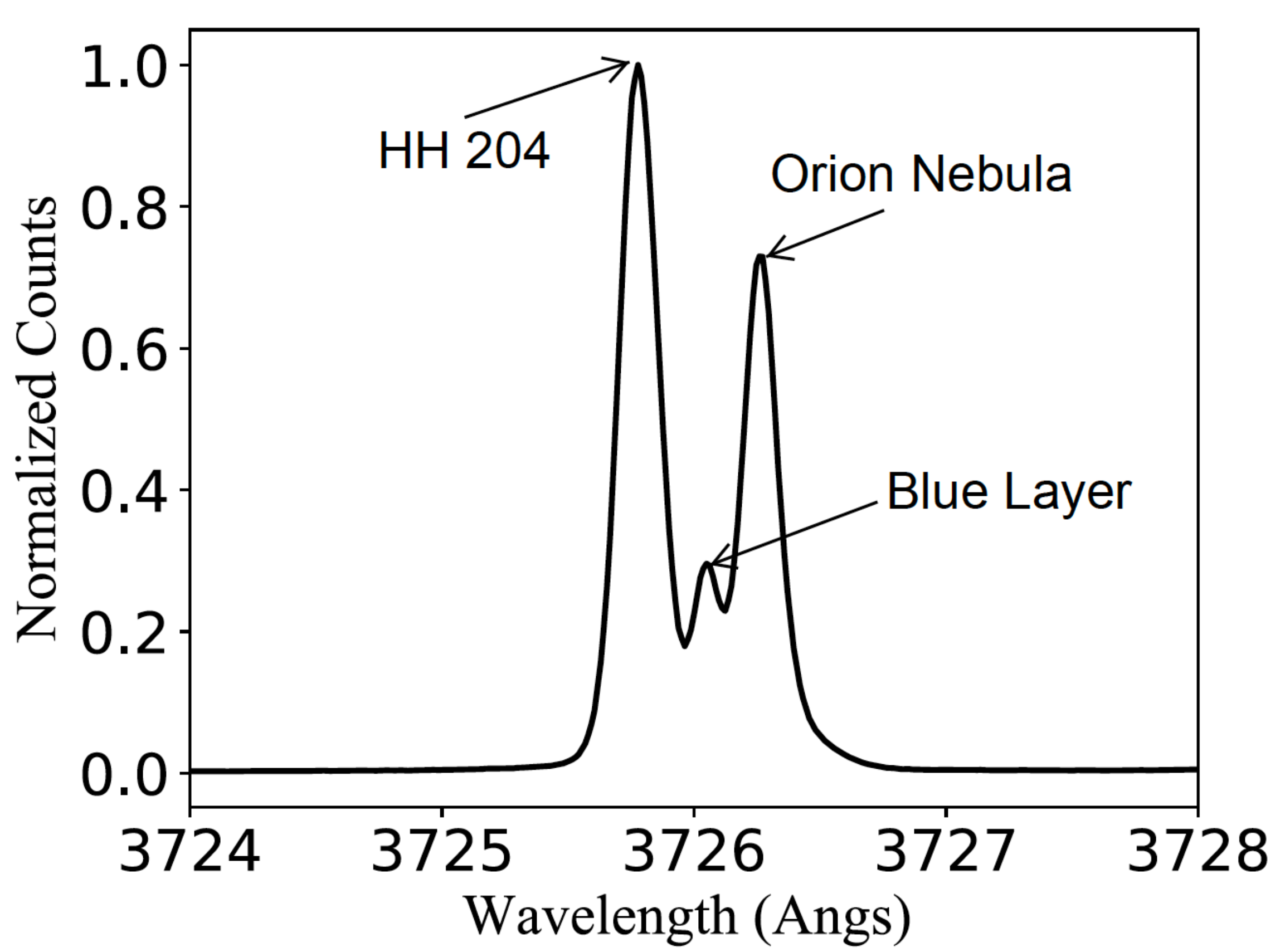}
\caption{Normalized $f(\text{[O\thinspace II] } \lambda 3727)$ in the spectrum that results from adding all the spatial pixels in the UVES slit. Each of the velocity components are identified.}
\label{fig:low_spectral_flux_3727}
\end{figure}

Following the usual procedure and assuming the physical conditions determined in the previous paragraph, we would determine the O$^{+}$ abundance using $T_{\rm e}(\text{[N\thinspace II]})$ and the O$^{2+}$ one with  $T_{\rm e}(\text{[O\thinspace III]})$, obtaining $\text{O}^{+}=8.15\pm 0.04$, $\text{O}^{2+}=7.63 \pm 0.02$ and a total abundance of $\text{O}=8.26 \pm 0.03$. This value of the O/H ratio is lower than the one determined for all the individual components. The only exception could be the DBL in cut~2, which shows $\text{O}=8.31 \pm 0.12$ (see Table~\ref{tab:total_abundances}), whose uncertainty is large enough to encompass the value obtained for the low resolution spectrum. However, this does not mean the DBL dominates the observed abundance of O, since it is the weakest component. This is demonstrated in Fig.~\ref{fig:low_spectral_flux_3727}, which shows the line profile of $f(\text{[O\thinspace II] } \lambda 3727)$, one of the most intense lines in the spectrum of the DBL.

The low O abundance obtained in the low resolution spectrum is due to the use of the classical diagnostics to estimate $n_{\rm e}$, which do not adequately account for the high density of HH~204. The critical densities of the levels involved in those diagnostics are below the density of HH~204 (see Table~D5 of Paper~I). Likewise, the sensitivity of $I(\text{[O\thinspace II] } \lambda 3726)/I(\text{[O\thinspace II] } \lambda 3729)$ and $I(\text{[S\thinspace II] } \lambda 6731)/I(\text{[S\thinspace II] } \lambda 6716)$ at $n_{\rm e} \sim 10^4 \text{ cm}^{-3}$ is much lower than at $n_{\rm e} \sim 10^2 - 10^3 \text{ cm}^{-3}$, the normal range of densities in H\thinspace II regions. The degree of ionization of each component also plays an important role. Although $I(\text{[Cl\thinspace III] } \lambda 5538)/I(\text{[Cl\thinspace III] } \lambda 5518)$ is more density sensitive than $I(\text{[O\thinspace II] } \lambda 3726)/I(\text{[O\thinspace II] } \lambda 3729)$ or $I(\text{[S\thinspace II] } \lambda 6731)/I(\text{[S\thinspace II] } \lambda 6716)$ at densities of around $n_{\rm e}\sim 10^{4} \text{ cm}^{-3}$, HH~204 --the component with the highest density-- has a very low degree of ionization. Therefore, in the combined emission of HH~204 and the Orion Nebula, the last  component has a greater weight in $n_{\rm e}(\text{[Cl \thinspace III]})$. On the other hand, $I(\text{[Fe\thinspace III] } \lambda 4658)/I(\text{[Fe\thinspace III] } \lambda 4702)$ is practically insensitive at densities smaller than $n_{\rm e}\sim 10^{3}\text{ cm}^{-3}$, and the critical density of this diagnostic is above $\sim 10^{6}\text{ cm}^{-3}$. In addition, most of the [Fe\thinspace III] emission comes from HH~204 due to its higher abundance of gaseous Fe with respect to the Orion Nebula and the DBL. These properties makes it an excellent indicator of the presence of high-density gas as in HH objects. In our case, the $n_{\rm e}(\text{[Fe\thinspace III]})=10790^{+3230} _{-2620}$  cm$^{-3}$ we obtain for the low resolution spectrum is rather close to the density of HH~204. This confirms the importance of the warning given by \citet{Morisset17} who, through photoionization models, predict large errors in the determination of the physical conditions and chemical abundances in nebulae if one assumes a single component when, in fact, there are several and some of them is composed by high-density gas. The exercise we present in this section is an observational confirmation.

%\cesar{Pienso que todo este parrafo no proporciona un resultado de suficiente interes. Propongo eliminarlo}\jorge{Estoy de acuerdo... No detectar las lineas de O I hace perder mucha fuerza al p\'arrafo.} Another interesting aspect is the behavior of the O RLs in the low-resolution spectrum. Unfortunately, the O\thinspace I RLs from multiplet 1 can not be studied in this spectrum since the nebular emission is deeply affected by sky features. On the other hand the O\thinspace II RLs do not have this problem. After our previous analysis, we know that the contribution of HH~204 and the DBL to the total O\thinspace II emission is negligible. However, in a low resolution spectrum we would not know this \textit{a priori}. This can be revealed by the resulting $n_{\rm e} (\text{O\thinspace II})$, obtained from the line ratios of O\thinspace II multiplet 1, which essentially will give the conditions of the Orion Nebula $n_{\rm e} (\text{O\thinspace II}) \sim 1500 \text{ cm}^{-3}$, a factor 2 lower than those already underestimated $n_{\rm e}(\text{[O\thinspace II]})$ and $n_{\rm e}(\text{[S\thinspace II]})$. In this low-resolution spectrum, since the O$^{2+}$ emission essentially comes from the Orion Nebula, and $T_{\rm e}(\text{[O\thinspace III]})$ is rather insensitive to the chosen density at values smaller than $\sim 10^{5} \text{ cm}^{-3}$, the ADF(O$^{2+}$) is practically the same as the one derived in the Orion Nebula. However, this is not the expected case for ADF(O$^{+}$), where the temperature for the ions with low ionization degree is density-dependent.

If instead of using the classical diagnostics to determine $n_{\rm e}$, we take the average of the densities obtained for each component (See Table~\ref{tab:pc}), weighted by their observed $F(\text{H}\beta)$, we get: $n_{\rm e}=6820 \pm 810$ cm$^{-3}$. This value is roughly between the predictions of classical  diagnostics and $n_{\rm e}(\text{[Fe\thinspace III]})$. Note that in Fig.~\ref{fig:plasma_mixed}, close to this value of density, $T_{\rm e}(\text{[O\thinspace II]})$ and $T_{\rm e}(\text{[S\thinspace II]})$ converge to $T_{\rm e}(\text{[N\thinspace II]})$.  Using that density, we obtain: $T_{\rm e}(\text{[O\thinspace II]})=8650^{+410} _{-520}$ K, $T_{\rm e}(\text{[S\thinspace II]})=9890^{+1100} _{-990}$ K, $T_{\rm e}(\text{[N\thinspace II]})=8850^{+210} _{-180}$ K, $T_{\rm e}(\text{[S\thinspace III]})=8800^{+250} _{-160}$ K and $T_{\rm e}(\text{[O\thinspace III]})=8490^{+90} _{-120}$ K. Calculating the ionic abundances of oxygen with these physical conditions, we obtain: $\text{O}^{+}=8.36^{+0.06} _{-0.05}$ and  $\text{O}^{2+}=7.64 \pm 0.02$, which implies $\text{O}=8.44 \pm 0.05$. These values are more consistent with those obtained in the analysis of the individual components. 
%Note that there is a difference of 0.18 dex between the first calculations and those obtained by using the known details of each of the components. 

It is clear that the discrepancy between the different density diagnostics is not necessarily an artifact of the atomic data used. Instead, each diagnostic may be revealing differently the changing conditions of the gas along the line of sight of the spectrum. Relying uncritically only on those density diagnostics that are consistent with each other could lead to significant systematic errors.
%In the case of our observations, the different density components have different radial velocities, which -- fortunately -- can be separated thanks to our high spectral resolution. However, one might wonder how many internal jets or gas clumps having velocities similar to that of the bulk of the ionized gas can exist in the Orion Nebula. These unresolved components -- that would in fact be most likely undetectable -- would be integrated along the line of sight and might affect the determination of the true physical and chemical properties of the nebula.   %\cesar{Ok con el parrafo, pero he cambiado algo la redaccion de las ultimas frases.}

\section{Origin of the jet that drives HH~204}
\label{sec:origin-jet-that}

\begin{figure*}
  \centering
  \includegraphics[width=\textwidth]{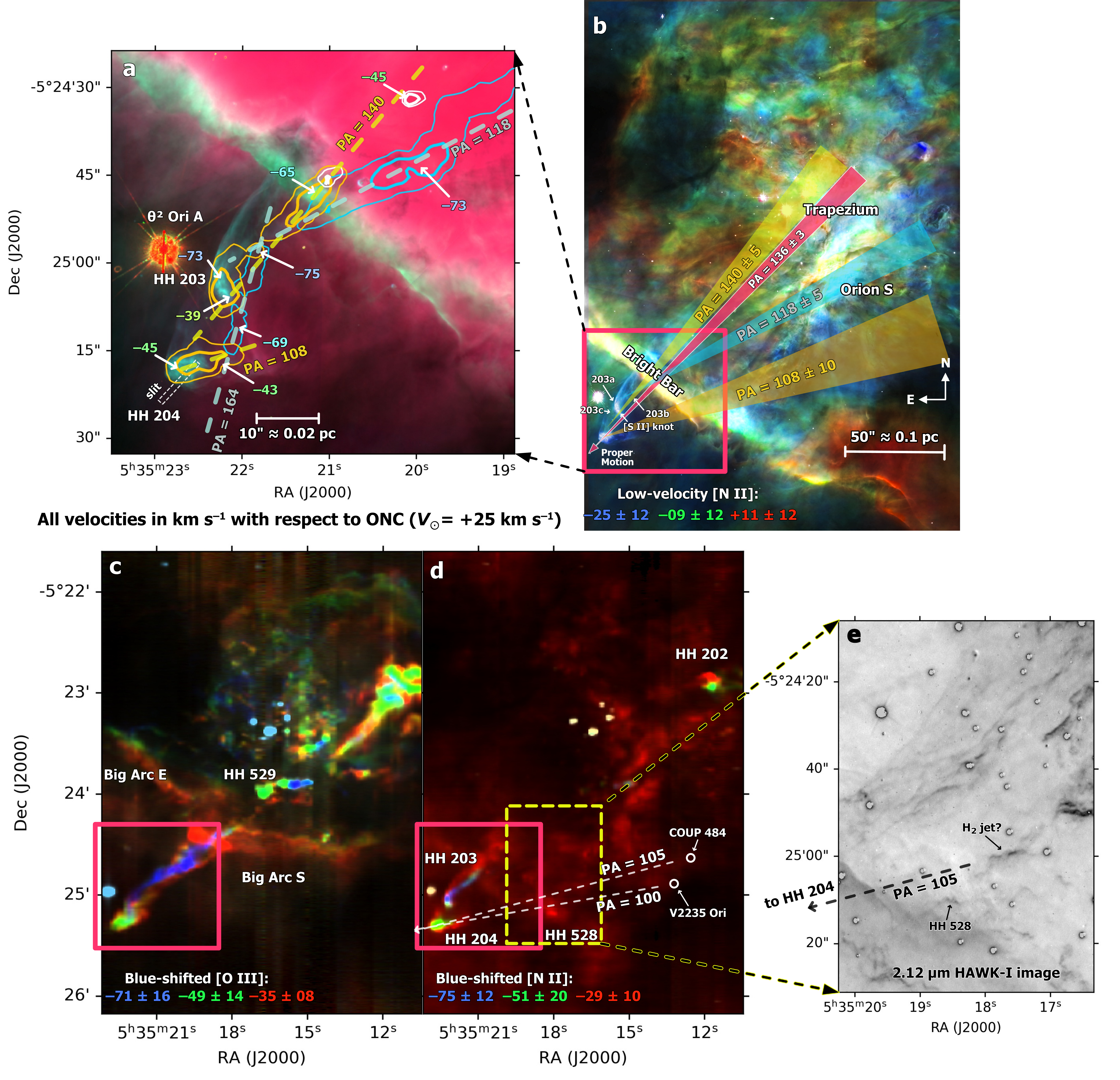}
  \caption{Location of HH~204 within the Orion Nebula.
    (a)~Same as Fig.~\ref{fig:hh204-finding-chart-simple}
    but showing an expanded view of the bow shocks and possible driving jets.
    Contours show highly blue-shifted emission of \oiii{}
    (cyan, centered on \SI{-70}{km.s^{-1}})
    and \nii{}
    (yellow, centered on \SI{-50}{km.s^{-1}}, and white, centered on \SI{-35}{km.s^{-1}}),
    derived from multiple longslit spectra
    \citetext{\citealp{Doi:2004a}
      as recalibrated in spectral atlas of \citealp{Garcia-Diaz:2008a}}.
    Mean velocities with respect to the Orion Nebular Cluster
    of particular features related to the HH objects are indicate by arrows.
    (b)~Location of HH~204 with respect to the inner Orion Nebula.
    The intensity of the image comes from the \nii{} \textit{HST} WFPC2 observations,
    but colorized according to the velocity of the slow-moving nebular gas
    (within about \SI{30}{km.s^{-1}} of the systemic velocity) as derived from the
    longslit spectra, see color key on figure.
    (c)~Highly blue-shifted \oiii{} emission for a field of view similar to that of panel~b.
    (d)~As panel~c but for highly blue-shifted \nii{} emission.
    Two candidate stellar sources along the back-projection of the PA108 flow
    are indicated by white circles (see discussion in text).
    (e)~Near-infrared HAWK-I imaging
    of the region outlined by a yellow dashed box in panel~d
    in the \SI{2.12}{\micro m} \chem{H_2} line \citep{Kissler-Patig:2008a},
    showing an emission filament that may be associated with HH~204.
    }
  \label{fig:hh204-finding-chart}
\end{figure*}

At least two different high-velocity flows converge on the general HH~203/204 region from the direction of the inner Orion Nebula (see Fig.~\ref{fig:hh204-finding-chart}), but it is not clear if either of them are directly responsible for driving the HH~204 bow shock. One flow is at a position angle (PA) of \(\approx \ang{118}\) and transitions from a high ionization state north-west of the Bright Bar (cyan contours in Fig.~\ref{fig:hh204-finding-chart}a), to a lower ionization state (yellow contours) to the south-east of the Bright Bar. The other is at \(\mathrm{PA} \approx \ang{140}\) and is of low ionization for its entire detected length. 
Both these flows give the appearance of driving HH~203, which implies that HH~203 may be a superposition of two unrelated bow shocks.
Such a superposition is consistent with the detection of two different velocity components (\num{-73} and \SI{-39}{km.s^{-1}}) at the head of the bow shock, and also with the complex structure apparent in high-resolution \textit{HST} images (see Fig.~\ref{fig:hh204-finding-chart}b).
\citet{ODell:2015a} noted that in addition to the main bow shock (HH~203a),
there appears to be a second faint bow shock (HH~203b), associated with
the PA118 flow. We also detect a third faint bow shock, which we denote HH~203c,
situated in front (SW) of HH~204a. Note that \citet{ODell:2015a} give position angles of \ang{124} and \ang{127}, respectively, for HH~203 and HH~204, which probably represent an average of the PA118 and PA140 flows. 

The southern portion of HH~203a, which we label as ``[\ion{S}{2}] knot'' in the figure, is particularly strong in the [\ion{S}{2}] and [\ion{O}{1}] filters
and coincides with the peak of the \SI{-39}{km.s^{-1}} feature. The spatial alignment and the similarity in velocity and ionization makes it likely that this knot is part of the PA140 flow. It is conceivable that this flow may extend farther to the SW and be driving the HH~204 bow shock, although there is no direct evidence for this. On the other hand, a third flow at \(\mathrm{PA} \approx \ang{108}\) is seen to feed into HH~204 from the west. This jet, first noted by \cite{Doi:2004a}, is very short and stubby, and can be traced back only \SI{10}{arcsec} (\SI{20}{mpc}) from the bow shock. There is another faint filament of high-velocity \oiii{} emission that extends between the HH~203 and HH~204 regions at \(\mathrm{PA} \approx \ang{108}\) (see Fig.~\ref{fig:hh204-finding-chart}a). This appears to provide a connecting bridge between the PA140 and PA108 flows, although the difference in velocity and ionization with respect to the PA108 flow argues against a physical association with HH~204. 

We have searched archival observations in other wavebands for any evidence of jets along the back projection of the PA108 axis. The most convincing association is with a molecular hydrogen filament seen in the \SI{2.12}{\micro m} line (see Fig.~\ref{fig:hh204-finding-chart}e). At the position of this filament, HH~204 is at \(\mathrm{PA} = \ang{105}\), which is well within the uncertainties, and the orientation of the filament is consistent with the same PA.\@
Unfortunately, no kinematic observations are currently available for this filament, so its association with HH~204 can only be tentative. The stellar source that best aligns with the \chem{H_2} filament is COUP~484, see Fig.~\ref{fig:hh204-finding-chart}e. However, this is a rather low luminosity star and therefore seems an unlikely candidate for driving such an impressive large-scale outflow. The star V2235~Ori is also marginally consistent within the uncertainty with the PA108 axis and is roughly 100 times brighter than COUP~484
in the K and L infrared bands \citep{Muench:2002a}, but its position is completely inconsistent with being the source of the \chem{H_2} filament.
There is also marginal evidence from MUSE observations \citep{Weilbacher:2015a} for a blue-shifted [\ion{Fe}{3}] filament that extends from the position of the \chem{H_2} filament towards HH~204, but the data are noisy. 

A further important line of evidence for the flow direction is provided by proper motion measurements. We have re-measured the proper motions using \textit{HST} images over an interval of 19 years (1996 to 2015) using the methodology described in section 1 of Paper I. For the ``nose'' of the HH~204 bow shock, we find a plane-of-sky velocity of (71 $\pm$ 9) km s$^{-1}$ at PA = (136 $\pm$ 3)$^{\circ}$. After correcting to a common distance of 417 pc the previous measurements of \citet{Doi:2002} are (83 $\pm$ 10) km s$^{-1}$ at PA = (137 $\pm$ 7)$^{\circ}$, which are consistent with our measurements within the uncertainties. The proper motion axis is shown by a large red arrow in Fig.~\ref{fig:hh204-finding-chart}b for comparison with the candidate axes from the high radial velocity jets. It is marginally consistent with the PA140 axis, but not all with the PA108, PA118, or PA164 axes.

In summary, convincing evidence for which large scale flow might be driving the
HH~204 bow shock is frustratingly absent. Although the PA108 flow is clearly associated with HH~204, its short length means that the exact orientation is very uncertain. The PA140 flow has a much better defined direction, but its extension beyond the position of the [\ion{S}{2}] knot in order to feed into the HH~204 bow shock is purely speculative. However, the close agreement between this flow direction and the proper motion axis is an additional argument in its favor.
The only thing that can be said with any degree of certainty is that the high-ionization PA118 flow is \emph{not} driving HH~204, but only HH~203. 

In Fig.~\ref{fig:hh204-finding-chart}b we show the back projection of all three of these flows into the core of the nebula,
assuming an uncertainty of \ang{\pm 10} for the PA108 flow and \ang{\pm 5} for the other two. The PA118 flow is consistent with an origin in the Orion~S
star forming region, as has been remarked many times previously \citep{ODell:1997a, Rosado:2002e, ODell:2003n}.
However, neither of the other flows are consistent with an origin in that region, unless the flow has suffered a relatively large-angle deviation.
The back-projection of the PA108 falls significantly to the south of the main Orion S region in an area with no convincing candidates for the driving source (see above discussion of the possible \chem{H_2} jet). The back projection of the PA140 flow intersects the Trapezium stars in the very center of the nebula, which raises the possibility that the source may be a proplyd, which are highly concentrated in that region. 
%The back-projection of the PA108 flow falls significantly to the south of the main Orion~S region.

\section{Discussion}
\label{sec:dic}

%In this second paper of our series dedicated to photoionized HH objects in the Orion Nebula, we study HH~204. Our spectroscopical observations correspond to the slit shown in Fig.~\ref{fig:hh204-finding-chart-simple}. 
The high spectral resolution of our data ($\lambda/\Delta \lambda \approx 6.5 \text{ km s}^{-1}$) allows to identify and properly separate 3 kinematical components of ionized gas: the Diffuse Blue Layer (DBL), the emission of the Orion Nebula and HH~204. In the following we will discuss in detail the results concerning each of these components. 

\subsection{The Diffuse Blue Layer}
\label{subsec:disc_blue_layer}

\begin{figure}
\centering
\includegraphics[width=\columnwidth]{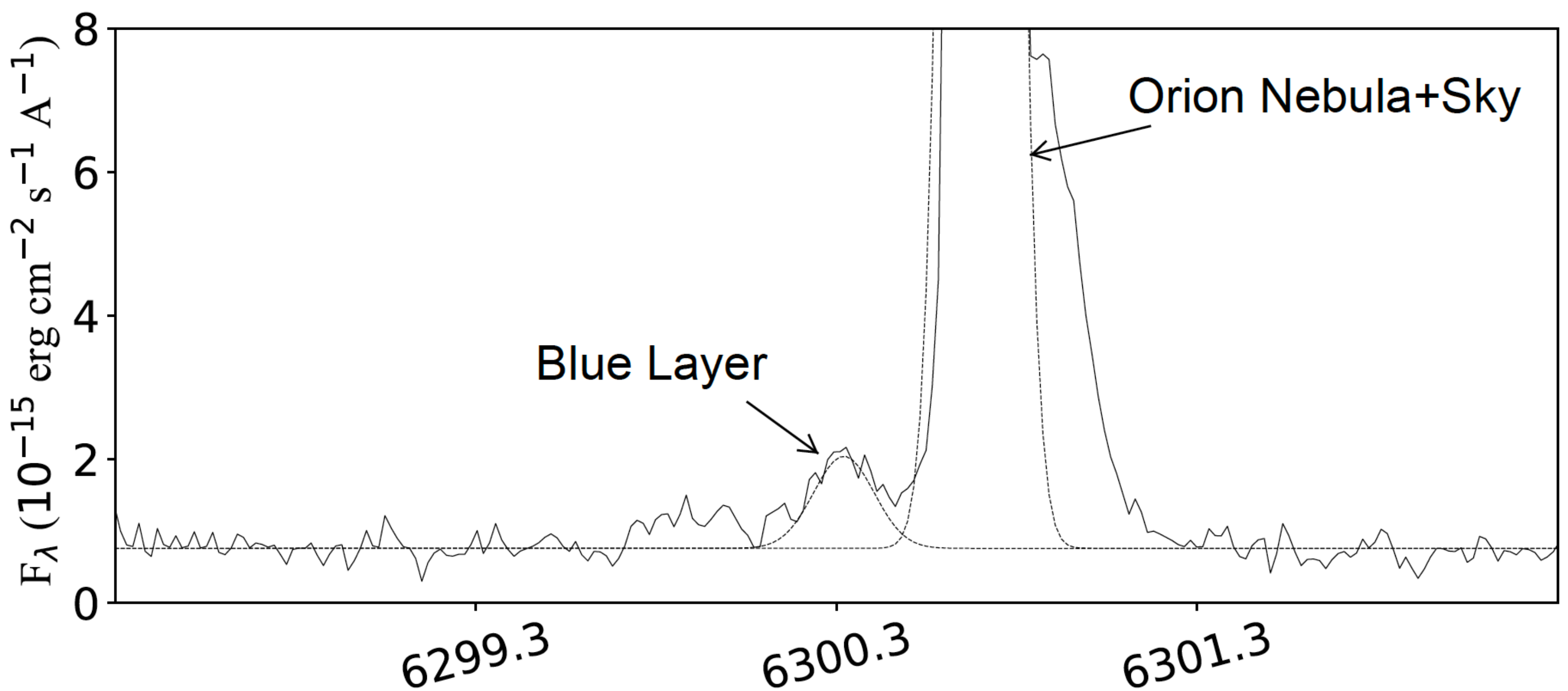}
\includegraphics[width=\columnwidth]{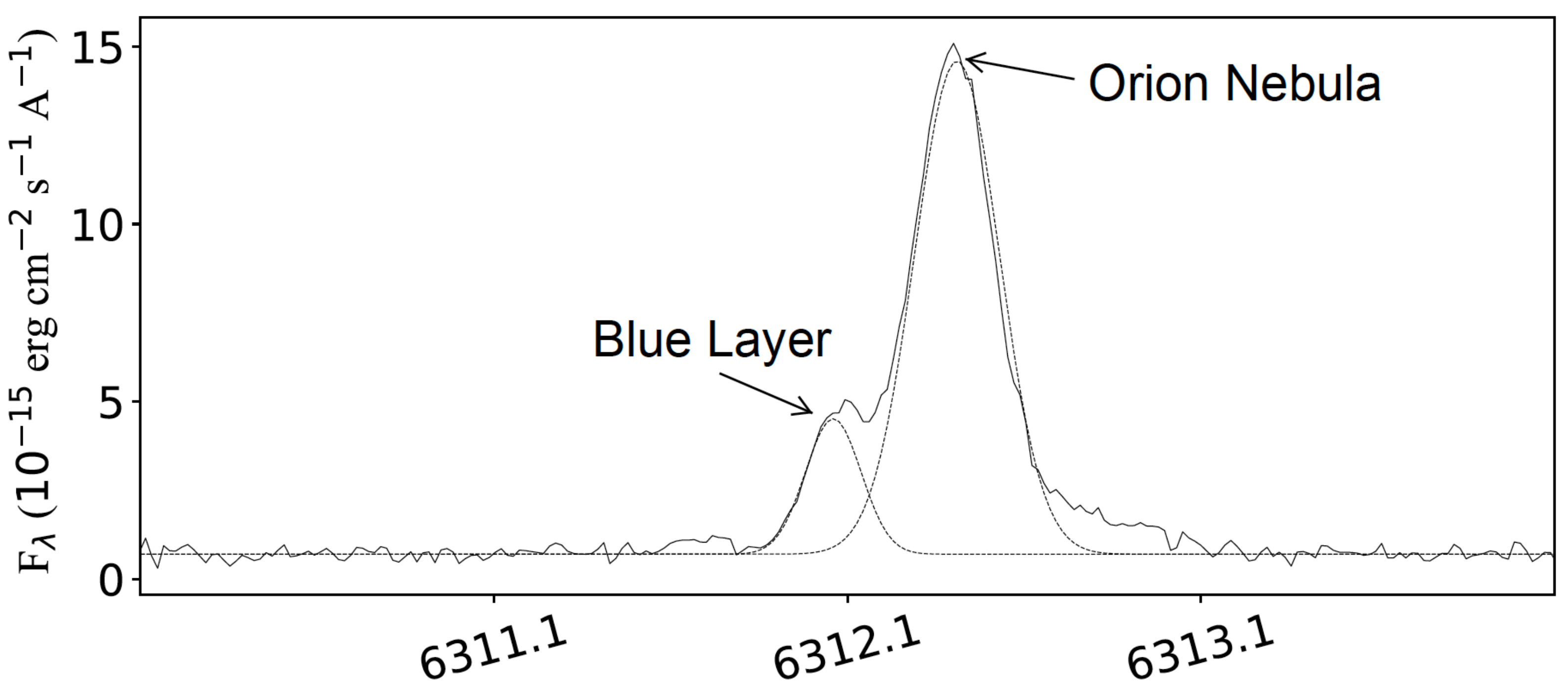}
\includegraphics[width=\columnwidth]{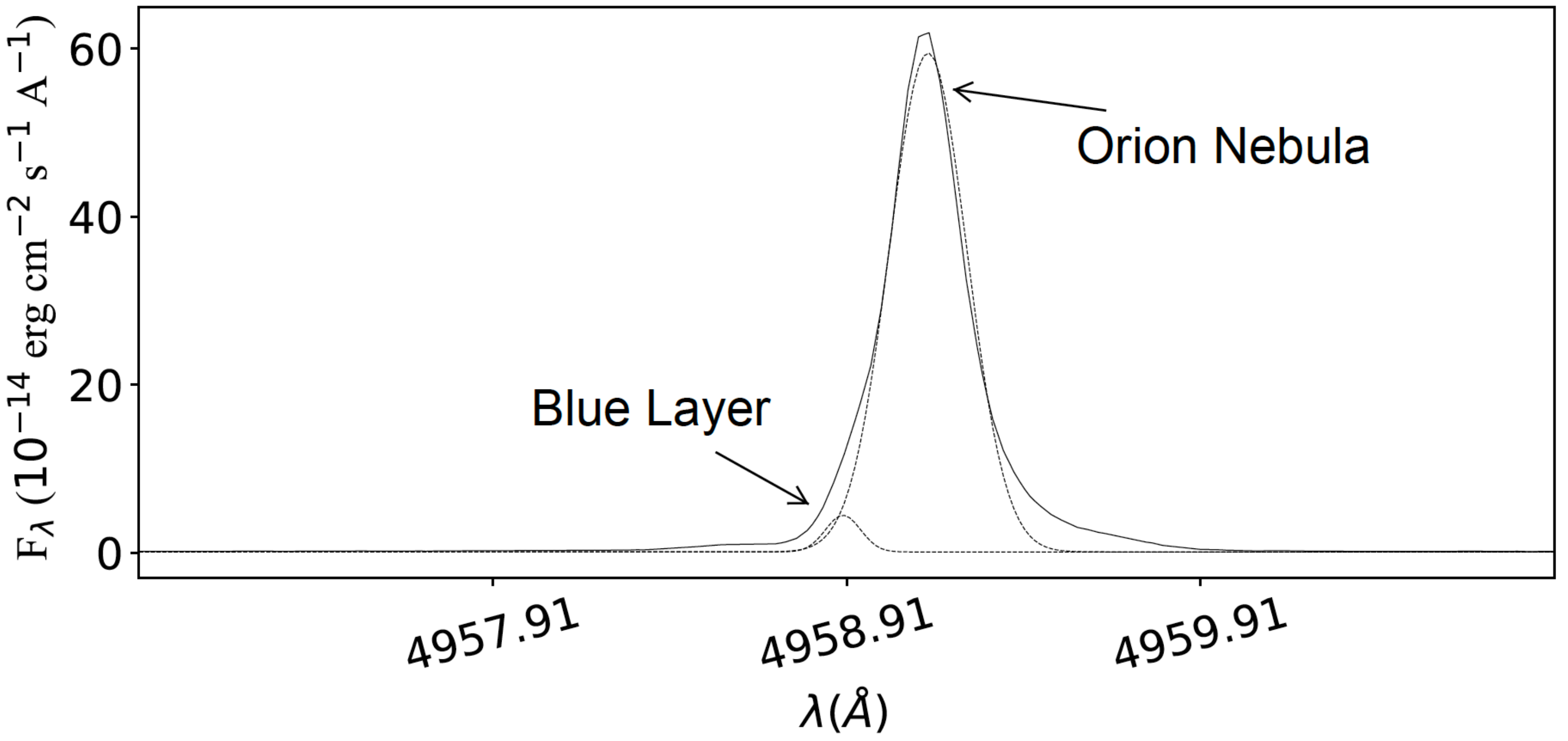}
\caption{Spectrum of cut~2 showing the emission of [O\thinspace I] $\lambda 6300$ (upper panel), [S\thinspace III] $\lambda 6312$ (middle panel) and [O\thinspace III] $\lambda 4959$ (lower panel) of the DBL and the Orion Nebula.}
\label{fig:cut2_6300_6312_4959}
\end{figure}

The component designated as the DBL was firstly reported by \citet{Deharveng73}, although it has been little studied, since high spectral resolution is required to separate its emission from that of the Orion Nebula. \citet{garciadiaz07} analyzed the velocity structure of the Orion Nebula through the emission of [O\thinspace I], [S\thinspace II] and [S\thinspace III] lines, using echelle spectroscopy. They detected the emission of the [S\thinspace II] doublet from the DBL, estimating a density of $\sim 400 \text{ cm}^{-3}$, which is in complete agreement with our estimates. These authors did not detect the emission of [O\thinspace I] or [S\thinspace III] in this component, although the emission of other low ionization ions such as [O\thinspace II] and [N\thinspace II] was detected in previous works \citep{Jones92,henneyodell99}. These limited  spectroscopical evidences lead to interpret the DBL as composed by fully ionized gas, whose ionizing radiation field was rather soft, probably coming from $\theta^{2} \text{ Ori A}$. We have detected all these lines along with [O\thinspace I] and [S\thinspace III] ones in the spectrum of this component extracted from cut~2 (see upper and middle panel of Fig.~\ref{fig:cut2_6300_6312_4959}). These emissions were also reported by \citet{odell18} in a later re-analysis of the atlas of lines of \citet[][]{garciadiaz07}. In addition, we detect a weak [O\thinspace III] emission, indicative of the presence of gas with a high degree of ionization as it is shown in the lower panel of Fig.~\ref{fig:cut2_6300_6312_4959}. 

Through observations of H\thinspace I 21-cm emission, \citet{vanderWerf13} determined the existence of several H\thinspace I velocity components in  the Orion Nebula. At the southeast, in the area where the DBL is located, these authors identified a blueshifted component named ``D'', interpreted as an expanding shell centered on $\theta^{2} \text{ Ori B}$, which is consistent with a scenario where this star ionizes the DBL. The observed [O\thinspace I] emission is consistent with the presence of an ionization front (IF) in this nebular feature. However, with the new information provided by the ionic abundances of the DBL -- estimated for the first time in this work -- the simple model where the gas is photoionized exclusively by $\theta^{2} \text{ Ori B}$ may not be correct. Although small, the contribution of O$^{2+}$ to the total abundance is not negligible, being around 10\%. On the other hand, assuming that the DBL should have a chemical composition similar to the Orion Nebula, this implies that the estimated N$^{+}$ abundance is approximately 75\% of the total nitrogen abundance, therefore N$^{2+}$ should be present in this component. Since $\theta^{2} \text{ Ori B}$ is a B0.7V star \citep{simondiaz10}, we do not expect such a star to emit a number of  photons capable of maintaining a significant proportion of highly ionized ions. This is reinforced by the spectroscopical results of Galactic  H\thinspace II regions ionized by B-type stars such as Sh~2-32, Sh~2-47, Sh~2-82, Sh~2-175, Sh~2-219, Sh~2-270, Sh~2-285, Sh~2-297 and IC~5146 \citep{garciarojas14,estebanygarciarojas18,arellanocordova2020B}. In all these regions, nitrogen is only once ionized and the contribution of O$^{2+}$ to the total oxygen is lower than 2\%, with the exception of the faint Sh~2-47, although the O$^{2+}$ abundance determination in this object is very uncertain. 

As we can see in the discussion above, the spectroscopical properties of the DBL suggest some ionization by radiation leakage from the Orion Nebula. \citet{simondiaz11} found abnormal emission of CELs of high-ionized species (mainly [O\thinspace III]) in the external zones of M~43, an H~II region ionized by a B0.5V star located at the northeast of the Orion Nebula. As those authors demonstrate, the spectral properties of this abnormal emission is consistent with contamination by scattered light from the Huygens Region. In our case,  we can discard the scattered nature of the emission of high-ionized ions in the DBL because (i) it has the same velocity as the lines of low-ionization ions, and (ii) we do not detect anomalies in the Balmer decrement of the spectrum of the DBL, that would be a signature of the presence of scattered emission \citep[see][]{simondiaz11}. Further observations with longer exposure time, similar spectral resolution and covering different areas of the Orion Nebula would shed more light on the extension and physical, chemical and geometrical properties of the DBL.
%\eduardo{This may indicate that there is an interaction with the photoionized region of the Orion Nebula, receiving a contribution of energetic photons. This could be a case analogous to M~43, an H~II region located to the northeast of the Orion Nebula which is ionized a B0.5V star, but receiving scattered light from the Huygens Region \citep{simondiaz11}. New observations with longer exposure time and similar spectral resolution devoted exclusively to study the Blue Layer would make possible to obtain chemical abundances with higher precision and being conclusive about this scenario.}

%\jorge{Lo mismo aqu\'i. Hay que recortar las figuras en el eje y par que aparezcan algo m\'as pegadas}
\subsection{The Nebular Component}
\label{subsec:disc_nebular}

There are notable differences in the degree of ionization and physical conditions of the gas of the nebular component studied in this work and in Paper~I. The degree of ionization in the area of the Orion Nebula observed in this paper is  $\text{O}^{2+}/\text{O}=0.42 \pm 0.04$, while $\text{O}^{2+}/\text{O} \sim 0.8$ in the area observed in Paper~I. This is an expected behavior considering the different distances of both areas with respect to the main ionizing star. The density in the nebular component in the direction of HH~204 is $n_{\rm e}=1440 \pm 170 \text{ cm}^{-3}$, significantly lower than the values of $n_{\rm e}\sim 6000\text{ cm}^{-3}$ obtained around HH~529~II and III in Paper I. This result, is again consistent with the more external position of HH~204 with respect to $\theta^{1} \text{ Ori C}$ and the center of a blister-shaped nebula. There is a remarkable consistency between the $T_{\rm e}(\text{[O\thinspace III]})$ and $T_{\rm e}(\text{[N\thinspace II]})$ values we obtain in this paper with the predictions of the radial distribution of those quantities given in eqs. 4 and 5 of  \citet{mesadelgado08}, confirming that the temperature decreases rather linearly with the radial distance from $\theta^{1} \text{ Ori C}$ in the Orion Nebula. 

As expected, the total abundances of O, N, S and Cl shown in Table~\ref{tab:total_abundances} are in good agreement with those included in Table~11 of Paper~I. However, the abundances of Ne and Ar are somewhat different because the use of different ICFs to estimate the contribution of unseen Ar$^{+}$ and Ne$^{+}$, which is larger due to the lower degree of ionization of the nebular component in the direction to HH~204. A similar situation occurs with the C abundance, which requires large corrections to estimate the important contribution of C$^+$. Although the total abundance of $\text{O}=8.42 \pm 0.04$ we obtain using CELs is consistent with the value of  $\text{O}=8.46 \pm 0.03$ derived in Paper~I, both are somewhat lower than the value of $\text{O}=8.51 \pm 0.03$ obtained by \citet{Esteban04} and \citet{mesadelgado09} in two different areas of the Orion Nebula. It is important to note that this difference seems to be correlated with the abundance discrepancy factor (ADF) of O$^{2+}$ estimated in each observed area. 
%The values of ADF(O$^{2+}$) are 0.36, 0.20, 0.13 and 0.11 dex for the nebular component of this work, Paper~I, \citet{Esteban04} and \citet{mesadelgado09}, respectively. If we correct the O$^{2+}$ abundances determined from CELs in this work and Paper~I by the difference between the measured ADF(O$^{2+}$) and the average value of the zones observed by \citet{Esteban04} and \citet{mesadelgado09}, 0.12 dex, we obtain $\text{O}=8.53 \pm 0.03$ for both nebular components, matching the values obtained by \citet{Esteban04} and \citet{mesadelgado09} within the uncertainties. Moreover, if we add the measured ADF(O$^{2+}$) to the O$^{2+}$ abundance from CELs, the four aforementioned determinations of the total oxygen abundance in Orion Nebula components converge to the same value of $\sim 8.6$. 
However, there are other explanations for the different O abundances obtained in different zones of the nebula. One can be related to a different depletion factor of O onto dust grains. This element may be trapped in the form of oxides, pyroxenes or olivines, compounds that would include atoms of metals such as Fe. However, the total abundance of Fe does not differ substantially between the aforementioned four zones of the the Orion Nebula and, unfortunately, the relatively large uncertainties associated with the Fe/H ratio do not permit to trace differences in depletion factors. %\cesar{Es esto lo que querias decir?}

%We remark the importance of this result noting that the observations of the 4 works were made with the same instrument and telescope, under similar observational conditions, covering different areas of the Orion Nebula with different physical and ionization conditions. \eduardo{ Since the chemical composition of the Orion Nebula must be the same in all observed areas, regardless of their degree of ionization or physical conditions, this result shows that the abundances of O based on CELs are not always reflecting the real abundances, and that it is possible that the problem lies in the estimation of the abundance of O$^{2+}$. However, this is not enough to show that the O$^{2+}$ abundances based on RLs are ``the good ones'' since, as we discussed previously, the total abundances of O are matched both when the ADF is 0.12 and when is zero. This is because as the abundance of O$^{2+}$ is increased, this ion dominates the total abundance of O, so additional increases to O$^{2+}$ directly increase the total abundance. This leaves open the possibility that the  O$^{2+}$ abundances based on RLs may be overestimated by a factor of around $\sim 0.12$ in the 4 zones observed.}

\subsection{HH~204}
\label{subsec:disc_hh204}

\subsubsection{Two-zone model for observed temperature structure}

\label{subsubsec:two_temps_model}
\begin{figure}
  \includegraphics[width=\linewidth]{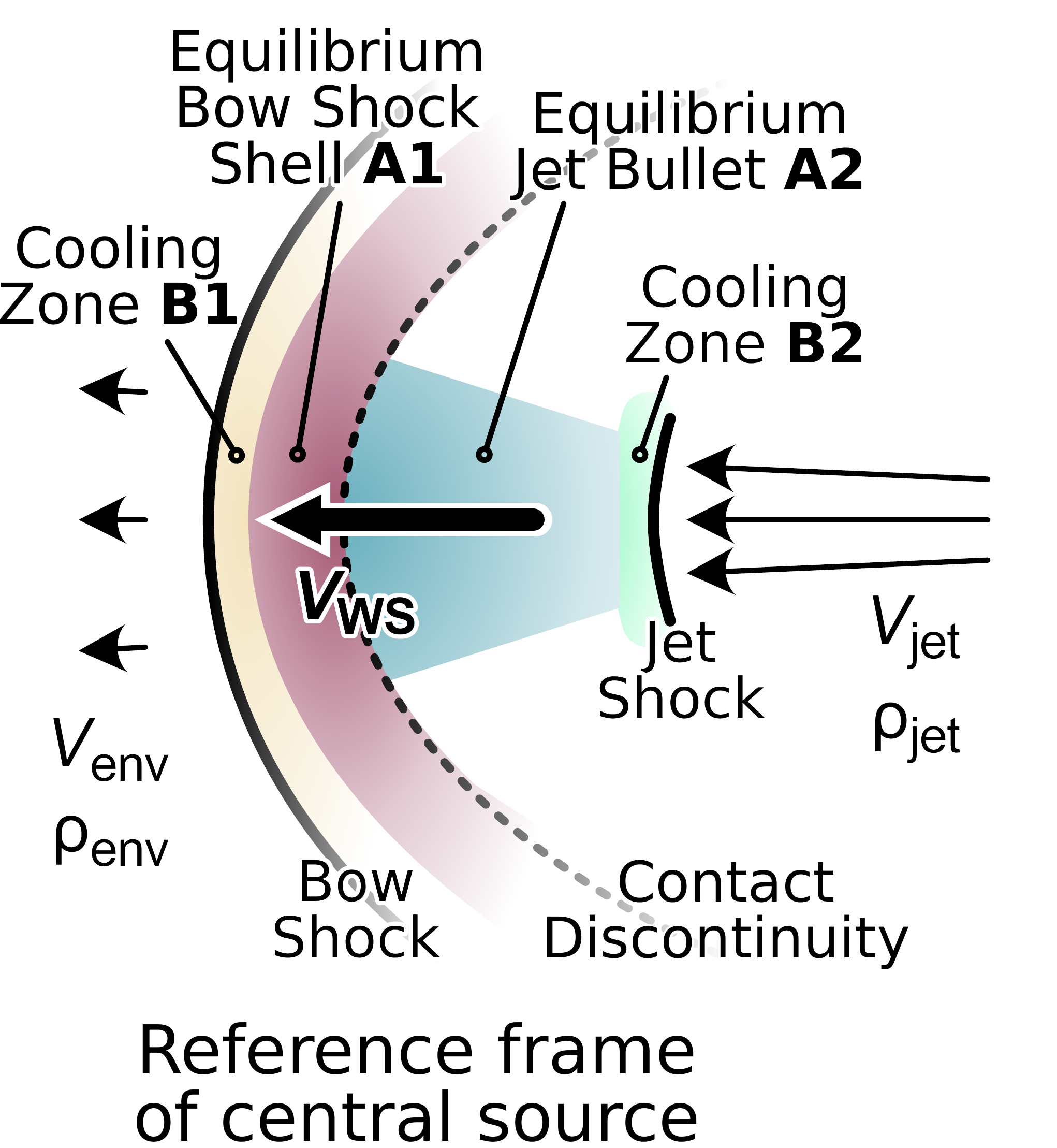}
  \caption{
    A simple model for the principal working surface of the HH~204 jet.
    The shocked gas can be divided conceptually into 4 zones:
    A1, A2, B1, and B2 (see text for details).
  }
  \label{fig:working-surface}
\end{figure}

Our spectroscopic observations allow us to analyze the physical conditions and ionic abundances of HH~204 with unprecedented detail. As  shown in Sec.~\ref{subsec:small_scale_pc},  the gas density is higher near the bow shock. On the other hand, only $T_{\rm e}(\text{[O\thinspace III]})$ seems affected by the shock, while $T_{\rm e}(\text{[N\thinspace II]})$ and $T_{\rm e}(\text{[S\thinspace III]})$ maintain their photoionization equilibrium values.
% $T_{\rm e}(\text{[N\thinspace II]})$ and $T_{\rm e}(\text{[S\thinspace III]})$ remain unaltered by the shock, being consistent with the expected values for a gas in photoionization equilibrium, while $T_{\rm e}(\text{[O\thinspace III]})$ seems to be altered by shock heating.
This result may be explained by the weakness of high-ionization emission from the densest post-shock gas in the bow shock and jet, allowing a greater relative contribution of the immediate post-shock cooling zone to the [O\thinspace III] lines.
% a larger contribution of the immediate post-shock cooling zone of the bowshock to the emission of high ionization ions, whose contribution to the emission is significantly lower in the compressed shell behind the bowshock and the main body of the jet --both in photoionization equilibrium-- owing to the low degree of ionization of the gas. 

\newcommand\zA{\ensuremath{_\mathrm{A}}}
\newcommand\zB{\ensuremath{_\mathrm{B}}}

At each position along the spectrograph slit, the line of sight will cross several zones with different physical conditions, as illustrated in Figure~\ref{fig:working-surface}:
% These may include:
\begin{description}
\item[A1] The compressed shell behind the bow-shock,
  which is in photoionization equilibrium;
\item[A2] The main body of the jet bullet, also in photoionization equilibrium;
\item[B1] The immediate post-shock cooling zone of the bow shock;
\item[B2] The post-shock cooling zone of the jet shock.
\end{description}
% In Fig.~\ref{fig:working-surface} we show the aforementioned areas.
In HH~204, the relative velocity between the unshocked jet and the working surface is very low (\(\approx \SI{15}{km.s^{-1}}\)), so the jet shock is much weaker than the bow shock, implying that the emission from zone B2 can be neglected compared with B1. Zones A1 and A2 should have similar conditions, and so can be merged into a single zone with density \(n\zA\) and temperature \(T\zA\). Although the zone B1 should have a range of temperatures, for simplicity we assume a single characteristic temperature \(T\zB\). The density of zone B is found by assuming pressure equilibrium with zone A: \(n\zB = n\zA T\zA / T\zB\). We define \(f\zB\) for a given ion as the fraction of the total ionic emission measure, \(\int n_{\mathrm{e}}\, n_{\mathrm{ion}}\,dz\), that comes from zone~B, with the remainder, \(f\zA = 1 - f\zB\), coming from zone~A.

The appropriate value of \(T\zB\) is rather uncertain, since it depends on the non-equilibrium evolution of ionization and temperature in the post-shock radiative relaxation layer. Most published shock models \citep{Cox:1985a, Sutherland:2017a} are calculated on the assumption that the far upstream and downstream ionization states are determined by the radiation from the shock itself. Care must therefore be exercised when translating their results to cases such as HH~204, where external irradiation from O~stars is a dominant factor. The curved bow shock in HH~204 should give a range of shock velocities, up to a maximum of \(V \approx \SI{84}{km.s^{-1}}\) (assuming the pre-shock medium is stationary). In principle, this corresponds to post-shock temperatures as high as \SI{2e5}{K}, but the gas at such temperatures will be too highly ionized to significantly emit optical lines. The cooling timescale is generally shorter than the recombination timescale, so the gas is over-ionized as it cools. It is only when the temperature falls below about \SI{50000}{K} that the abundance of \chem{O^{++}} becomes significant \citetext{e.g., Fig.~11 of \citealp{Allen:2008a}}, allowing the emission of the optical [\ion{O}{3}] lines. A similar situation is seen in middle-aged supernova remnants, such as the Cygnus Loop \citep{Raymond:2020a}. 

\begin{figure}
  \textbf{\Large a}\\
  \includegraphics[width=\linewidth]{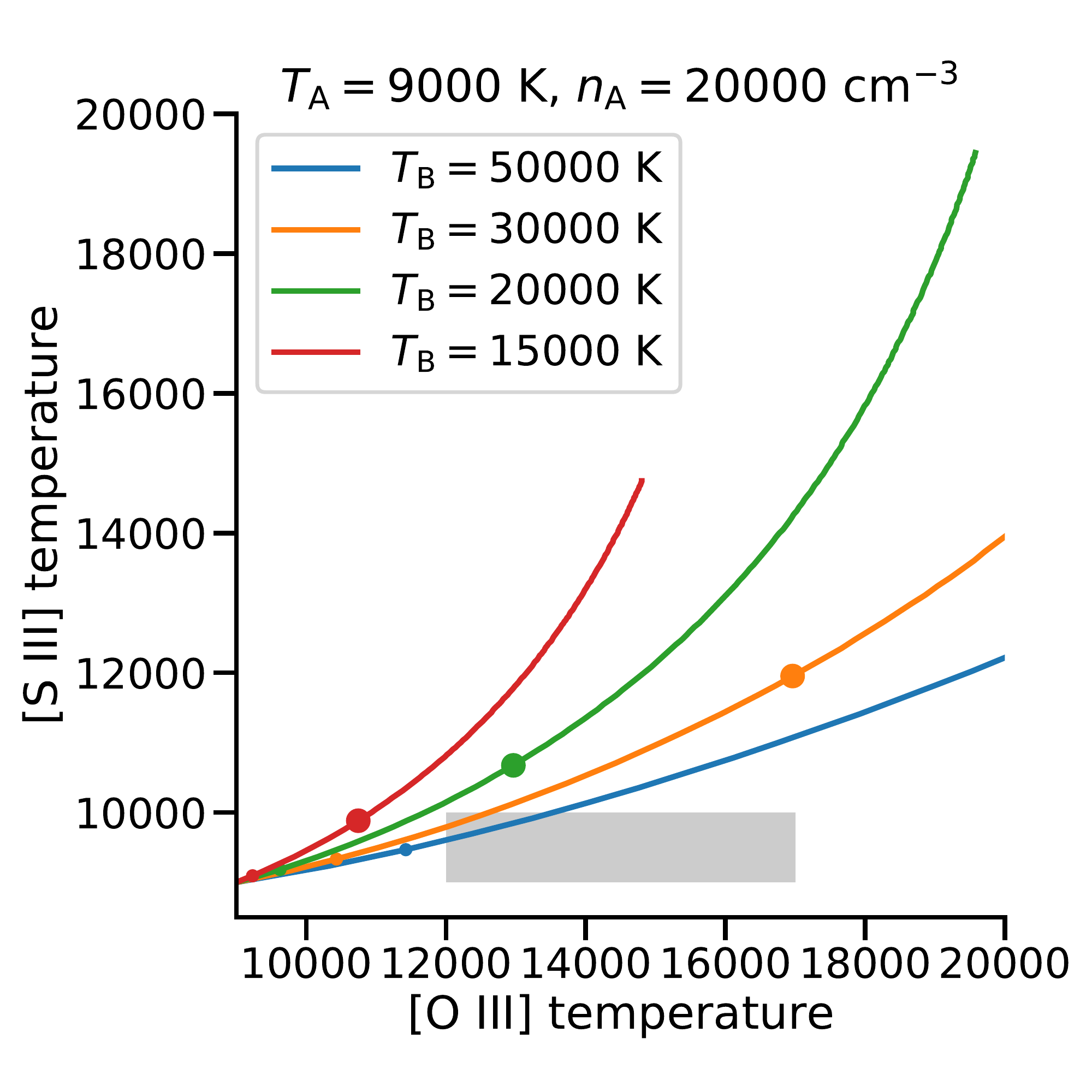}\\
  \textbf{\Large b}\\
  \includegraphics[width=\linewidth]{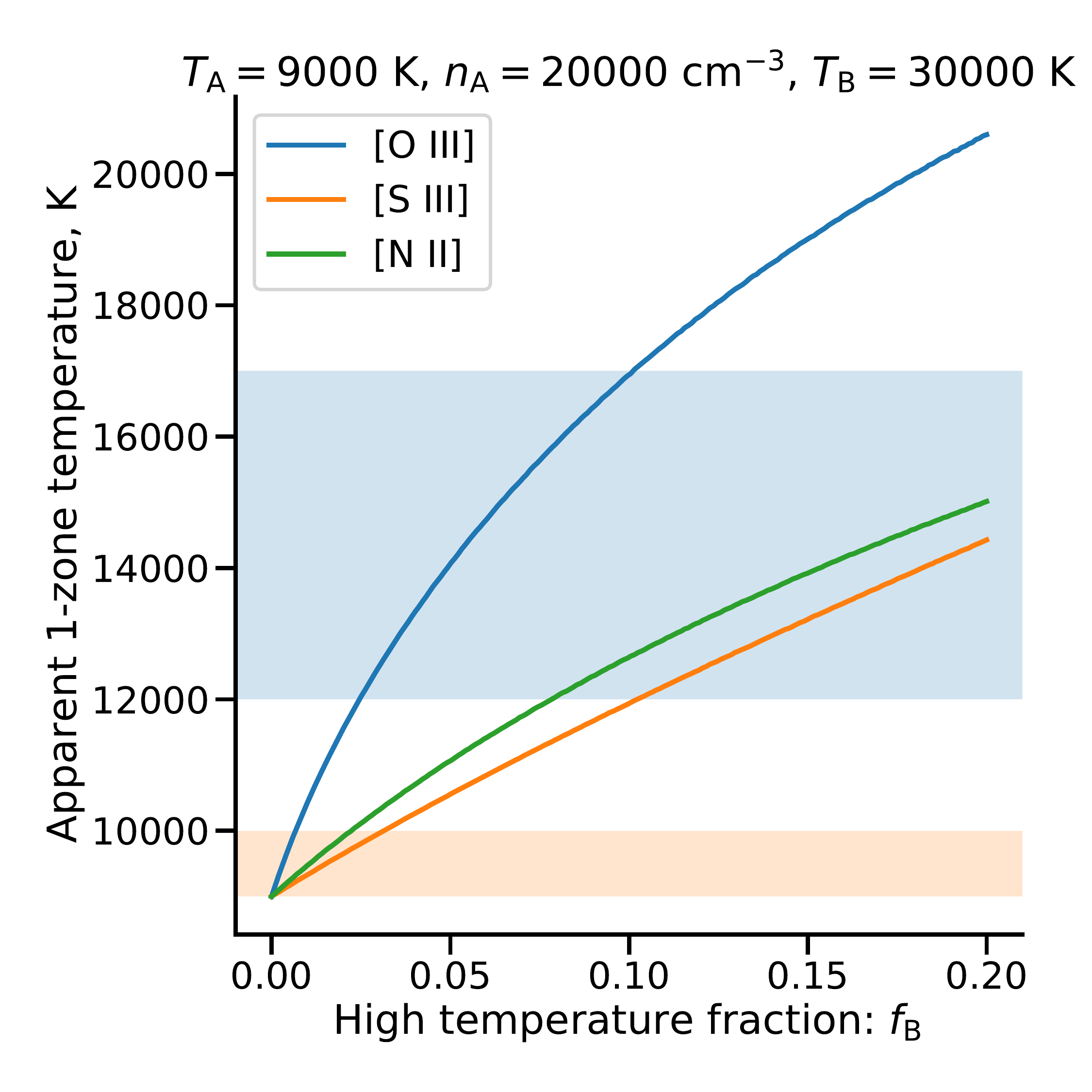}
  \caption{
    Simple two-zone model for spatial variations in temperature diagnostics.
    (a)~Correlation between derived \(\Te\) from \oiii{} and \siii{} lines,
    assuming that the fraction \(f\zB\) of ionic emission measure that arises
    in the hot component (with temperature \(T\zB\)) is the same for both ions.
    Values of \(f\zB = 0.01\) (small dots) and \(f\zB = 0.1\) (large dots)
    are indicated on each curve.
    The gray rectangle shows the observed range of values,
    which cannot be explained under this assumption (see text). 
    (b)~Derived \(\Te\) for \oiii{}, \siii{}, and \nii{} lines as a function
    of \(f\zB\), assuming \(T\zB = \SI{30000}{K}\).
    Colored bands show the observed ranges, which imply that \(f\zB(\oiii)\)
    must be larger than for the other ions (see text).
  }
  \label{fig:two-zone}
\end{figure}

We look for solutions where both \(T\zA\) and \(T\zB\) are constant along the slit, so that any spatial variation in the temperature diagnostics is driven primarily by variation in \(f\zB\). Although the density diagnostics do show a gradient with position, both \(T(\oiii)\) and \(T(\siii)\) are relatively insensitive to density, so for simplicity we assume \(n\zA\) is constant. We use the Python library PyNeb to calculate the per-zone emission coefficients, \(j(T\zA, n\zA)\) and \(j(T\zB, n\zB)\), for each emission line. For a given diagnostic line pair, 1 and 2, the ratio is calculated as

\begin{equation}
  \label{eq:2-zone-ratio}
  R_{12} = \frac
  {(1 - f\zB) \, j_1(T\zA, n\zA) + f\zB\, j_1(T\zB, n\zB)}
  {(1 - f\zB) \, j_2(T\zA, n\zA) + f\zB\, j_2(T\zB, n\zB)}.
\end{equation}

This is then fed into PyNeb's \texttt{getTemDen} function to find the equivalent single-zone temperature that would give the same ratio (assuming a density of \(n\zA\)). It is clear from Eq.~(\ref{eq:2-zone-ratio}) that for \(f\zB = 0\) one must recover \(\Te = T\zA\) and that for \(f\zB = 1\) one must recover \(\Te = T\zB\). But for intermediate values of \(f\zB\), the derived temperature will differ between ions because of variations in the temperature sensitivity of the diagnostic ratios.

We first investigate the case of a common \(f\zB\) for all ions, but we find that this is unable to reproduce the observations. This is demonstrated in Fig.~\ref{fig:two-zone}a, which shows the relation between \(\Te(\oiii)\) and \(\Te(\siii)\) for 4 different values of \(T\zB\) between \num{15000} and \SI{50000}{K}. We set \(T\zA = \SI{9000}{K}\) and \(n\zA = \SI{20000}{cm^{-3}}\) in all cases and \(f\zB\) increases from left to right along each curve.The gray rectangle shows the observed range of temperatures along the spectrograph slit (Fig.~\ref{fig:small_scale_physical_conditions}): \(\Te(\oiii)\) shows a systematic decline from \(\approx \SI{17000}{K}\) near the bow shock to \(\approx \SI{12000}{K}\) further away, while \(\Te(\siii)\) is roughly constant at \num{9000} to \SI{10000}{K}, with no apparent correlation with \(\Te(\oiii)\). The two-zone models with \(T\zB \ge \SI{30000}{K}\) all show \(\Te(\oiii) > \Te(\siii)\) as \(f\zB\) increases, but this is insufficient to explain the observations. For example, in order to achieve \(\Te(\oiii) = \SI{17000}{K}\) the models predict \(\Te(\siii) > \SI{11000}{K}\), which is significantly higher than observed.

In Fig.~\ref{fig:two-zone}b, we relax the assumption of a common \(f\zB\) for all ions, showing separately the predicted values of \(\Te(\nii)\), \(\Te(\siii)\), and \(\Te(\oiii)\) as a function of \(f\zB\), assuming \(T\zB = \SI{30000}{K}\). The ranges of observed values are shown by colored bands, blue for \oiii{} and orange for \nii{} and \siii{}. From the figure it is apparent that a decline from \(f\zB(\oiii) \approx 0.1\) at \(x = 0\) to \(f\zB(\oiii) \approx 0.02\) for \(x > \SI{5}{mpc}\) is required to explain the \(\Te(\oiii)\) profile, whereas \(f\zB(\siii) < 0.01\) and \(f\zB(\nii) < 0.01\) is required at all positions.

It is not surprising that \(f\zB\) should vary between ions since the photoionization equilibrium ion fraction of \chem{O^{2+}} from zone~A is much lower than that of \chem{N^{+}} or \chem{S^{2+}}. Assuming \(f\zB \ll 1\), then the ionic abundances given in the ``Cut~1, HH~204'' column of Table~\ref{tab:cels_abundances} correspond to zone~A. These yield \chem{O^{2+} / \chem{O} = 0.005} and \chem{S^{2+} / \chem{S} = 0.61} if the abundances of unobserved ion stages are negligible. The lack of [\ion{N}{3}] lines means that \chem{N^{+} / \chem{N}} cannot be estimated directly, but is likely of order unity. The fact that \chem{O^{2+}} is only present in trace amounts in the photoionization equilibrium gas means that the relative contribution from the post-shock cooling zone is much larger than for \chem{S^{2+}} and \chem{N^{+}}. This is confirmed by emission line imaging of HH~204 \citep{Weilbacher:2015a}, which shows a morphology in \siii{} and \nii{} that is clearly dominated by the compact jet bullet, whereas the emission in \oiii{} is more diffuse within the parabolic envelope of the bow shock.

Although part of this high-ionization degree gas may be out of photoionization equilibrium, fortunately, its impact is negligible in the global abundance analysis of HH~204. The fact that $T_{\rm e}(\text{[N\thinspace II]})$ and $T_{\rm e}(\text{[S\thinspace III]})$ are kept in balance in HH~204 proves that the low and medium-ionization degree gas, which comprises more than $\sim 99\%$ of the total, is in photoionization equilibrium.

\subsubsection{A trapped ionization front}
\label{subsubsec:trapped_IF}
%In HH~204, the velocity separation between deuterium and hydrogen lines coincides with the theoretical isotopic shift, implying that HH~204 should  actually contain such ionization front.

The detection of emission lines of neutral elements such as [O\thinspace I] and [N\thinspace I] and the high density and low degree of ionization of HH~204 suggest that it contains an ionization front. In previous studies, the detection of these lines has been interpreted as product of the interaction of HH~204 with neutral material, such as that found in the Orion's Veil \citep[][]{ODell:1997a,odell97,takami02}. However, there are several arguments against this scenario and in favor of the existence of a trapped ionization front. (i) The spatial distribution of the [O\thinspace I] emission, shown in Fig.~\ref{fig:cuts}, is more concentrated than that of [O\thinspace II] or [O\thinspace III], located at the southeast of HH~204, in the opposite direction to $\theta^{1} \text{ Ori C}$, consistent with a zone shielded from the ionizing radiation. (ii) As we discussed in Sec.~\ref{subsec:deuterium}, deuterium lines are produced by fluorescence excitation areas beyond an ionization front. Finally, (iii) the combination of the tangential and radial motions of HH~204 allows to know the 3D-trajectory of its associated jet. From its apparent distance to Orion-S (its likely origin), \citet{Doi:2004a} estimate that HH~204 has moved $\sim 0.15 \text{ pc }$ radially towards the observer. Although \citet[][]{vanderWerf13} argued that the Orion Veil lays $\sim 0.3 \text{ pc }$ apart from Orion-S, \citet[][]{abel16} established that the distance must be significantly larger and therefore a direct interaction between HH~204 and the Veil is unlikely. If those distance estimations are correct, HH~204 would be located within the main ionized gas volume of the Orion Nebula or interacting with the Nearer Ionized Layer  \citep[NIL, see][]{Abel19,Odell20}. %\cesar{He cambiado el orden de los items, creo que los del [OI] y las lineas de DI son evidencias bastante directas y la que ahora he puesto como (iii) es mas circunstancial y no me parece que se pueda interpretar como un indicio claro de frente de ionizacion} %(iv) The $T_{\rm e} (\text{[O\thinspace I]})$ we obtain for HH~204 is incompatible with entrained material from a neutral region such as the Veil, which would have a temperature of a few hundred K. \citep[][]{Abel06}.

\subsubsection{The ADF and the ``true" O abundance}
\label{subsubsec:ADF_and_O}

The origin of the AD problem has been related to  temperature, density or chemical inhomogeneities in the nebulae or fluorescence effects on the intensity of RLs \citep[see ][and references therein]{Peimbert67, torrespeimbert80, liu01, Pequignot91, garciarojas07,Escalante12}. As mentioned in Sec.~\ref{subsec:small_scale_ca}, the O$^+$ abundances calculated with RLs and CELs are equal in HH~204. 
Since practically all oxygen is singly ionized in this object, this implies that HH~204, contrary to what is usually found in ionized nebulae, does not show an ADF in total O abundance. Therefore, the ``true" O abundance  should be $\sim 8.6$ in this object, slightly lower  than the recommended solar O abundance \citep[$8.73 \pm 0.07$,][]{lodders19}. In this regard, there are three properties of HH~204 that we want to highlight: (i) In Sec.~\ref{subsec:small_scale_pc}, we show that the spatial distribution of $T_{\rm e}(\text{[N\thinspace II]})$ is constant, i.~e. there are no significant temperature fluctuations in the plane of the sky that may be translated into fluctuations in the line of sight for ions of low degree of ionization (See Sec.~\ref{subsubsec:two_temps_model}). The presence of temperature fluctuations would produce the underestimation of the O$^+$ abundance based on CELs. (ii) In Sec.~\ref{subsec:ionic_abundances_fe_ni}, we show that the effects of starlight fluorescence are negligible in the determination of the abundances of Ni$^{+}$ and Fe$^{+}$ due to the large distance between HH~204 and the ionizing source in addition to the high density of the HH object. Thus, if there is any mechanism in which the continuum pumping can affect the population of the levels of multiplet 1 of O\thinspace I, this may be diminished in a similar way. (iii) The jet-geometry of HH~204, with a relatively small angle of $ 32 \pm 6 ^{\circ}$ with respect to the plane of the sky (see Sec.~\ref{sec:origin-jet-that}), implies that any gradient in the electron density of the gas along the jet axis should be separated in the plane of the sky. This is a more favorable geometry for analysis than the case of a jet flowing directly towards the observer where different zones will overlap the same line of sight. Therefore, chemical or density inhomogeneities in the line of sight appear not to be present in HH~204. Unfortunately, we can not perform a similar analysis for the ADF(O$^{2+}$) because O\thinspace II RLs are not detected in HH~204.

In Table~\ref{tab:oxy_abundances} we compile the O abundances obtained in all chemical abundance studies of the Orion Nebula based on deep echelle spectroscopy taken with UVES. We include determinations based on both, RLs and CELs, assuming $t^2$ = 0 in the last case. 
A first note of caution should be given concerning the fraction of O depleted onto dust grains,
which may be different in different parts of the nebula. 
\citet{mesadelgado09} estimated this fraction to be typically $\sim$ 0.12 dex but it may be lower in HH objects due to destruction of dust grains in shock fronts. 
This implies a maximum extra uncertainty of $\sim$0.1 dex in any given O abundance measurement due to depletion variations.
%One the set of RL determinations shows a dispersion ($\sigma$ = 0.10 dex) larger than the set of abundances derived from CELs ($\sigma$ = 0.06 dex).

%If we assume the value of $\text{O} \sim 8.6$ obtained for HH~204 as representative of the Orion Nebula, this would imply that RLs overestimate the O abundances in several zones. 

If we assume that the O abundances based on RLs are the ``true" ones for all objects, then HH~529~II and III 
% and other zones of the Orion Nebula QUITAR WJH 
show higher O/H ratios than the rest. 
In Paper~I we discuss the possibility of having a slight overmetallicity in HH~529~II and III, due to the entrainment of material from the accretion disk of the stellar source of the jets. On the other hand, the O abundances based on RLs found in the nebular components studied in Paper~I and \citet[][]{Esteban04} are also marginally higher than what is found in HH~204. These discrepancies may have  explanations of a different nature in each case, apart from dust depletion variations.
%Since the O abundance in HH~204 should be slightly higher than in the nebular component, either for the same reasons mentioned for HH~529~II and HH~529~III or due to higher destruction of dust with O content, this would require an additional explanation, probably a different one for each particular case.

The low measured CEL abundance values found in the more highly ionized regions of the nebula could be reconciled with the HH~204 value by considering different small proportions of O depletion onto dust grains in addition to small contributions from other phenomena as, for example, temperature fluctuations. In this context, if we assume that half of the difference between $\sim 8.6$ -- considering that the O abundance obtained in HH~204 is the true one of the Orion Nebula-- and the O abundance based on CELs obtained by  \citet[][]{Esteban04} is due to dust depletion and the rest to temperature fluctuations, this would be compatible with $t^2\sim 0.008$, a value considerable smaller than the $t^2\sim 0.022$ necessary to match the O abundances from RLs obtained in the same spectrum. In this case, the relevant question is why the RLs are giving higher O abundances in all cases except HH~204. An important difference between the determination of O abundance in HH~204 and the other zones or objects included in Table~\ref{tab:oxy_abundances} is that, in HH~204, the contribution of O$^{2+}$ to the total abundance is negligible. It is important to say that \citet{mesadelgado09} also obtained an  ADF(O$^{+}$) equal to zero in both, HH~202~S and the nebular component. However, the contribution of O$^{2+}$ is much larger in those spectra and their ADF(O$^{2+}$) are not zero. This result suggests that the AD problem is affecting specially to lines used to derive O$^{2+}$ abundances and perhaps related to unaccounted effects on the intensity of O\thinspace II RLs.  It is still premature to draw any conclusions in this regard but we will explore this important issue further in future papers of this series. %\jorge{Da la sensaci\'on de que la discusi\'on se queda un poco a medias... Qu\'e puede implicar en cualquiera de los escenarios propuestos que el ADF(O+)=0 y el de O++ no lo sea? Es inconsistente con fluctuaciones de temnperatura, pero sería consistente con los efectos de fluorescencia en las líneas de O II, no?} \eduardo{Si, lo es, pero creo que para decir que es fluorescencia tendriamos que tener alguna evidencia de su presencia o de mas o menos como es que esta funcionando.}

%The low ionization degree of HH~204 is an advantage for determining the abundance of certain elements. In areas close to the bowshock, where the ionization degree drops to practically zero, we can obtain the total abundance of N, S, Cl, Fe and Ni without the use of an ICF, simply adding the ionic abundances determined from our optical spectrum. This makes it possible to eliminate the generally most important source of uncertainty and thus obtain very precise abundances, which can be used as representative of the Orion Nebula. %The downside of the low ionization degree of HH~204 is that elements showing only high ionization potential ion optical lines, as He or Ne, need very large ICFs, and their total abundances cannot be derived with good precision. \jorge{Yo creo que esto es m\'as un p\'arrafo para las conclusiones que para esta secci\'on, ya que esto se ha comentado ya previamente.}

%The elements where only ions with a high degree of ionization are observed require large corrections in HH~204. For example, in the case of He, the obtained He$^{+}$ abundance implies that 62\% of the helium is neutral \citep[adopting a total He abundance of 10.95][]{mendez20,mendez2021}. Therefore, since their total abundances would be rather uncertain, we do not estimate them. On the other hand, HH~204 is the ideal place to analyze the abundances of low and intermediate degree of ionization ions. 

From Table~\ref{tab:total_abundances}, it is clear that the Fe abundance in HH~204 is higher than in the other components due to dust destruction at the bowshock. Following the same procedure as in Paper~I, comparing the observed Fe/O values in HH~204 and the nebular component with the expected solar value \citep{lodders19}, we estimate that $\sim 6\%$ of the total Fe is in the gaseous phase in the nebular component, while this fraction goes up to 21\% in HH~204, representing an increase of a factor 3.5. A similar factor can be assumed for Ni.

\begin{deluxetable*}{ccccccccccccc}
\tablecaption{Oxygen abundances in the Orion Nebula based on UVES spectroscopy. \label{tab:oxy_abundances}}
\tablewidth{0pt}
\tablehead{
 & \multicolumn{3}{c}{RLs} & \multicolumn{3}{c}{CELs}& \\
Region & O$^{+}$ & O$^{2+}$ & O  & O$^{+}$ & O$^{2+}$ & O & Reference
}
\startdata
\multirow{4}{*}{Orion Nebula} & $8.15 \pm 0.13$ & $8.57\pm 0.01$ & $8.71 \pm 0.03$ & $7.76 \pm 0.15$ & $8.43 \pm 0.01$ & $8.51 \pm 0.03$& \citet{Esteban04}\\
 & $8.01 \pm 0.12$ & $8.46\pm 0.03$ & $8.59 \pm 0.05$ & $8.00 \pm 0.06$ & $8.35 \pm 0.03$ & $8.51 \pm 0.03$& \citet{mesadelgado09}\\
 & $8.25 \pm 0.06$ & $8.52\pm 0.02$ & $8.70 \pm 0.03$ & $7.83 \pm 0.05$ & $8.35 \pm 0.03$ & $8.46 \pm 0.03$& \citet{mendez2021}\\
 & - & $8.40\pm 0.03$ & $8.60 \pm 0.03^{*}$& $8.18 \pm 0.06$ & $8.04 \pm 0.02$ & $8.42 \pm 0.04$& This work\\
 HH~202~S & $8.25\pm 0.16$ & $8.44\pm 0.03$ & $8.65 \pm 0.05$& $8.29 \pm 0.06$ & $8.08 \pm 0.03$ & $8.50 \pm 0.04$& \citet{mesadelgado09}\\
 HH~529~II & $<7.91$ & $8.83\pm 0.07$ & $8.83 \pm 0.07$& $7.36 \pm 0.12$ & $8.54 \pm 0.03$ & $8.57 \pm 0.03$& \citet{mendez2021}\\
 HH~529~III & $<7.95$ & $8.84\pm 0.09$ & $8.84 \pm 0.09$& $7.51 \pm 0.22$ & $8.48 \pm 0.03$ & $8.53 \pm 0.03$& \citet{mendez2021}\\
 HH~204 & $8.57\pm 0.03$ & $<7.54$ & \boldmath{$8.57 \pm 0.03$}& $8.62 \pm 0.05$ &$6.34\pm 0.02$ &\boldmath{$8.62 \pm 0.05$} & This Work \\
\enddata
\tablecomments{Abundances in units of 12+log(X$^{\text{n}+}$/H$^+$) or 12+log(X/H).\\
$^*$ Using the O$^{+}$ abundance based on CELs.}
\end{deluxetable*}

\subsection{On the presence of high-density inclusions}
\label{subsec:highdensity}

Last but not least we want to discuss the influence of the presence of an unrecognized high-density component in the spectrum of a photoionized region. We have studied this scenario in Sec.~\ref{sec:mixing_things} adding the nebular emission from the Orion Nebula, the DBL and HH~204, which would be obtained when observing with a velocity resolution  lower than $\sim$54 km s$^{-1}$ or $R \approx 5550$. 
In this case, the classical density diagnostics based on ratios of [O\thinspace II], [S\thinspace II] and [Cl\thinspace III] do not adequately detect the high density of HH~204. The biased low density values determined with these diagnostics lead to an overestimate of $T_{\rm e}(\text{[N\thinspace II]})$ and a subsequent underestimate of abundances of some elements. 
In the case of O$^{+}$, the underestimate would be $\sim 0.2 \text{ dex}$ (see Sec.~\ref{sec:mixing_things}), producing a similar impact on the total O abundance, as O$^{+}$ is the dominant ion. 
In addition, there is an indirect effect on other elements in which total abundance is derived from lines of highly ionized ions through the ICFs.
This is because they depend on the degree of ionization, parameterized by   O$^{2+}$/(O$^{+}$+O$^{2+}$). 
The impact of high-density inclusions on the abundances will depend on their contribution to the integrated volume, ionization degree, and density.

Previous studies of the area of HH~204 -- all based on lower spectral resolution spectroscopy -- reported localized peaks of $T_{\rm e}(\text{[N\thinspace II]})$ \citep{mesadelgado08,nunezdiaz12,odell17}, which were interpreted as the product of shock heating. 
The results presented in  Sec.~\ref{sec:small_scale} demonstrate that this interpretation is not correct and can be noted in Fig.~2 of \citet{odell17}. That figure shows that $n_{\rm e}(\text{[S\thinspace II]})$ increases when  approaching the bowshock from the direction of the jet, reaching a zone where its value stabilizes around $\sim 5000\text{ cm}^{-3}$ and decreases again when moving outwards. However, when using $n_{\rm e}(\text{[Fe\thinspace III]})$ -- as shown in Fig.~\ref{fig:small_scale_physical_conditions} --  instead of stabilizing when approaching the bowshock, the density steadily increases up to $\sim 20000 \text{ cm}^{-3}$ at the bowshock of HH~204. Considering that $T_{\rm e}(\text{[N\thinspace II]})$ diagnostic tends to be density sensitive for values larger than $\sim 1000 \text{ cm}^{-3}$, an underestimate of $n_{\rm e}$ implies an overestimate of $T_{\rm e}$ and consequently, we will obtain significantly lower ionic abundances based on CELs, which intensity is strongly dependent on temperature.

%The apparent peaks of $T_{\rm e}(\text{[N\thinspace II]})$ are artifacts produced by the use of $n_{\rm e}$ values that are below the true ones. $T_{\rm e}(\text{[O\thinspace III]})$ does not show such peaks due to the very low ionization degree of HH~204 and the higher critical densities of the levels that produce the nebular and auroral [O\thinspace III] lines. 
% Considering the used atomic data for Fe$^{2+}$, this ratio is practically  insensitive at densities lower than $\sim 10^3 \text{ cm}^{-3}$, while its critical density is above $\sim 10^6 \text{ cm}^{-3}$. In the case of a gas inclusion with density of around or larger than $\sim 10^4 \text{ cm}^{-3}$ immersed in a region with an overall density of 10$^2$ $-$ 10$^3$ cm$^{-3}$,

In Table~\ref{tab:pc} we can see that, for HH~204, even the classical density diagnostics give values consistent with those obtained from the ratio of [Fe\thinspace III] lines. This is because HH~204, due to its orientation and the spectral resolution of the observations, can be interpreted basically as a single slab of high-density gas. This would be different in the case of an HH object moving directly towards us and observed with low resolution spectroscopy. We would most likely have a density gradient in the line of sight, because it will cross the compressed gas at the bowshock and the less denser material traveling behind along the jet axis. In situations like this, a way to detect the presence of high-density inclusions -- as HH objects -- can be the use of the $I(\text{[Fe\thinspace III] } \lambda 4658)/I(\text{[Fe\thinspace III] } \lambda 4702)$ ratio as a density diagnostic. This diagnostic will be biased to the higher density component, while classical ones will be biased in the opposite direction. A significant discrepancy between the [Fe\thinspace III] diagnostic and classical ones in a region of apparently low $n_{\rm e}$ may serve as an indicator of this type of situation. However, factors such as the degree of ionization of the gas and the relative volume occupied by each mixed component can mask density inhomogeneities. It is advisable to analyze each available density diagnostic even if they are discrepant with the others, as such discrepancies can indicate the presence of real inhomogeneities.   

\section{Conclusions}
\label{sec:conc}

We have studied the physical conditions and chemical composition of the photoionized Herbig-Haro object HH~204 through deep high-spectral resolution UVES spectroscopy and \textit{HST} imaging.  
Our spectral resolution allows us to cleanly separate HH~204 from the various kinematic components of the Orion Nebula along the same line of sight.

We have analyzed the distribution of the physical conditions of HH~204 along the slit with sub-arcsecond spatial resolution. 
We find a steady increase of $n_{\rm e}$ from $\sim 10000 \text{ cm}^{-3}$ at $\sim 13 \text{ mpc}$ behind the bowshock to $\sim 20000 \text{ cm}^{-3}$ close  to it. 
The temperature determined from the most abundant ion stages,
such as $T_{\rm e} \text{([N\thinspace II])}$ and $T_{\rm e} \text{([S\thinspace III])}$ is approximately constant at \SI{9000 \pm 500}{K} along the slit. 
In contrast, $T_{\rm e} \text{([O\thinspace III])}$ is generally higher and shows a pronounced gradient from \(\approx \SI{17000}{K}\) close to the bow shock to \(\approx \SI{12000}{K}\) at distances \(> \SI{5}{mpc}\).
We interpret this in terms of a two-zone model 
(Sec.~\ref{subsubsec:two_temps_model}).
Zone~A represents gas that is at the photoionization equilibrium temperature,
and which contributes the overwhelming majority of the low and intermediate-ionization emission.  
Zone~B is a higher temperature cooling layer behind the bow shock, and this contributes a significant fraction of the [\ion{O}{3}] emission but contributes negligibly to the other ions.

%\jorge{Una frase resumiendo la secci\'on 9.3.1?}%Based on the observed distribution of $T_{rm e} \text{([O\thinspace III])}$ and assuming the most likely 3D geometry of the observed area of HH~204, we perform what we think may be the most plausible estimate of the Peimbert's $t^2$ representative of the integrated gas volume of a nebular object.

We estimate that around $\sim$ 99\% of the gas in the observed area of HH~204 is composed of low and intermediate ionization stages (ionization potential \(< \SI{25}{eV}\)). 
Based on the intensity of CELs, we determine the ionic abundances of O$^{+}$, O$^{2+}$, N$^{+}$, Ne$^{2+}$, S$^{+}$, S$^{2+}$, Cl$^{+}$, Cl$^{2+}$, Ar$^{2+}$, Fe$^{+}$, Fe$^{2+}$, Ni$^{+}$, Ni$^{2+}$, Ca$^{+}$ and Cr$^{+}$ 
We also calculate the ionic abundances of He$^{+}$, O$^{+}$ and C$^{2+}$ from the relative intensity of RLs. 
In HH~204, we find no difference when determining the O$^{+}$ abundance using CELs or RLs. Since practically all O is O$^{+}$ in this object, we can say that the abundance discrepancy (AD) is virtually zero for HH~204, contrary to what is found in essentially all ionized nebulae. 
Both, CELs and RLs provide an O abundance of $\sim 8.60 \pm 0.05$, slightly lower than the solar value of $\text{O}=8.73\pm 0.07$ recommended by \citet{lodders19}, but consistent with many other independent determinations for the Orion Nebula.

Due to the low degree of ionization of HH~204, we can derive the O, N, S, Cl, Fe and Ni abundances without ICFs. In principle, those O, N, S and Cl abundances should be representative of the Orion Nebula ones as well.
Fe and Ni abundances of HH~204 are a factor 3.5 higher than in the Orion Nebula due to the destruction of dust grains at the bowshock. We also found direct evidences of the presence of an ionization front trapped in HH~204 such as the detection of deuterium lines produced by non-ionizing far-UV photons.

From  archival \textit{HST} imaging with a higher spatial resolution than our spectra,
we find a narrow border of high [O\thinspace III]/H$\alpha$ that traces the leading edge of the bow shock in HH~204 
(Fig.~\ref{fig:ratio-hst-oiii-ha} and sec.~\ref{sec:high-resol-imag}). 
We identify this with the post-shock cooling layer, with a width \(\approx \SI{0.1}{mpc}\). 
This is the same as the high-temperature zone~B, which we invoked in order to explain the spatial profile of \(T_\mathrm{e}(\oiii)\) in the UVES spectra. 
Note however that this layer is much narrower than can be spatially resolved in our spectroscopic observations, 
which means that the effects on temperature diagnostics are diluted. 
We predict that much higher values of \(T_\mathrm{e}(\oiii) \approx \SI{30000}{K}\) would be found if the \(\lambda 4363 / \lambda 5007\) ratio were to be observed at a spatial resolution of \SI{0.05}{arcsec}.

We investigate the origin of the driving jets of both HH~204 and the nearby HH~203 using both proper motions and channel maps of highly blueshifted emission
(Sec.~\ref{sec:origin-jet-that}).
We find that HH~203 is the superposition of two flows: a high ionization and high velocity flow at PA118, which originates in the Orion~S region, plus a low-ionization and lower-velocity flow at PA140, which originates near the Trapezium.
The proper motion of the HH~204 bow shock is closely aligned with the PA140 flow, suggesting that HH~204 may also be driven by this same jet, but there is little evidence of such a connection from the blue-shifted channel maps. 
Instead, there is evidence for a third flow at PA108 that appears to be feeding into HH~204 and which may be connected to a molecular hydrogen filament originating in the region to the south of Orion~S. 

Our observations allow us to separate and analyze the spectrum of the Diffuse Blue Layer, an ionized gas component with a radial velocity different from that of the Orion Nebula and HH~204. We have estimated its physical conditions -- its $T_{\rm e}$ for the first time --, revealing that it has a density lower than the Orion Nebula. We have calculated its chemical composition for the first time.% This component seems to be ionized by leaking photons from the Orion Nebula. 

Our analysis of the spectrum of the kinematic component corresponding to the Orion Nebula reveals a lower ionization degree and $n_{\rm e}$  with respect to the results of Paper~I. This comparison also indicates that $T_{\rm e}$ in the Orion Nebula decrease with the radial distance from $\theta^{1} \text{ Ori C}$. The chemical composition of the nebular component is similar to that found in Paper~I although there seems to be a slightly lower O abundance (less than 0.04 dex), perhaps related to different depletion factors onto dust grains of this element.

We carry out the exercise of simulating a spectrum with lower spectral and spatial resolution, where the spectra of the different kinematic components are mixed. We find that the analysis of this integrated spectrum can lead to erroneous physical conditions and chemical abundances. For example, the estimation of $n_{\rm e}$ by averaging $n_{\rm e} \text{([O\thinspace II])}$, $n_{\rm e} \text{([S\thinspace II])}$ and $n_{\rm e} \text{([Cl\thinspace III])}$ underestimates the true density, resulting in an overestimation of the temperature of the low ionization ions, which constitute an important fraction of the gas in HH~204, the dominant component of the integrated spectrum. This fact leads to an underestimate of the abundances and to obtaining a mistaken average degree of ionization, the parameter on which most ICF schemes are based. Therefore, the determination of the chemical abundances would be wrong in practically all elements. Indicators of density such as  $I(\text{[Fe\thinspace III] } \lambda 4658)/I(\text{[Fe\thinspace III] } \lambda 4702)$ may be used to detect the presence of high density clumps associated with HH objects or shocks. 
A similar point is made by \citet{ODell:2021l} with respect to unrecognized heterogeneity in physical conditions leading to misleading results, and we echo the warning of that paper.

\section*{Acknowledgements}

This work is based on observations collected at the European Southern Observatory, Chile, proposal number ESO 092.C-0323(A). We are grateful to the anonymous referee for his/her helpful comments. We acknowledge support from the Agencia Estatal de Investigaci\'on del Ministerio de Ciencia e Innovaci\'on (AEI-MCINN) under grant {\it Espectroscop\'\i a de campo integral de regiones \ion{H}{2} locales. Modelos para el estudio de regiones \ion{H}{2} extragal\'acticas} with reference 10.13039/501100011033. WJH is grateful for financial support provided by Direcci\'on General de Asuntos del Personal Acad\'emico, Universidad Nacional Aut\'onoma de M\'exico, through grant Programa de Apoyo a Proyectos de Investigaci\'on e Inovaci\'on Tecnol\'ogica IN107019. JG-R acknowledges support from Advanced Fellowships under the Severo Ochoa excellence programs SEV-2015-0548 and CEX2019-000920-S. JEM-D thanks the support of the Instituto de Astrof\'isica de Canarias under the Astrophysicist Resident Program and acknowledges support from the Mexican CONACyT (grant CVU 602402). The authors acknowledge support under grant P/308614 financed by funds transferred from the Spanish Ministry of Science, Innovation and Universities, charged to the General State Budgets and with funds transferred from the General Budgets of the Autonomous Community of the Canary Islands by the MCIU.  

%%%%%%%%%%%%%%%%%%%%%%%%%%%%%%%%%%%%%%%%%%%%%%%%%%

%%%%%%%%%%%%%%%%%%%% REFERENCES %%%%%%%%%%%%%%%%%%

\bibliography{Mendez}{}
\bibliographystyle{aasjournal}

\newpage
%%%%%%%%%%%%%%%%%%%%%%%%%%%%%%%%%%%%%%%%%%%%%%%%%%

%%%%%%%%%%%%%%%%% APPENDICES %%%%%%%%%%%%%%%%%%%%%

\appendix

\section{How reliable are the atomic data of [Fe\thinspace III] that we use?}
\label{sec:atomic_data_fe3}

Generally, the discrepancy between the physical conditions derived from diagnostics based on [Fe\thinspace III] lines and those estimated from other ions has been interpreted as a result of errors in the transition probabilities and/or in the collision strengths of the Fe$^{2+}$ ion \citep[See the introduction of][]{laha17}. With HH~204 we have an excellent opportunity to test the reliability of the atomic data we use for this ion for the following reasons: (i) we have enough spectral resolution to separate the emission of HH~204 from that of the Orion Nebula. (ii) Owing to its geometry and 3D trajectory, we do not expect significant inhomogeneities in the physical conditions within the line of sight for ions of low and intermediate ionization stages. Due to this, all the density diagnostics used in Table~\ref{tab:pc} are consistent with each other, while the global gas temperature, due to the low degree of ionization, is well represented by $T_{\rm e}(\text{[N\thinspace II]})$. (iii) The [Fe\thinspace III] emission is enhanced owing to the destruction of dust grains containing Fe atoms in the shock, which allows to have a good signal to noise even for some weak lines that are difficult to detect. 

We have used a set of transition probabilities compiled in PyNeb, which includes the data from  \citet[][]{Quinet96} and those from \citet[][]{Johansson00} for $^5\text{D}- ^5\text{S}_2$ transitions. However, these transitions produce lines out of the spectral range covered by our observation, so we finally only use the calculations from \citet[][]{Quinet96}. Table~\ref{tab:fe3_ratios_theo} shows that the transition probabilities we use are in good agreement with the observed intensity ratios of lines arising from the same upper level in the case of lines used to determine physical conditions. However, the intensity ratios between lines that arise from different upper levels do depend on $n_{\rm e}$ and $T_{\rm e}$. As discussed above and in Sec.~\ref{subsec:highdensity}, in HH~204 there are no significant density inhomogeneities that may produce a bias in some diagnostics, contrary to the case analyzed in Sec.~\ref{sec:mixing_things}. Thus, all the density diagnostics included in Table~\ref{tab:pc} give consistent results, and the average of $n_{\rm e}\text{([O\thinspace II])}$, $n_{\rm e}\text{([S\thinspace II])}$, $n_{\rm e}\text{([Cl\thinspace III])}$ and $n_{\rm e}\text{([Fe\thinspace II])}$ is $n_{\rm e}=13330 \pm 550$. Using this density for its calculation, $T_{\rm e}(\text{[N\thinspace II]})$ remains practically unchanged from what is shown in Table~\ref{tab:pc}. Considering these values of $n_{\rm e}$ and $T_{\rm e}(\text{[N\thinspace II]})$, we can check the validity of [Fe\thinspace III] atomic data by applying the procedure that we describe below.  Firstly, we take into account all the observed [Fe\thinspace III] lines that are not affected by blending with other lines, sky features or telluric absorptions. Then we normalize their emission with respect to $I(\text{[Fe\thinspace III] } \lambda 4658)/I(\text{H}\beta)=1000$. We discard the [Fe\thinspace III] $\lambda \lambda 3355.50, 7078.22$ lines, since their FWHM are much wider than the rest of [Fe\thinspace III] lines, which is indicative of line blending. [Fe\thinspace III] $\lambda 9203.85$ is also discarded because it shows a radial velocity of $\sim 10\text{ km s}^{-1}$,  larger than the velocities of the rest of [Fe\thinspace III] lines, which may be indicative of a doubtful identification. We also discard [Fe\thinspace III] $\lambda 8838.14$ because, although we deblend it from a very close sky feature, its intensity may not be completely reliable. Once we have the set of [Fe\thinspace III] lines with confident observed intensity ratios, they are compared with the predictions of the atomic data for the assumed physical conditions and considering error propagation. The results are shown in Table~\ref{tab:test_atomic_data_fe3}.

%By taking these physical conditions, derived entirely from other different ions \cesar{NO entiendo que quieres decir, a que condiciones fisicas te refieres?},\eduardo{a las que derivo del promedio de los diagnosticos clasicos, que menciono antes, que vienen de otros iones que no son feIII} we can check the validity of [Fe\thinspace III] atomic data following : (i) We take all the observed [Fe\thinspace III] lines not affected by observational issues such as blend with other lines or with sky features or telluric absorptions. Then we normalize their emission with $I(\text{[Fe\thinspace III] } \lambda 4658)/I(\text{H}\beta)=1000$. We discard the following lines: $\lambda \lambda 3355.50, 7078.22$, since their observed FWHM are much wider than other [Fe\thinspace III] lines, which is indicative of line blending; $\lambda 9203.85$, that has a higher radial velocity of around $\sim 10\text{ km s}^{-1}$ than the rest of [Fe\thinspace III] lines, which may be indicative of a doubtful identification and $\lambda 8838.14$ since it was deblended from a very close sky feature and their intensity may not be reliable. (ii) Then, the observations were compared with the predictions of the atomic data under the adopted physical conditions, considering the propagation of uncertainties. The results are shown in Table~\ref{tab:test_atomic_data_fe3}.

Table~\ref{tab:test_atomic_data_fe3} does not include $5\text{D}-7\text{S}$ transitions ([Fe\thinspace III] $\lambda \lambda 3322.47, 3371.35, 3406.18$ lines) because their transition probabilities are not calculated in the reference of the atomic data used (their ``Predicted'' and ``Difference'' columns are empty). However, their measured intensities can be used to check other atomic data sets that do include them. In general,  Table~\ref{tab:test_atomic_data_fe3} shows good agreement between  predicted and observed intensity ratios of  [Fe\thinspace III] lines. Only 4 lines ($\lambda \lambda 4008.34, 4079.69, 4985.88, 7088.46$) show differences larger than 10\%, exceeding the error bars. This can be attributed to errors in their atomic data. The first two lines arise from the same $^3\text{G}_4$ upper level, so their intensity ratio only depends on their transition probabilities. Although the $I(\text{[Fe\thinspace III] } \lambda 4008.34)/I(\text{[Fe\thinspace III] } \lambda 4079.69)$ ratio is not included in Table~\ref{tab:fe3_ratios_theo} -- these lines were not used to determine physical conditions -- its intensity ratio of 4.43 $\pm$ 0.30 is larger than the theoretical one of 3.92. Therefore it is plausible that part of the observed discrepancy is due to incorrect transition probabilities. The largest differences reported in Table~\ref{tab:test_atomic_data_fe3} are for [Fe\thinspace III] $\lambda \lambda 4985.88, 7088.46$ lines, but we can not find an obvious explanation for this.
In addition to the atomic data used, we have checked other sets. For transition probabilities: \citet[][]{NP96,BBQ10}. For collision strengths: \citet[][]{BBQ10,BB14}. We have tried all possible combinations of these data. Of the 9 combinations, the atomic data we use in this paper minimizes the difference between the predicted and measured intensity ratios. The results of this appendix indicate that the atomic data used in this work for [Fe\thinspace III] lines contribute little to errors in the derived physical conditions and the Fe$^{2+}$ abundances, at least for the conditions of HH~204. As we discuss in Sec.~\ref{sec:mixing_things} the discrepancy normally found between $n_{\rm e}(\text{[Fe\thinspace III]})$ and the classical diagnostics -- as $n_{\rm e}(\text{[O\thinspace II]})$ or $n_{\rm e}(\text{[S\thinspace II]})$ -- may be rather indicative of the presence of high-density inclusions within the line of sight. For a complete test of the atomic data, similar studies would be necessary in different ranges of physical conditions. We will continue investigating this topic in other HH objects in future papers of this series.

\section{Supporting material}
\label{sec:sup_mat}

\begin{figure*}
\centering
  \begin{minipage}{7.5cm}
    %\centering\includegraphics[height=4cm,width=\columnwidth]{gatito-cesped_0.jpg}
    \centering\includegraphics[height=4cm,width=\columnwidth]{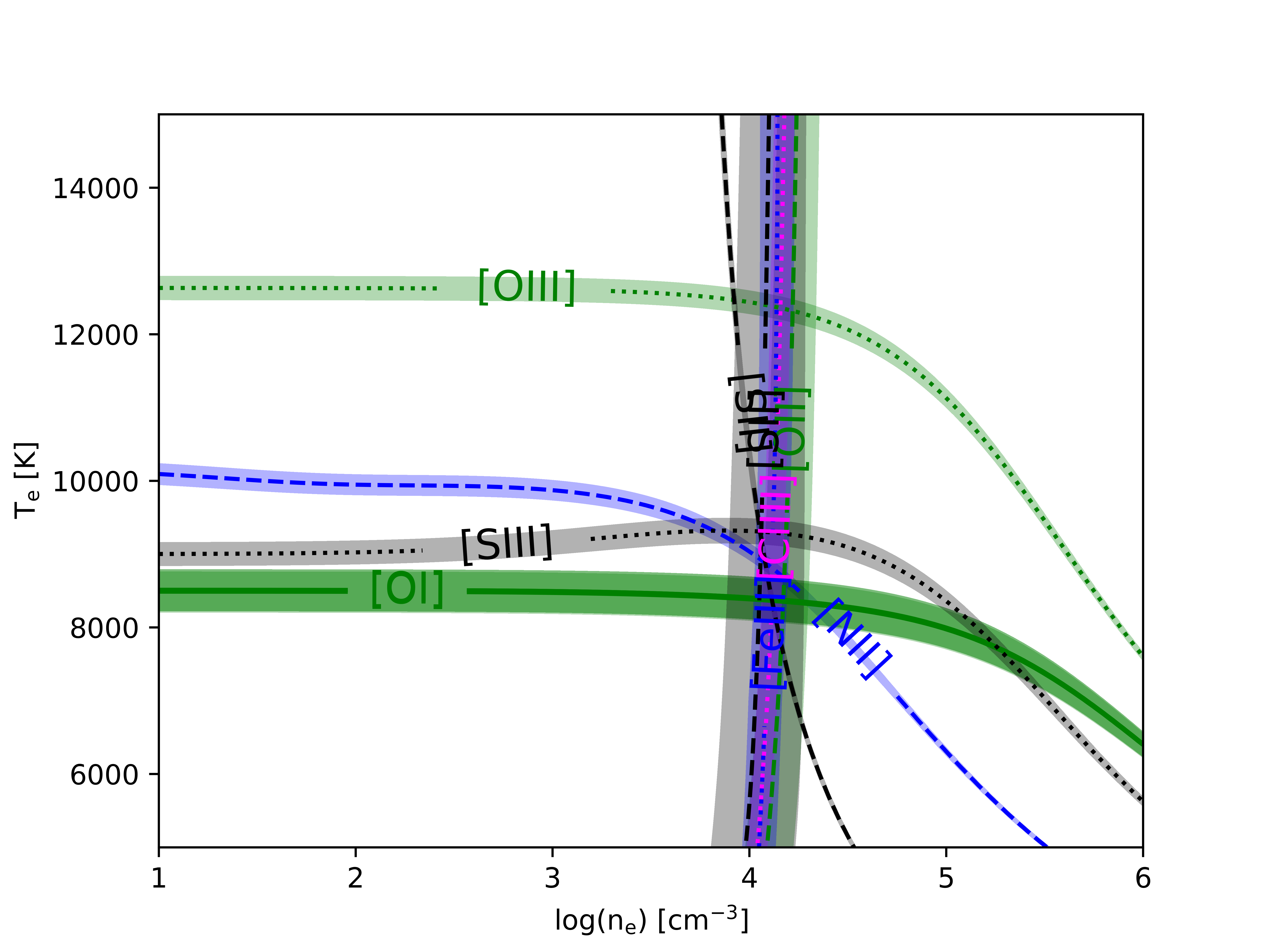}

    \centerline{(a) Cut 1, HH~204.}
    \smallskip
  \end{minipage}
  \begin{minipage}{7.5cm}
     %\centering\includegraphics[height=4cm,width=\columnwidth]{gatito-cesped_0.jpg}
    \centering\includegraphics[height=4cm,width=\columnwidth]{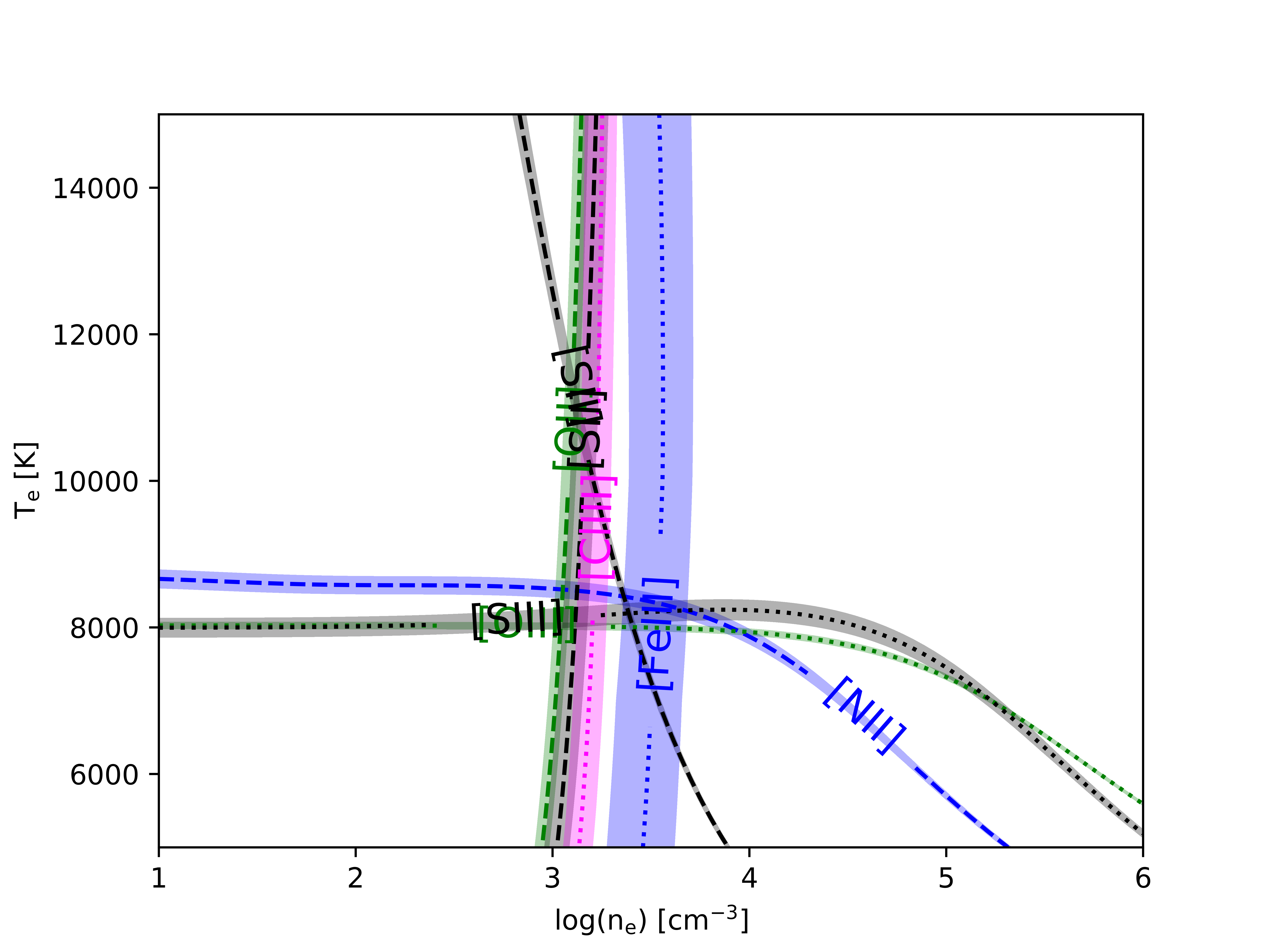}
    \centerline{(b) Cut 1, nebular emission plus Blue Layer emission.}
    \smallskip
  \end{minipage}
 
  \begin{minipage}{7.5cm}
   %\centering\includegraphics[height=4cm,width=\columnwidth]{gatito-cesped_0.jpg}
    \centering\includegraphics[height=4cm,width=\columnwidth]{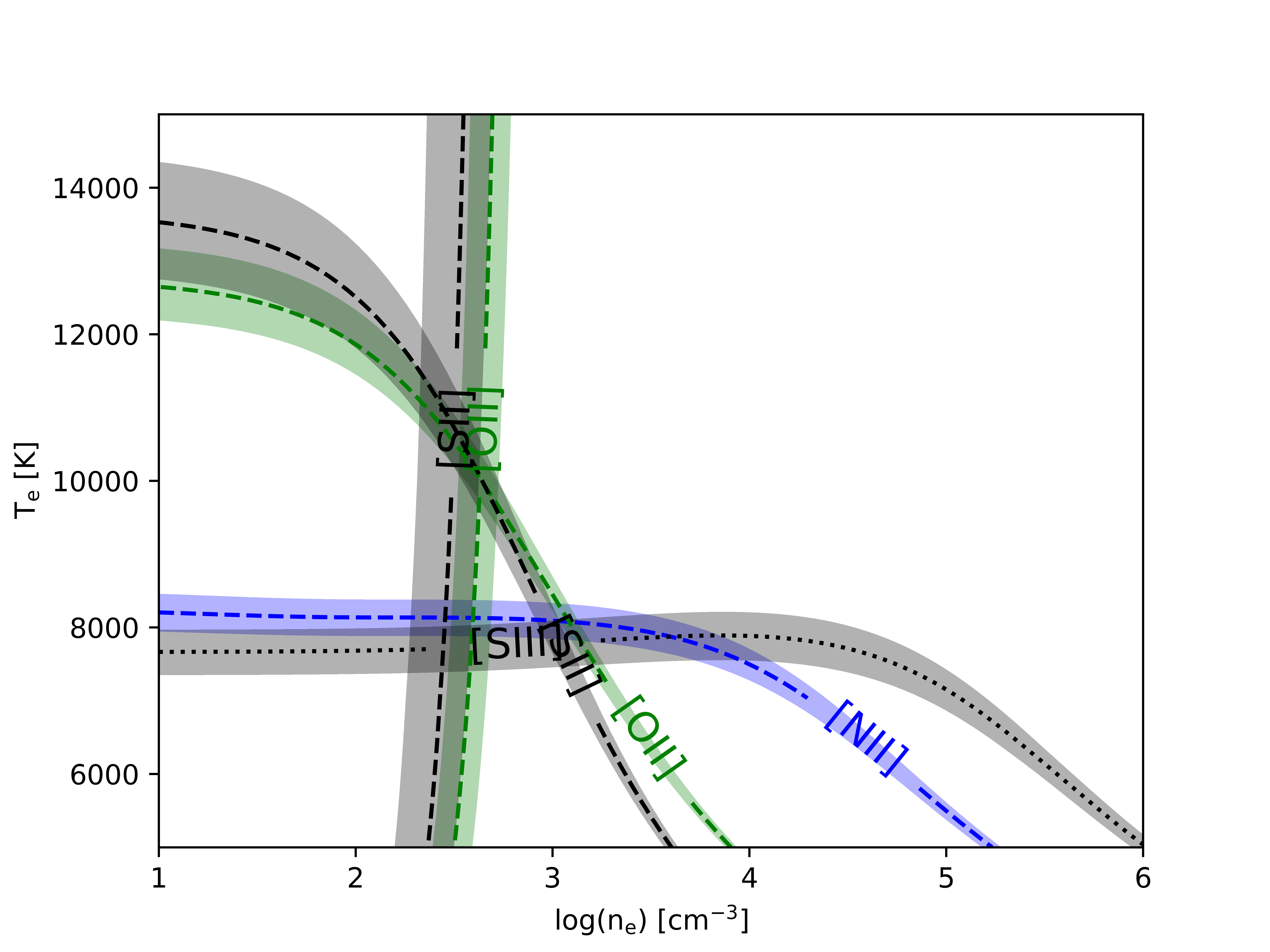}
    \centerline{(c) Cut 2, Blue Layer.}
    \smallskip
  \end{minipage}
  \begin{minipage}{7.5cm}
   % \centering\includegraphics[height=4cm,width=\columnwidth]{gatito-cesped_0.jpg}

    \centering\includegraphics[height=4cm,width=\columnwidth]{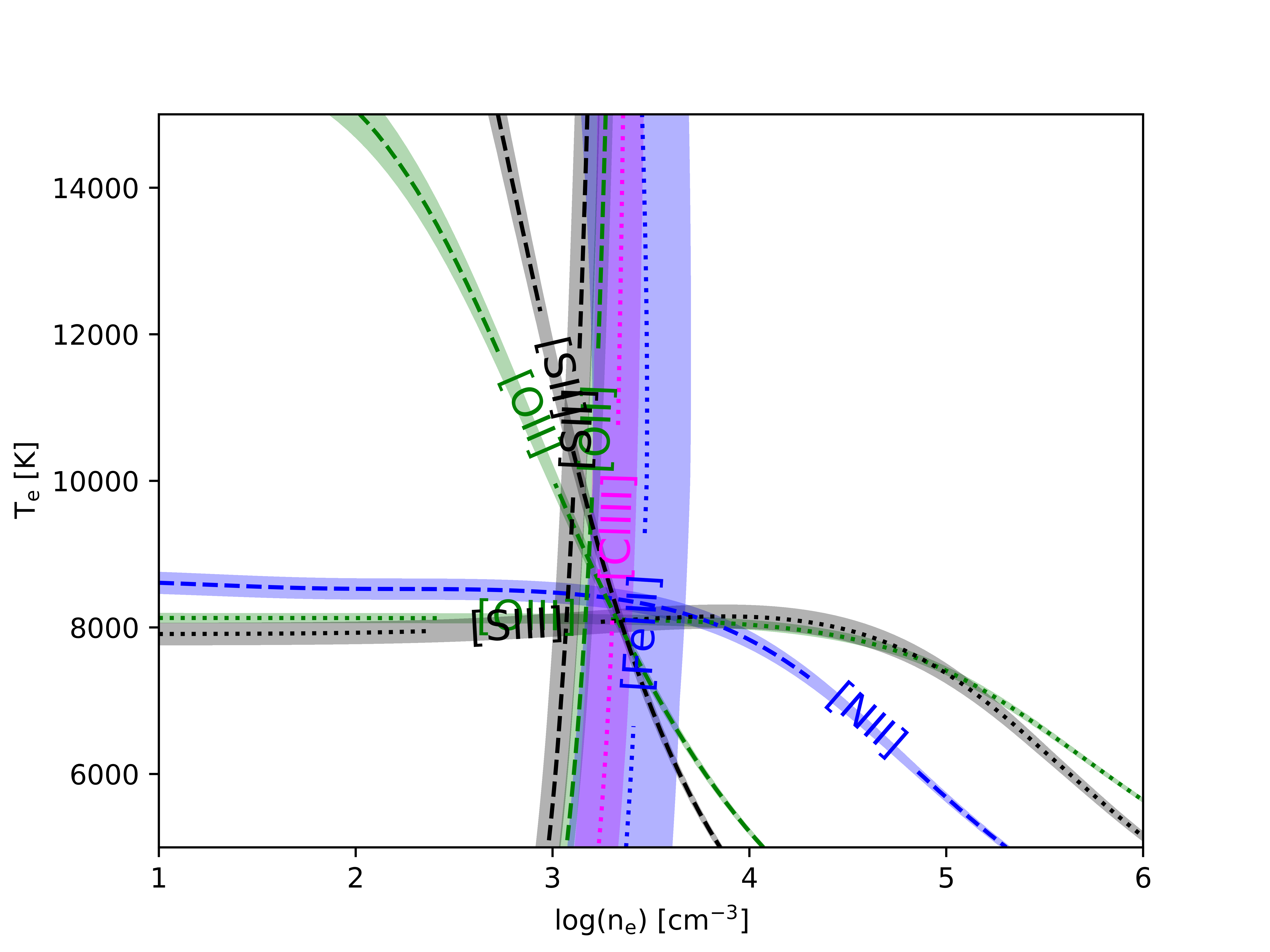}
    \centerline{(d) Cut 2, nebular component.}
    \smallskip
  \end{minipage}

  \caption{Plasma diagnostic plots for the individual analyzed components. The labeled diagnostics correspond to those discussed in Sec.~\ref{subsec:physical_cond}.}
\label{fig:plasma}
\end{figure*}

In this appendix we include the following material:

\begin{itemize}
    \item Fig.~\ref{fig:plasma}: Plasma diagnostics for the individual components analyzed in this work.
    
    \item Table~\ref{tab:sample_of_lines}: Sample of lines of the spectra of cut~1 as found in the online material.

    \item Table~\ref{tab:atomic_data}: Atomic data set used for CELs.
    
    \item Table~\ref{tab:rec_atomic_data}: Atomic data set used for RLs.
    
    \item Table~\ref{tab:fe3_ratios_theo}: Measured and predicted
    [Fe\thinspace III] intensity ratios from lines that arise from a common upper level. 
    
    \item Table~\ref{tab:fe2_ratios_theo}: Measured and predicted [Fe\thinspace II] intensity ratios from lines that arise from a common upper level. 
    
    \item Table~\ref{tab:test_atomic_data_fe3}: Measured and predicted [Fe\thinspace III] intensity ratios for all detected lines, using the atomic data chosen in this work.
    
    \item Table~\ref{tab:big_detail_blue}: Pixel-to-pixel spatial distribution of the physical conditions and ionic abundances of HH~204 in the UVES blue arm spectra.
    
    \item Table~\ref{tab:big_detail_red1}: Pixel-to-pixel spatial distribution of the physical conditions and ionic abundances of HH~204 in the UVES red arm spectra.

    \item Table~\ref{tab:big_detail_red2}: Pixel-to-pixel spatial distribution of ionic abundances of HH~204 in the UVES red arm spectra.

\end{itemize}

\begin{table*}
\centering
\caption{Sample of 15 lines from the spectra of cut~1.} 
\label{tab:sample_of_lines}
%\begin{adjustbox}{width=\textwidth}
\resizebox{\textwidth}{!}{%
\begin{tabular}{@{}ccccccccccccccccc@{}}

\hline
& &\multicolumn{6}{c}{HH~204}& \multicolumn{6}{c}{The Orion Nebula+The Diffuse Blue Layer}&\\
$\lambda_0$ ( \AA ) & Ion & $\lambda_{obs}$ & Vel$\left( \lambda_0 \right)$ (Km s$^{-1}$) & FWHM (Km s$^{-1}$) & F$\left( \lambda \right)$/F$\left( \mbox{H}\beta \right)$ & I$\left( \lambda \right)$/I$\left( \mbox{H}\beta \right)$ & Err \% & $\lambda_{obs}$ & Vel$\left( \lambda_0 \right)$ (Km s$^{-1}$) & FWHM (Km s$^{-1}$) & F$\left( \lambda \right)$/F$\left( \mbox{H}\beta \right)$ & I$\left( \lambda \right)$/I$\left( \mbox{H}\beta \right)$ & Err \% & Notes \\
\hline
4701.64 & $\mbox{[Fe}\thinspace \mbox{III]}$ & 4701.25 & -24.76 & 18.30 $\pm$ 0.02 & 1.424 & 1.460 & 2 & 4701.83 & 12.22 & 23.27 $\pm$ 0.33 & 0.193 & 0.197 & 3 &   \\
4713.14 & $\mbox{He}\thinspace \mbox{I}$ & 4712.80 & -21.48 & 18.57 $\pm$ 0.31 & 0.206 & 0.211 & 3 & 4713.35 & 13.51 & 27.03 $\pm$ 0.30 & 0.436 & 0.443 & 3 &  \\
4728.07 & $\mbox{[Fe}\thinspace \mbox{II]}$ & 4727.75 & -20.08 & 13.57 $\pm$ 0.36 & 0.072 & 0.073 & 4 & 4728.45 & 24.31 & 8.18 $\pm$ 4.17 & 0.004 & 0.004 & 30 &  \\
4734.00 & $\mbox{[Fe}\thinspace \mbox{III]}$ & 4733.57 & -27.00 & 18.18 $\pm$ 0.05 & 0.634 & 0.647 & 2 & 4734.15 & 9.74 & 23.81 $\pm$ 1.68 & 0.067 & 0.068 & 6 &   \\
4740.17 & $\mbox{[Ar}\thinspace \mbox{IV]}$ & * & * & * & * & * & * & 4740.35 & 11.65 & 14.86 $\pm$ 6.68 & 0.006 & 0.006 & 29 &  \\
4754.81 & $\mbox{[Fe}\thinspace \mbox{III]}$ & 4754.42 & -24.90 & 18.29 $\pm$ 0.04 & 0.800 & 0.813 & 2 & 4755.00 & 11.67 & 22.95 $\pm$ 0.58 & 0.127 & 0.128 & 4 &   \\
4769.53 & $\mbox{[Fe}\thinspace \mbox{III]}$ & 4769.14 & -24.77 & 18.23 $\pm$ 0.04 & 0.505 & 0.512 & 2 & 4769.73 & 12.32 & 22.75 $\pm$ 0.72 & 0.064 & 0.065 & 4 &   \\
4774.73 & $\mbox{[Fe}\thinspace \mbox{II]}$ & 4774.42 & -19.69 & 13.19 $\pm$ 0.51 & 0.070 & 0.071 & 4 & * & * & * & * & * & * &  \\
4777.70 & $\mbox{[Fe}\thinspace \mbox{III]}$ & 4777.38 & -20.30 & 17.82 $\pm$ 0.21 & 0.304 & 0.308 & 3 & 4777.97 & 16.73 & 24.34 $\pm$ 2.76 & 0.032 & 0.032 & 10 &   \\
4803.29 & $\mbox{N}\thinspace \mbox{II}$ & * & * & * & * & * & * & 4803.46 & 10.50 & 18.10 $\pm$ 2.08 & 0.019 & 0.019 & 8 &  \\
4814.54 & $\mbox{[Fe}\thinspace \mbox{II]}$ & 4814.23 & -19.37 & 14.38 $\pm$ 0.04 & 0.393 & 0.396 & 2 & * & * & * & * & * & * &  \\
4874.50 & $\mbox{[Fe}\thinspace \mbox{II]}$ & 4874.18 & -19.51 & 13.84 $\pm$ 0.54 & 0.051 & 0.051 & 4 & * & * & * & * & * & * &  \\
4861.32 & $\mbox{H}\thinspace \mbox{I}$ & 4860.97 & -21.47 & 24.67 $\pm$ 0.01 & 100.000 & 100.000 & 2 & 4861.52 & 12.45 & 30.59 $\pm$ 0.01 & 100.000 & 100.000 & 2 &  \\
4861.32 & $\mbox{H}\thinspace \mbox{I}$ & 4859.66 & -102.26 & 14.31 $\pm$ 1.90 & 0.106 & 0.106 & 10 & * & * & * & * & * & * &  Deuterium \\
4874.50 & $\mbox{[Fe}\thinspace \mbox{II]}$ & 4874.18 & -19.51 & 13.84 $\pm$ 0.54 & 0.051 & 0.051 & 4 & * & * & * & * & * & * &  \\
4881.07 & $\mbox{[Fe}\thinspace \mbox{III]}$ & 4880.71 & -21.92 & 18.00 $\pm$ 0.01 & 2.251 & 2.245 & 2 & 4881.30 & 14.32 & 20.70 $\pm$ 0.17 & 0.248 & 0.247 & 3 &   \\
\hline
\end{tabular}%
    }
%\end{adjustbox}
\end{table*}

\begin{deluxetable*}{lcc}
\tablecaption{Atomic data set used for collisionally excited lines. \label{tab:atomic_data}}
\tablewidth{0pt}
\tablehead{
\multicolumn{1}{l}{Ion} & \multicolumn{1}{c}{Transition Probabilities} &
\multicolumn{1}{c}{Collision Strengths} 
}
\startdata
O$^{0}$   &  \citet{Wiese96} & \citet{Bhatia95}\\
O$^{+}$   &  \citet{Fischer04} & \citet{Kisielius09}\\
O$^{2+}$  &  \citet{Wiese96}, \citet*{Storey00} & \citet{Storey14}\\
N$^{+}$   &  \citet{Fischer04} & \citet{Tayal11}\\
Ne$^{2+}$  &  \citet{McLaughlin11} & \citet{McLaughlin11}\\
S$^{+}$   &  \citet{Podobedova09} & \citet{Tayal10}\\
S$^{2+}$  &  \citet{Podobedova09} & \citet{Grieve14}\\
Cl$^{+}$ &  \citet{Mendoza83} & \citet{Tayal04}\\
Cl$^{2+}$ &  \citet{Fritzsche99} & \citet{Butler89}\\
Ar$^{2+}$ &   \citet{Mendoza83_2}, \citet{Kaufman86}  & \citet*{Galavis95}\\
Ar$^{3+}$ &   \citet{Mendoza82b}  & \citet{Ramsbottom97}\\
Fe$^{+}$ & \citet{bautista15} & \citet{bautista15}\\
Fe$^{2+}$ & \citet{Quinet96} , \citet{Johansson00} & \citet{Zhang96}\\
Fe$^{3+}$ & \citet{Fischer08} & \citet{Zhang97}\\
Ni$^{+}$ & \citet{Quinet96Ni}, \citet{Nussbaumer82} & \citet{Bautista04}\\
Ni$^{2+}$ & \citet{Bautista01} & \citet{Bautista01}\\
Ca$^{+}$ & \citet{melendez07} & \citet{melendez07}\\
Cr$^{+}$ & \citet{Tayal20} & \citet{Tayal20} \\
\enddata
\end{deluxetable*}

\begin{deluxetable}{lc}
\tablecaption{Effective recombination coefficients used for recombination lines. \label{tab:rec_atomic_data}}
\tablewidth{0pt}
\tablehead{
\multicolumn{1}{l}{Ion} & \multicolumn{1}{c}{Reference}  
}
\startdata
H$^{+}$   & \citet{Storey95}\\
He$^{+}$   & \citet{Porter12,Porter13}\\
O$^{+}$   & \citet{Pequignot91}\\
O$^{2+}$   & \citet{Storey17}\\
C$^{2+}$   & \citet{Davey00}\\
\enddata
\end{deluxetable}

\begin{deluxetable}{lcccccccccccc}
\tablecaption{Comparison of the observed [Fe\thinspace III] intensity ratios in HH~204 and theoretical ones predicted by the transition probabilities of \citet{Quinet96} and \citet{Johansson00}. \label{tab:fe3_ratios_theo}}
\tablewidth{0pt}
\tablehead{
Line ratio &  HH~204 & Prediction
}
\startdata
3240/3286 & $3.63 \pm 0.81$ & 3.60 \\
3240/3319 & $3.63 \pm 0.86$ &5.06\\
3240/8729$^{*}$ & $11.35 \pm 1.06$ &11.87 \\
3335/3357 & $1.16 \pm 0.20$ & 1.18 \\
3335/8838$^{**}$ & $6.15 \pm 1.00$ & 4.93 \\
4607/4702 & $0.18 \pm 0.01$ & 0.17 \\
4607/4770 & $0.51 \pm 0.01$ & 0.51 \\
4667/4734 & $0.29 \pm 0.01$ & 0.28 \\
4667/4778 & $0.60 \pm 0.03$ & 0.57 \\
4658/4755 & $5.33 \pm 0.15$ & 5.49 \\
4881/4987 & $6.07 \pm 0.17$ & 5.76 \\
5011/5085 & $5.85 \pm 0.32$ & 5.94\\
5271/5412 & $10.75 \pm 0.39$ &   11.01 \\
\enddata
\tablecomments{$^*$ The emission of [Fe\thinspace III]$\lambda 8728.84$ from HH~204 was deblended from the nebular component of [C\thinspace I]$\lambda 8727.13$. \\
$^{**}$ The emission of [Fe\thinspace III]$\lambda 8838.14$ was deblended from a sky feature.
}
\end{deluxetable}

\begin{deluxetable}{lcccccccccccc}
\tablecaption{Comparison of the observed [Fe\thinspace II] intensity ratios in HH~204 and theoretical ones predicted by the transition probabilities of \citet{bautista15}. \label{tab:fe2_ratios_theo}}
\tablewidth{0pt}
\tablehead{
Line ratio &  HH~204 & Prediction
}
\startdata
9052/9399 & $5.45 \pm 0.48$ & $5.49$ & \\
9052/7927 & $18.84 \pm 1.99$ & $6.91$ &  \\
8892/9227 & $1.71 \pm 0.11$ & $1.80$ &  \\
8892/7874 & $28.58 \pm 4.15$ & $10.64$ &  \\
8892/7687 & $3.85 \pm 0.22$ & $1.48$ &  \\
9268/9034$^{*}$ & $1.33 \pm 0.09$ & $1.28$ &  \\
9268/7733$^{*}$ & $11.72 \pm 1.37$ & $5.08$ & \\
\enddata
\tablecomments{$^*$ [Fe\thinspace II]$\lambda 9267.56$ was deblended from  a sky emission.
}
\end{deluxetable}

\begin{deluxetable*}{ccccccccccccc}
\tablecaption{Comparison between predicted and measured [Fe\thinspace III] intensity ratios with the chosen atomic data. The intensities are normalized to $I(\text{[Fe\thinspace III]}) \lambda 4658=1000$. \label{tab:test_atomic_data_fe3} }
\tablewidth{0pt}
\tablehead{
$\lambda$ ( \AA ) & Predicted $I(\lambda)$/$I(4658)$ & Measured $I(\lambda)$/$I(4658)$ & Difference 
}
\startdata
3239.79 & $90.3 \pm 3.1$ & $84.6\pm 6.3$ & -7\% $\pm$ 8\% & \\
3286.24 & $25.1 \pm 0.9$ & $22.3\pm 4.9$ & -13\% $\pm$ 30\% & \\
3319.27 & $17.8 \pm 0.6$ & $22.9\pm 4.5$ & 23\% $\pm$ 16\% & \\
3322.47 & - & $103.0\pm 7.0$ & - & \\
3334.95 & $32.1 \pm 1.1$ & $28.8\pm 3.8$ & -14\% $\pm$ 17\% & \\
3356.59 & $27.2 \pm 0.9$ & $24.5\pm 2.0$ & -9\% $\pm$ 9\% & \\
3366.22 & $14.6 \pm 0.5$ & $18.5\pm 2.5$ & 21\% $\pm$ 12\% & \\
3371.35 & - & $67.9\pm 4.4$ & - & \\
3406.18 & - & $39.1\pm 3.3$ & - & \\
4008.34 & $57.4 \pm 0.7$ & $50.0\pm 1.6$ & -14\% $\pm$ 3\% & \\
4046.49 & $8.2 \pm 0.1$ & $7.7\pm 0.7$ & -6\% $\pm$ 10\% & \\
4079.69 & $14.6 \pm 0.2$ & $11.2\pm 0.8$ & -30\% $\pm$ 7\% & \\
4096.68 & $3.2 \pm 0.1$ & $2.3\pm 0.4$ & -37\% $\pm$ 30\% & \\
4607.12 & $58.5 \pm 0.3$ & $60.3\pm 1.5$ & 2\% $\pm$ 2\% & \\
4667.11 & $40.4 \pm 0.5$ & $43.1\pm 1.6$ & 6\% $\pm$ 3\% &   \\
4701.64 & $338.5 \pm 1.9$ & $336.9\pm 10.3$ & 0\% $\pm$ 3\% &   \\
4734.00 & $146.1 \pm 1.6$ & $150.8\pm 3.4$ & 3\% $\pm$ 2\% &   \\
4754.81 & $182.1 \pm 0.0$ & $187.3\pm 5.7$ & 3\% $\pm$ 3\% &   \\
4769.53 & $115.3 \pm 0.7$ & $118.2\pm 3.2$ & 2\% $\pm$ 2\% &   \\
4777.70 & $70.3 \pm 0.8$ & $71.4\pm 2.4$ & 0\% $\pm$ 3\% &   \\
4881.07 & $484.3 \pm 2.1$ & $519.0\pm 15.3$ & 7\% $\pm$ 2\% &   \\
4924.66 & $6.6 \pm 0.2$ & $6.8\pm 0.5$ & 1\% $\pm$ 7\% & \\
4930.64 & $40.3 \pm 0.6$ & $43.6\pm 1.7$ & 7\% $\pm$ 3\% &   \\
4985.88 & $15.5 \pm 0.7$ & $8.1\pm 0.6$ & -91\% $\pm$ 15\% & \\
4987.29 & $84.1 \pm 0.4$ & $85.2\pm 2.3$ & 1\% $\pm$ 2\% & \\
5011.41 & $143.1 \pm 1.3$ & $147.4\pm 4.5$ & 1\% $\pm$ 2\% & \\
5084.85 & $24.1 \pm 0.2$ & $25.0\pm 1.3$ & 4\% $\pm$ 5\% & \\
5270.57 & $487.8 \pm 3.8$ & $525.0\pm 14.5$ & 6\% $\pm$ 2\% &   \\
5412.06 & $44.3 \pm 0.3$ & $48.9\pm 1.9$ & 9\% $\pm$ 3\% &   \\
7088.46 & $1.1 \pm 0.1$ & $1.8\pm 0.2$ & 39\% $\pm$ 6\% & \\
8728.84 & $7.6 \pm 0.3$ & $7.3\pm 0.4$ & -3\% $\pm$ 7\% &   \\
9701.87 & $24.6 \pm 0.7$ & $24.8\pm 1.1$ & 0\% $\pm$ 5\% & \\
9942.38 & $15.9 \pm 0.5$ & $18.3\pm 1.4$ & 13\% $\pm$ 6\% & \\
\enddata
\end{deluxetable*}

\begin{deluxetable*}{ccccccccccccc}
\tablecaption{Spatial distribution of physical conditions and ionic abundances along HH~204 as a function of the distance from the bowshock. Values derived from the blue arm spectrum. \label{tab:big_detail_blue} }
\tablewidth{0pt}
\tablehead{
Distance & $n_{\rm e}\text{([Fe\thinspace III])}$ & $T_{\rm e}\text{([O\thinspace III])}$&  & & & \\
(mpc) & ($\text{cm}^{-3}$)& (K) &  He$^{+}$ ($\lambda 4471$) & O$^{+}$& O$^{2+}$& Ne$^{2+}$ 
}
\startdata
0.00 & $21180 \pm 5900$ & $16790 \pm 700$ & $10.50 \pm 0.02$ & $8.72 \pm 0.14$ & $6.13 \pm 0.05$& -\\
0.49 & $19020 \pm 5160$ & $17200 \pm 600$ & $10.46 \pm 0.01$ & $8.64 \pm 0.12$ & $6.02 \pm 0.04$& -\\
0.98 & $19610 \pm 4530$ & $18510 \pm 560$ & $10.46 \pm 0.01$ & $8.72 \pm 0.10$ & $5.89 \pm 0.03$& -\\
1.47 & $21810 \pm 3940$ & $15620 \pm 470$ & $10.46 \pm 0.01$ & $8.67 \pm 0.10$ & $6.05 \pm 0.04$& -\\
1.96 & $15930 \pm 4160$ & $15200 \pm 580$ & $10.43 \pm 0.01$ & $8.61 \pm 0.09$ & $6.07 \pm 0.04$& -\\
2.44 & $16170 \pm 3370$ & $15880 \pm 580$ & $10.41 \pm 0.01$ & $8.62 \pm 0.07$ & $5.94 \pm 0.04$& -\\
2.93 & $14040 \pm 2220$ & $14730 \pm 460$ & $10.39 \pm 0.01$ & $8.56 \pm 0.07$ & $5.99 \pm 0.04$& -\\
3.42 & $17480 \pm 3310$ & $14660 \pm 420$ & $10.38 \pm 0.01$ & $8.65 \pm 0.08$ & $5.99 \pm 0.03$& -\\
3.91 & $15800 \pm 2990$ & $13990 \pm 350$ & $10.39 \pm 0.01$ & $8.59 \pm 0.10$ & $6.03 \pm 0.03$& -\\
4.40 & $13600 \pm 2530$ & $13480 \pm 460$ & $10.41 \pm 0.01$ & $8.56 \pm 0.08$ & $6.06 \pm 0.04$& -\\
4.89 & $14800 \pm 2630$ & $12430 \pm 380$ & $10.38 \pm 0.01$ & $8.62 \pm 0.08$ & $6.17 \pm 0.04$& -\\
5.38 & $13680 \pm 2760$ & $12020 \pm 390$ & $10.38 \pm 0.01$ & $8.57 \pm 0.08$ & $6.22 \pm 0.05$& -\\
5.87 & $12210 \pm 2730$ & $11910 \pm 360$ & $10.41 \pm 0.01$ & $8.52 \pm 0.08$ & $6.25 \pm 0.04$& -\\
6.36 & $13180 \pm 2480$ & $11300 \pm 340$ & $10.41 \pm 0.01$ & $8.56 \pm 0.06$ & $6.32 \pm 0.04$& -\\
6.85 & $12270 \pm 2580$ & $11600 \pm 310$ & $10.44 \pm 0.01$ & $8.57 \pm 0.09$ & $6.31 \pm 0.04$& $4.99 \pm 0.07$\\
7.33 & $13740 \pm 2610$ & $11500 \pm 390$ & $10.45 \pm 0.01$ & $8.61 \pm 0.08$ & $6.34 \pm 0.05$& $5.12 \pm 0.07$\\
7.82 & $13440 \pm 2520$ & $11470 \pm 290$ & $10.45 \pm 0.01$ & $8.56 \pm 0.08$ & $6.36 \pm 0.04$& $5.14 \pm 0.06$\\
8.31 & $14040 \pm 2450$ & $11630 \pm 340$ & $10.45 \pm 0.01$ & $8.58 \pm 0.07$ & $6.33 \pm 0.04$& $5.14 \pm 0.07$\\
8.80 & $11630 \pm 2200$ & $11260 \pm 250$ & $10.44 \pm 0.01$ & $8.51 \pm 0.07$ & $6.39 \pm 0.03$& $5.16 \pm 0.06$\\
9.29 & $13490 \pm 2450$ & $11450 \pm 260$ & $10.47 \pm 0.01$ & $8.51 \pm 0.08$ & $6.41 \pm 0.04$& $5.19 \pm 0.05$\\
9.78 & $11960 \pm 2490$ & $11230 \pm 280$ & $10.52 \pm 0.01$ & $8.54 \pm 0.07$ & $6.50 \pm 0.03$& $5.25 \pm 0.06$\\
10.27 & $11400 \pm 2390$ & $11620 \pm 220$ & $10.58 \pm 0.01$ & $8.52 \pm 0.07$ & $6.51 \pm 0.03$& $5.23 \pm 0.05$\\
10.76 & $11420 \pm 2190$ & $12020 \pm 280$ & $10.64 \pm 0.01$ & $8.51 \pm 0.07$ & $6.49 \pm 0.03$& $5.24 \pm 0.05$\\
11.25 & $9490 \pm 1970$ & $11530 \pm 310$ & $10.70 \pm 0.01$ & $8.48 \pm 0.06$ & $6.62 \pm 0.04$& $5.41 \pm 0.06$\\
11.74 & $10400 \pm 1980$ & $11710 \pm 250$ & $10.75 \pm 0.01$ & $8.53 \pm 0.07$ & $6.63 \pm 0.03$& $5.41 \pm 0.06$\\
12.22 & $10330 \pm 2020$ & $11550 \pm 200$ & $10.80 \pm 0.01$ & $8.53 \pm 0.07$ & $6.71 \pm 0.02$& $5.54 \pm 0.04$\\
12.71 & $9740 \pm 1970$ & $11030 \pm 240$ & $10.84 \pm 0.01$ & $8.50 \pm 0.07$ & $6.84 \pm 0.03$& $5.61 \pm 0.05$\\
\enddata
\tablecomments{Abundances in units of 12+log(X$^{\text{n}+}$/H$^+$)}
\end{deluxetable*}

\begin{deluxetable*}{ccccccccccccc}
\tablecaption{Spatial distribution of physical conditions and ionic abundances along HH~204 as a function of the distance from the bowshock. Values derived from the red arm spectrum. \label{tab:big_detail_red1} }
\tablewidth{0pt}
\tablehead{
Distance & $T_{\rm e}\text{([N\thinspace II])}$ & $T_{\rm e}\text{([S\thinspace III])}$ &   & & & & & \\
(mpc) & (K) & (K) &  He$^{+}$ ($\lambda 5876$)& He$^{+}$ ($\lambda 6678$)& N$^{+}$ & O$^{+}$ (RLs) & S$^{+}$& S$^{2+}$}
\startdata
0.29 & $8780 \pm 310$ & $9150 \pm 190$ & $10.41 \pm 0.01$ & $10.57 \pm 0.03$ & $7.79 \pm 0.06$ & $8.68 \pm 0.15$ & $6.82 \pm 0.09$ & $6.84 \pm 0.03$ \\
0.66 & $8970 \pm 340$ & $8780 \pm 180$ & $10.43 \pm 0.01$ & $10.51 \pm 0.03$ & $7.75 \pm 0.05$ & $8.87 \pm 0.11$ & $6.77 \pm 0.08$ & $6.90 \pm 0.03$ \\
1.02 & $8670 \pm 280$ & $9140 \pm 210$ & $10.43 \pm 0.01$ & $10.50 \pm 0.02$ & $7.80 \pm 0.05$ & - & $6.82 \pm 0.09$ & $6.83 \pm 0.03$ \\
1.38 & $8920 \pm 300$ & $9500 \pm 220$ & $10.42 \pm 0.01$ & $10.48 \pm 0.02$ & $7.75 \pm 0.05$ & $8.75 \pm 0.12$ & $6.78 \pm 0.08$ & $6.77 \pm 0.03$ \\
1.74 & $9100 \pm 290$ & $9390 \pm 190$ & $10.42 \pm 0.01$ & $10.43 \pm 0.02$ & $7.71 \pm 0.04$ & - & $6.73 \pm 0.07$ & $6.78 \pm 0.02$ \\
2.10 & $8860 \pm 290$ & $9050 \pm 200$ & $10.41 \pm 0.01$ & $10.48 \pm 0.02$ & $7.73 \pm 0.05$ & $8.57 \pm 0.15$ & $6.73 \pm 0.08$ & $6.82 \pm 0.02$ \\
2.46 & $8980 \pm 220$ & $9290 \pm 210$ & $10.39 \pm 0.01$ & $10.45 \pm 0.02$ & $7.73 \pm 0.04$ & $8.80 \pm 0.09$ & $6.74 \pm 0.06$ & $6.82 \pm 0.03$ \\
2.83 & $8850 \pm 220$ & $9420 \pm 170$ & $10.37 \pm 0.01$ & $10.43 \pm 0.02$ & $7.73 \pm 0.03$ & $8.70 \pm 0.09$ & $6.72 \pm 0.07$ & $6.78 \pm 0.02$ \\
3.19 & $8750 \pm 230$ & $9250 \pm 180$ & $10.37 \pm 0.01$ & $10.39 \pm 0.02$ & $7.75 \pm 0.04$ & $8.72 \pm 0.07$ & $6.76 \pm 0.07$ & $6.77 \pm 0.03$ \\
3.55 & $8790 \pm 190$ & $9410 \pm 180$ & $10.36 \pm 0.01$ & $10.43 \pm 0.02$ & $7.75 \pm 0.04$ & $8.68 \pm 0.07$ & $6.79 \pm 0.07$ & $6.75 \pm 0.02$ \\
3.91 & $8860 \pm 260$ & $9410 \pm 180$ & $10.36 \pm 0.01$ & $10.42 \pm 0.02$ & $7.75 \pm 0.04$ & $8.66 \pm 0.07$ & $6.78 \pm 0.07$ & $6.74 \pm 0.03$ \\
4.27 & $8900 \pm 230$ & $9390 \pm 190$ & $10.37 \pm 0.01$ & $10.42 \pm 0.02$ & $7.74 \pm 0.04$ & $8.74 \pm 0.06$ & $6.75 \pm 0.06$ & $6.74 \pm 0.03$ \\
4.64 & $8800 \pm 230$ & $9430 \pm 190$ & $10.36 \pm 0.01$ & $10.41 \pm 0.02$ & $7.75 \pm 0.04$ & $8.79 \pm 0.06$ & $6.72 \pm 0.06$ & $6.73 \pm 0.02$ \\
5.00 & $8760 \pm 230$ & $9400 \pm 190$ & $10.34 \pm 0.01$ & $10.40 \pm 0.02$ & $7.74 \pm 0.04$ & $8.74 \pm 0.07$ & $6.72 \pm 0.06$ & $6.72 \pm 0.03$ \\
5.36 & $8960 \pm 220$ & $9480 \pm 190$ & $10.36 \pm 0.01$ & $10.41 \pm 0.02$ & $7.71 \pm 0.04$ & $8.72 \pm 0.06$ & $6.66 \pm 0.06$ & $6.71 \pm 0.02$ \\
5.72 & $8880 \pm 210$ & $9310 \pm 190$ & $10.36 \pm 0.01$ & $10.42 \pm 0.02$ & $7.72 \pm 0.04$ & $8.71 \pm 0.06$ & $6.63 \pm 0.08$ & $6.74 \pm 0.03$ \\
6.08 & $8800 \pm 220$ & $9330 \pm 180$ & $10.36 \pm 0.01$ & $10.42 \pm 0.02$ & $7.74 \pm 0.04$ & $8.73 \pm 0.05$ & $6.67 \pm 0.06$ & $6.74 \pm 0.03$ \\
6.44 & $8810 \pm 220$ & $9280 \pm 180$ & $10.38 \pm 0.01$ & $10.42 \pm 0.02$ & $7.73 \pm 0.04$ & $8.64 \pm 0.06$ & $6.66 \pm 0.07$ & $6.76 \pm 0.02$ \\
6.81 & $8760 \pm 210$ & $9240 \pm 210$ & $10.39 \pm 0.01$ & $10.43 \pm 0.02$ & $7.72 \pm 0.04$ & $8.65 \pm 0.06$ & $6.61 \pm 0.07$ & $6.76 \pm 0.03$ \\
7.17 & $8790 \pm 240$ & $9290 \pm 200$ & $10.40 \pm 0.01$ & $10.44 \pm 0.02$ & $7.70 \pm 0.04$ & $8.73 \pm 0.04$ & $6.61 \pm 0.06$ & $6.75 \pm 0.03$ \\
7.53 & $8840 \pm 220$ & $9480 \pm 170$ & $10.42 \pm 0.01$ & $10.45 \pm 0.02$ & $7.70 \pm 0.03$ & $8.56 \pm 0.06$ & $6.57 \pm 0.06$ & $6.71 \pm 0.02$ \\
7.89 & $8830 \pm 250$ & $9500 \pm 180$ & $10.40 \pm 0.01$ & $10.45 \pm 0.01$ & $7.68 \pm 0.04$ & $8.57 \pm 0.06$ & $6.56 \pm 0.07$ & $6.71 \pm 0.02$ \\
8.25 & $8800 \pm 270$ & $9410 \pm 190$ & $10.41 \pm 0.01$ & $10.46 \pm 0.02$ & $7.71 \pm 0.04$ & $8.62 \pm 0.06$ & $6.58 \pm 0.07$ & $6.73 \pm 0.02$ \\
8.61 & $8830 \pm 220$ & $9360 \pm 200$ & $10.41 \pm 0.01$ & $10.46 \pm 0.02$ & $7.70 \pm 0.04$ & $8.64 \pm 0.05$ & $6.56 \pm 0.07$ & $6.75 \pm 0.02$ \\
8.98 & $8870 \pm 230$ & $9370 \pm 160$ & $10.42 \pm 0.02$ & $10.47 \pm 0.02$ & $7.70 \pm 0.03$ & $8.56 \pm 0.10$ & $6.53 \pm 0.06$ & $6.77 \pm 0.02$ \\
9.34 & $8920 \pm 210$ & $9220 \pm 180$ & $10.44 \pm 0.01$ & $10.48 \pm 0.02$ & $7.70 \pm 0.03$ & $8.61 \pm 0.06$ & $6.53 \pm 0.06$ & $6.80 \pm 0.02$ \\
9.70 & $8860 \pm 210$ & $9410 \pm 200$ & $10.45 \pm 0.01$ & $10.49 \pm 0.01$ & $7.70 \pm 0.04$ & $8.63 \pm 0.05$ & $6.52 \pm 0.06$ & $6.77 \pm 0.03$ \\
10.06 & $8800 \pm 210$ & $9570 \pm 160$ & $10.47 \pm 0.01$ & $10.51 \pm 0.01$ & $7.69 \pm 0.03$ & $8.60 \pm 0.05$ & $6.47 \pm 0.05$ & $6.74 \pm 0.02$ \\
10.42 & $8730 \pm 210$ & $9740 \pm 210$ & $10.49 \pm 0.01$ & $10.53 \pm 0.01$ & $7.69 \pm 0.04$ & $8.62 \pm 0.06$ & $6.41 \pm 0.07$ & $6.70 \pm 0.03$ \\
10.79 & $8850 \pm 210$ & $9900 \pm 200$ & $10.54 \pm 0.01$ & $10.59 \pm 0.01$ & $7.65 \pm 0.04$ & $8.56 \pm 0.07$ & $6.34 \pm 0.07$ & $6.68 \pm 0.03$ \\
11.15 & $8770 \pm 210$ & $9890 \pm 220$ & $10.59 \pm 0.01$ & $10.62 \pm 0.02$ & $7.65 \pm 0.04$ & $8.56 \pm 0.08$ & $6.27 \pm 0.07$ & $6.67 \pm 0.03$ \\
11.51 & $8820 \pm 210$ & $9680 \pm 210$ & $10.63 \pm 0.01$ & $10.65 \pm 0.01$ & $7.64 \pm 0.04$ & $8.77 \pm 0.11$ & $6.23 \pm 0.06$ & $6.72 \pm 0.02$ \\
11.87 & $8720 \pm 220$ & $9690 \pm 190$ & $10.68 \pm 0.01$ & $10.71 \pm 0.01$ & $7.64 \pm 0.04$ & $8.60 \pm 0.07$ & $6.24 \pm 0.06$ & $6.73 \pm 0.03$ \\
12.23 & $8740 \pm 200$ & $9840 \pm 200$ & $10.73 \pm 0.01$ & $10.75 \pm 0.02$ & $7.64 \pm 0.04$ & $8.75 \pm 0.06$ & $6.24 \pm 0.08$ & $6.72 \pm 0.02$ \\
12.59 & $8860 \pm 190$ & $9750 \pm 210$ & $10.76 \pm 0.01$ & $10.79 \pm 0.01$ & $7.60 \pm 0.03$ & $8.68 \pm 0.09$ & $6.18 \pm 0.06$ & $6.73 \pm 0.03$ \\
12.96 & $8980 \pm 200$ & $9720 \pm 190$ & $10.80 \pm 0.01$ & $10.82 \pm 0.01$ & $7.57 \pm 0.03$ & $8.41 \pm 0.13$ & $6.12 \pm 0.07$ & $6.74 \pm 0.02$ \\
13.32 & $8320 \pm 270$ & $9880 \pm 200$ & $10.82 \pm 0.01$ & $10.85 \pm 0.01$ & $7.53 \pm 0.03$ & $8.67 \pm 0.09$ & $6.07 \pm 0.05$ & $6.70 \pm 0.02$ \\
\enddata
\tablecomments{Abundances in units of 12+log(X$^{\text{n}+}$/H$^+$)}
\end{deluxetable*}

\begin{deluxetable*}{ccccccccccccc}
\tablecaption{Further ionic abundances along HH~204 as a function of the distance from the bowshock. Values derived from the red arm spectrum. \label{tab:big_detail_red2} }
\tablewidth{0pt}
\tablehead{
 Distance& & & & & & & & & \\
 (mpc) &   Cl$^{+}$& Cl$^{2+}$& Ar$^{2+}$ & Ca$^{+}$ & Cr$^{+}$& Fe$^{+}$& Fe$^{2+}$ & Ni$^{+}$& Ni$^{2+}$}
\startdata
0.29 & $4.93 \pm 0.05$ & $4.63 \pm 0.08$ & $5.66 \pm 0.05$ & $3.64 \pm 0.05$ & $4.57 \pm 0.06$ & $6.35 \pm 0.04$ & $6.46 \pm 0.06$  & $5.08 \pm 0.03$ & $5.05 \pm 0.07$\\
0.66 &$4.92 \pm 0.04$ & $4.83 \pm 0.08$ & $5.66 \pm 0.04$ & $3.66 \pm 0.05$ & $4.59 \pm 0.06$ & $6.35 \pm 0.04$ & $6.42 \pm 0.06$  & $5.09 \pm 0.03$ & $5.07 \pm 0.06$\\
1.02 &$4.94 \pm 0.03$ & $4.74 \pm 0.05$ & $5.70 \pm 0.04$ & $3.64 \pm 0.05$ & $4.71 \pm 0.05$ & $6.32 \pm 0.05$ & $6.49 \pm 0.05$  & $5.08 \pm 0.03$ & $5.10 \pm 0.05$\\
1.38 &$4.84 \pm 0.04$ & $4.67 \pm 0.06$ & $5.64 \pm 0.04$ & $3.47 \pm 0.06$ & $4.59 \pm 0.05$ & $6.27 \pm 0.05$ & $6.45 \pm 0.06$  & $5.00 \pm 0.03$ & $5.06 \pm 0.06$\\
1.74 &$4.80 \pm 0.04$ & $4.72 \pm 0.04$ & $5.59 \pm 0.03$ & $3.54 \pm 0.05$ & $4.49 \pm 0.04$ & $6.17 \pm 0.04$ & $6.40 \pm 0.05$  & $4.97 \pm 0.03$ & $5.00 \pm 0.06$\\
2.10 &$4.82 \pm 0.04$ & $4.86 \pm 0.05$ & $5.61 \pm 0.04$ & $3.46 \pm 0.04$ & $4.53 \pm 0.04$ & $6.23 \pm 0.05$ & $6.43 \pm 0.05$  & $5.00 \pm 0.03$ & $5.04 \pm 0.05$\\
2.46 &$4.84 \pm 0.04$ & $4.84 \pm 0.04$ & $5.60 \pm 0.03$ & $3.53 \pm 0.05$ & $4.55 \pm 0.04$ & $6.26 \pm 0.04$ & $6.42 \pm 0.04$  & $5.03 \pm 0.03$ & $5.07 \pm 0.05$\\
2.83 &$4.82 \pm 0.03$ & $4.83 \pm 0.04$ & $5.59 \pm 0.03$ & $3.51 \pm 0.04$ & $4.58 \pm 0.03$ & $6.28 \pm 0.03$ & $6.46 \pm 0.04$  & $5.02 \pm 0.03$ & $5.03 \pm 0.04$\\
3.19 &$4.83 \pm 0.03$ & $4.80 \pm 0.05$ & $5.58 \pm 0.03$ & $3.52 \pm 0.04$ & $4.59 \pm 0.04$ & $6.28 \pm 0.03$ & $6.49 \pm 0.05$  & $5.07 \pm 0.03$ & $5.08 \pm 0.05$\\
3.55 &$4.83 \pm 0.03$ & $4.71 \pm 0.05$ & $5.56 \pm 0.03$ & $3.57 \pm 0.04$ & $4.58 \pm 0.04$ & $6.28 \pm 0.03$ & $6.46 \pm 0.04$  & $5.07 \pm 0.02$ & $5.05 \pm 0.04$\\
3.91 &$4.83 \pm 0.03$ & $4.73 \pm 0.04$ & $5.54 \pm 0.04$ & $3.56 \pm 0.04$ & $4.53 \pm 0.04$ & $6.29 \pm 0.03$ & $6.47 \pm 0.05$  & $5.06 \pm 0.03$ & $5.06 \pm 0.05$\\
4.27 &$4.81 \pm 0.03$ & $4.69 \pm 0.05$ & $5.55 \pm 0.03$ & $3.54 \pm 0.04$ & $4.49 \pm 0.04$ & $6.30 \pm 0.03$ & $6.46 \pm 0.05$  & $5.05 \pm 0.03$ & $5.09 \pm 0.05$\\
4.64 &$4.81 \pm 0.03$ & $4.73 \pm 0.04$ & $5.56 \pm 0.03$ & $3.54 \pm 0.03$ & $4.44 \pm 0.03$ & $6.32 \pm 0.03$ & $6.49 \pm 0.04$  & $5.03 \pm 0.02$ & $5.11 \pm 0.04$\\
5.00 &$4.80 \pm 0.03$ & $4.71 \pm 0.04$ & $5.56 \pm 0.03$ & $3.55 \pm 0.04$ & $4.46 \pm 0.04$ & $6.32 \pm 0.04$ & $6.50 \pm 0.04$  & $5.02 \pm 0.03$ & $5.11 \pm 0.05$\\
5.36 &$4.76 \pm 0.04$ & $4.68 \pm 0.04$ & $5.53 \pm 0.03$ & $3.50 \pm 0.04$ & $4.41 \pm 0.03$ & $6.27 \pm 0.03$ & $6.46 \pm 0.04$  & $4.99 \pm 0.03$ & $5.07 \pm 0.04$\\
5.72 &$4.77 \pm 0.03$ & $4.75 \pm 0.04$ & $5.55 \pm 0.03$ & $3.54 \pm 0.04$ & $4.36 \pm 0.04$ & $6.27 \pm 0.03$ & $6.49 \pm 0.04$  & $5.00 \pm 0.02$ & $5.12 \pm 0.04$\\
6.08 &$4.76 \pm 0.03$ & $4.71 \pm 0.05$ & $5.57 \pm 0.03$ & $3.55 \pm 0.03$ & $4.37 \pm 0.04$ & $6.29 \pm 0.03$ & $6.51 \pm 0.04$  & $5.00 \pm 0.02$ & $5.14 \pm 0.04$\\
6.44 &$4.77 \pm 0.03$ & $4.76 \pm 0.04$ & $5.58 \pm 0.03$ & $3.56 \pm 0.04$ & $4.34 \pm 0.03$ & $6.27 \pm 0.03$ & $6.50 \pm 0.04$  & $4.99 \pm 0.02$ & $5.13 \pm 0.04$\\
6.81 &$4.75 \pm 0.03$ & $4.76 \pm 0.04$ & $5.60 \pm 0.03$ & $3.58 \pm 0.04$ & $4.30 \pm 0.04$ & $6.26 \pm 0.03$ & $6.51 \pm 0.04$  & $4.96 \pm 0.02$ & $5.12 \pm 0.04$\\
7.17 &$4.71 \pm 0.03$ & $4.79 \pm 0.04$ & $5.59 \pm 0.03$ & $3.55 \pm 0.04$ & $4.23 \pm 0.04$ & $6.20 \pm 0.04$ & $6.48 \pm 0.05$  & $4.92 \pm 0.03$ & $5.11 \pm 0.05$\\
7.53 &$4.69 \pm 0.03$ & $4.76 \pm 0.04$ & $5.60 \pm 0.03$ & $3.50 \pm 0.03$ & $4.18 \pm 0.03$ & $6.15 \pm 0.03$ & $6.49 \pm 0.04$  & $4.88 \pm 0.02$ & $5.12 \pm 0.04$\\
7.89 &$4.65 \pm 0.03$ & $4.71 \pm 0.04$ & $5.59 \pm 0.03$ & $3.52 \pm 0.03$ & $4.22 \pm 0.04$ & $6.12 \pm 0.04$ & $6.48 \pm 0.05$  & $4.87 \pm 0.02$ & $5.13 \pm 0.04$\\
8.25 &$4.67 \pm 0.03$ & $4.76 \pm 0.04$ & $5.61 \pm 0.03$ & $3.55 \pm 0.03$ & $4.18 \pm 0.04$ & $6.17 \pm 0.04$ & $6.52 \pm 0.05$  & $4.90 \pm 0.03$ & $5.15 \pm 0.04$\\
8.61 &$4.72 \pm 0.03$ & $4.77 \pm 0.03$ & $5.61 \pm 0.03$ & $3.58 \pm 0.04$ & $4.25 \pm 0.04$ & $6.20 \pm 0.03$ & $6.50 \pm 0.04$  & $4.92 \pm 0.02$ & $5.14 \pm 0.04$\\
8.98 &$4.75 \pm 0.04$ & $4.74 \pm 0.03$ & $5.61 \pm 0.03$ & $3.59 \pm 0.04$ & $4.23 \pm 0.04$ & $6.21 \pm 0.04$ & $6.51 \pm 0.04$  & $4.92 \pm 0.02$ & $5.18 \pm 0.04$\\
9.34 &$4.74 \pm 0.03$ & $4.81 \pm 0.03$ & $5.61 \pm 0.03$ & $3.60 \pm 0.03$ & $4.21 \pm 0.03$ & $6.21 \pm 0.04$ & $6.49 \pm 0.04$  & $4.92 \pm 0.02$ & $5.16 \pm 0.04$\\
9.70 &$4.70 \pm 0.03$ & $4.79 \pm 0.04$ & $5.63 \pm 0.03$ & $3.56 \pm 0.03$ & $4.20 \pm 0.04$ & $6.19 \pm 0.03$ & $6.51 \pm 0.04$  & $4.90 \pm 0.02$ & $5.17 \pm 0.04$\\
10.06 &$4.67 \pm 0.03$ & $4.78 \pm 0.03$ & $5.64 \pm 0.03$ & $3.53 \pm 0.04$ & $4.13 \pm 0.04$ & $6.13 \pm 0.03$ & $6.52 \pm 0.04$  & $4.84 \pm 0.02$ & $5.18 \pm 0.04$\\
10.42 &$4.58 \pm 0.04$ & $4.76 \pm 0.04$ & $5.65 \pm 0.03$ & $3.48 \pm 0.04$ & $3.98 \pm 0.04$ & $6.01 \pm 0.03$ & $6.54 \pm 0.05$  & $4.75 \pm 0.02$ & $5.15 \pm 0.04$\\
10.79 &$4.48 \pm 0.03$ & $4.68 \pm 0.04$ & $5.66 \pm 0.03$ & $3.43 \pm 0.04$ & $3.72 \pm 0.07$ & $5.82 \pm 0.03$ & $6.51 \pm 0.04$  & $4.64 \pm 0.02$ & $5.12 \pm 0.04$\\
11.15 &$4.44 \pm 0.03$ & $4.74 \pm 0.04$ & $5.72 \pm 0.03$ & $3.39 \pm 0.04$ & $3.66 \pm 0.09$ & $5.72 \pm 0.05$ & $6.53 \pm 0.04$  & $4.52 \pm 0.03$ & $5.16 \pm 0.04$\\
11.51 &$4.42 \pm 0.04$ & $4.80 \pm 0.03$ & $5.76 \pm 0.03$ & $3.34 \pm 0.04$ & $3.47 \pm 0.10$ & $5.65 \pm 0.05$ & $6.52 \pm 0.05$  & $4.45 \pm 0.03$ & $5.18 \pm 0.04$\\
11.87 &$4.42 \pm 0.04$ & $4.81 \pm 0.03$ & $5.82 \pm 0.03$ & $3.24 \pm 0.05$ & - & $5.58 \pm 0.05$ & $6.53 \pm 0.05$  & $4.38 \pm 0.03$ & $5.17 \pm 0.04$\\
12.23 &$4.37 \pm 0.03$ & $4.78 \pm 0.04$ & $5.86 \pm 0.04$ & $3.26 \pm 0.05$ & - & $5.62 \pm 0.05$ & $6.52 \pm 0.05$  & $4.34 \pm 0.02$ & $5.17 \pm 0.04$\\
12.59 &$4.38 \pm 0.04$ & $4.76 \pm 0.04$ & $5.88 \pm 0.03$ & $3.26 \pm 0.04$ & $3.50 \pm 0.12$ & $5.54 \pm 0.05$ & $6.52 \pm 0.03$  & $4.28 \pm 0.02$ & $5.17 \pm 0.04$\\
12.96 &$4.33 \pm 0.12$ & $4.82 \pm 0.04$ & $5.90 \pm 0.03$ & $3.20 \pm 0.05$ & $3.43 \pm 0.13$ & $5.58 \pm 0.05$ & $6.49 \pm 0.04$  & $4.24 \pm 0.03$ & $5.13 \pm 0.04$\\
13.32 &$4.29 \pm 0.04$ & $4.73 \pm 0.04$ & $5.93 \pm 0.02$ & $3.22 \pm 0.05$ & - & $5.42 \pm 0.07$ & $6.47 \pm 0.03$  & $4.23 \pm 0.03$ & $5.14 \pm 0.04$\\
\enddata
\tablecomments{Abundances in units of 12+log(X$^{\text{n}+}$/H$^+$)}
\end{deluxetable*}

%% This command is needed to show the entire author+affiliation list when
%% the collaboration and author truncation commands are used.  It has to
%% go at the end of the manuscript.
%\allauthors

%% Include this line if you are using the \added, \replaced, \deleted
%% commands to see a summary list of all changes at the end of the article.
%\listofchanges

\end{document}